\documentclass[12pt]{sourcebook}
\usepackage{graphicx}
\usepackage{epsfig}
\usepackage{fancyhdr}
\usepackage{multind}

\newcommand{\beq}{\begin{equation}}
\newcommand{\eeq}[1]{\label{#1}\end{equation}}
\newcommand{\eeqn}{\end{equation}}
\newenvironment{Eqnarray}{\arraycolsep 0.14em\begin{eqnarray}}{\end{eqnarray}}
\newcommand{\beqa}{\begin{Eqnarray}}
\newcommand{\eeqa}[1]{\label{#1}\end{Eqnarray}}
\newcommand{\eeqan}{\end{Eqnarray}}

\newcommand{\leqn}[1]{(\ref{#1})}
\renewcommand{\bar}[1]{\overline{#1}}
\newcommand{\etal}{{\it et al.}}
\newcommand{\ie}{{\it i.e.}}
\newcommand{\eg}{{\it e.g.}}
\newcommand{\etc}{{\it etc.}}

\newcommand{\lsim}{\mathrel{\raise.3ex\hbox{$<$\kern-.75em\lower1ex\hbox{$\sim$}}}}
\newcommand{\gsim}{\mathrel{\raise.3ex\hbox{$>$\kern-.75em\lower1ex\hbox{$\sim$}}}}
\newcommand{\A}{{\cal A}}

\newcommand{\half}{\frac{1}{2}}

\newcommand{\ee}{e^+e^-}
\newcommand{\sstw}{\sin^2\theta_w}

\newcommand{\mz}{m_Z}

\newcommand{\msb}{{\bar{\scriptscriptstyle M \kern -1pt S}}}

\newcommand{\ch}[1]{\widetilde\chi^+_{#1}}

\newcommand{\neu}[1]{\widetilde\chi^0_{#1}}
\newcommand{\s}[1]{\widetilde{#1}}

\makeatletter
\def\section{\@startsection{section}{0}{\z@}{5.5ex plus .5ex minus
 1.5ex}{2.3ex plus .2ex}{\large\bf}}
\def\subsection{\@startsection{subsection}{1}{\z@}{3.5ex plus .5ex minus
 1.5ex}{1.3ex plus .2ex}{\normalsize\bf}}
\def\subsubsection{\@startsection{subsubsection}{2}{\z@}{-3.5ex plus
-1ex minus  -.2ex}{2.3ex plus .2ex}{\normalsize\sl}}

\renewcommand{\@makecaption}[2]{%
   \vskip 10pt
   \setbox\@tempboxa\hbox{\small #1: #2}
   \ifdim \wd\@tempboxa >\hsize     % IF longer than one line:
       \small #1: #2\par          %   THEN set as ordinary paragraph.
     \else                        %   ELSE  center.
       \hbox to\hsize{\hfil\box\@tempboxa\hfil}
   \fi}

 \def\citenum#1{{\def\@cite##1##2{##1}\cite{#1}}}
\def\citea#1{\@cite{#1}{}}
 
\newcount\@tempcntc
\def\@citex[#1]#2{\if@filesw\immediate\write\@auxout{\string\citation{#2}}\fi
  \@tempcnta\z@\@tempcntb\m@ne\def\@citea{}\@cite{\@for\@citeb:=#2\do
    {\@ifundefined
       {b@\@citeb}{\@citeo\@tempcntb\m@ne\@citea\def\@citea{,}{\bf ?}\@warning
       {Citation `\@citeb' on page \thepage \space undefined}}%
    {\setbox\z@\hbox{\global\@tempcntc0\csname b@\@citeb\endcsname\relax}%
     \ifnum\@tempcntc=\z@ \@citeo\@tempcntb\m@ne
       \@citea\def\@citea{,}\hbox{\csname b@\@citeb\endcsname}%
     \else
      \advance\@tempcntb\@ne
      \ifnum\@tempcntb=\@tempcntc
      \else\advance\@tempcntb\m@ne\@citeo
      \@tempcnta\@tempcntc\@tempcntb\@tempcntc\fi\fi}}\@citeo}{#1}}
\def\@citeo{\ifnum\@tempcnta>\@tempcntb\else\@citea\def\@citea{,}%
  \ifnum\@tempcnta=\@tempcntb\the\@tempcnta\else
  {\advance\@tempcnta\@ne\ifnum\@tempcnta=\@tempcntb \else\def\@citea{--}\fi
    \advance\@tempcnta\m@ne\the\@tempcnta\@citea\the\@tempcntb}\fi\fi}
\makeatother

\newcommand{\tev}{{\rm\,TeV}}

\newcommand{\gev}{{\rm\,GeV}}

\newcommand{\peff}  {{\cal P}_{\rm{eff}}}

\newcommand{\BC}{\begin{center}}
\newcommand{\EC}{\end{center}}

\newcommand{\alr}{$A_{LR} \,$}
\newcommand{\alrm}{A_{LR}}
\newcommand{\pelum}{$P_e^{\rm lum} \,$}                    
\newcommand{\slashchar}[1]{\setbox0=\hbox{$#1$}           
  \dimen0=\wd0                 
 \setbox1=\hbox{/} \dimen1=\wd1               
 \ifdim\dimen0>\dimen1                        
 \rlap{\hbox to \dimen0{\hfil/\hfil}}      
 #1                                        
 \else                                        
 \rlap{\hbox to \dimen1{\hfil$#1$\hfil}}   
 /                                         
 \fi}                    
\newcommand{\eml}{e^-_L}
\newcommand{\emr}{e^-_R}
\newcommand{\epl}{e^+_L}
\newcommand{\epr}{e^+_R}
\newcommand{\pem}{{\cal P}_{-}}
\newcommand{\pep}{{\cal P}_{+}}
\newcommand{\st}   {{\widetilde t}}
\newcommand{\smu}{\widetilde \mu}
\newcommand{\smul}{\widetilde \mu_L}
\newcommand{\smur}{\widetilde \mu_R}
\newcommand{\stau}{\widetilde \tau}
\newcommand{\sel}{\widetilde e_L}
\newcommand{\ser}{\widetilde e_R}
\newcommand{\nino}{\widetilde \chi^0}
\newcommand{\cinop}{\widetilde \chi^+}
\newcommand{\cinom}{\widetilde \chi^-}

\newcommand{\bit}{\begin{itemize}}         
\newcommand{\eit}{\end{itemize}}

\newcommand{\ifb}{{\rm fb}^{-1}}

\def\D0{D\O\ \hskip-0.5mm}

\def\vs{{\it vs.}}
\def\sci#1#2{#1\times10^{#2}}
\def\sci#1#2{#1\times10^{#2}}

\begin{document}

\pagestyle{fancy}
\thispagestyle{empty}

\setcounter{footnote}{0}
\renewcommand{\thefootnote}{\fnsymbol{footnote}}

\begin{flushright}
{\small
  BNL--52627, 
           CLNS 01/1729, 
           FERMILAB--Pub--01/058-E,  \\
           LBNL--47813,  
           SLAC--R--570,  
           UCRL--ID--143810--DR\\   
             LC--REV--2001--074--US \\
            hep-ex/0106058\\
             June 2001}  
\end{flushright}

\bigskip
\begin{center}
{\bf\LARGE
 Linear Collider Physics Resource Book\\[1ex] for Snowmass 2001\\[4ex]
Part 4: Theoretical, Accelerator, \\[1.5ex]
and Experimental Options}
\\[6ex]
{\it American Linear Collider Working Group}
\footnote{Work supported in part by the US Department of Energy under
contracts DE--AC02--76CH03000,
DE--AC02--98CH10886, DE--AC03--76SF00098, DE--AC03--76SF00515, and
W--7405--ENG--048, and by the National Science Foundation under
contract PHY-9809799.}
\medskip
\end{center}

\vfill

\begin{center}
{\bf\large
Abstract }
\end{center}

This Resource Book reviews the physics opportunities of a next-generation
$e^+e^-$ linear collider and discusses options for the experimental program.
Part 4 discusses options for the linear collider program, at a number of
levels.  First, it presents a broad review of physics beyond the
Standard Model, indicating how the linear collider is relevant to each
possible pathway.  Next, it surveys options for the accelerator and
experimental plan, including the questions of the running scenario, the
issue of one or two interaction regions, and the options for positron
polarization, photon-photon collisions, and $e^-e^-$ collisions.  Finally,
it reviews the detector design issues for the linear collider and presents
three possible detector designs.

\vfill
\vfill

\newpage
\emptyheads
\blankpage \thispagestyle{empty}
\fancyheads
\emptyheads

 \frontmatter\setcounter{page}{1}

\hbox to\hsize{\null}
\thispagestyle{empty}
\vfill
\begin{figure}[hp]
\begin{center}
\epsfig{file=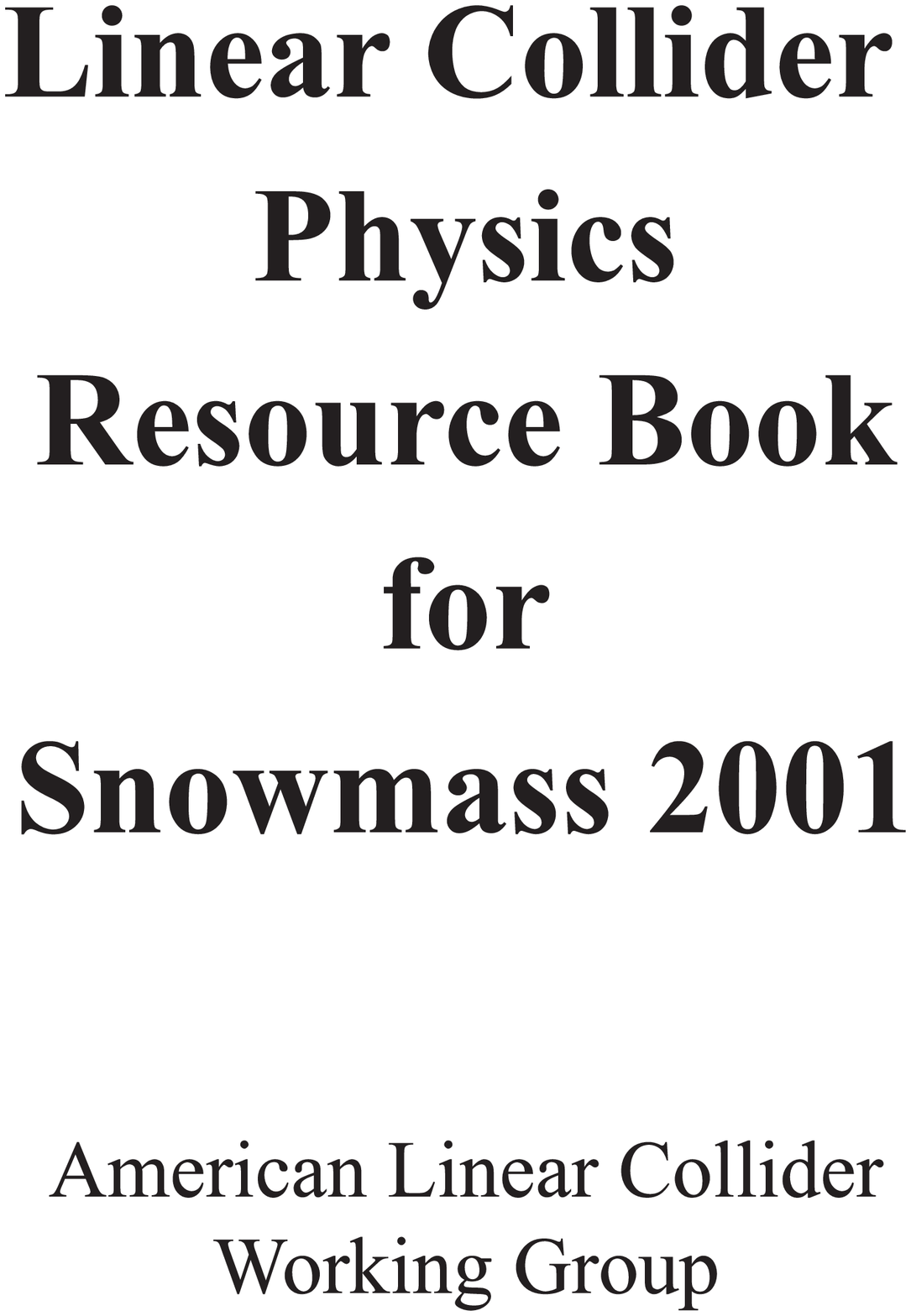,height=6.5in}
\end{center}
\end{figure}
\vfill
\newpage

\begin{center} 
           BNL--52627, 
           CLNS 01/1729, 
           FERMILAB--Pub--01/058-E,  \\
           LBNL--47813,  
           SLAC--R--570,  
           UCRL--ID--143810--DR\\[2ex]   
            LC--REV--2001--074--US \\[2ex]
            June 2001  
\end{center}

\vfill

\begin{center}
\fbox{\parbox{5in}
{This document, and the material and data contained therein, was developed
under sponsorship of the United States Government.  Neither the United
States nor the Department of Energy, nor the Leland Stanford Junior University, 
nor their employees, nor their respective contractors, subcontractors, or
their employees, makes any warranty, express or implied, or assumes any
liability of responsibility for accuracy, completeness or usefulness of any
information, apparatus, product or process disclosed, or represents that its
use will not infringe privately owned rights.  Mention of any product, its
manufacturer, or suppliers shall not, nor is intended to imply approval,
disapproval, or fitness for any particular use.  A royalty-free,
nonexclusive right to use and disseminate same for any purpose
whatsoever, is expressly reserved to the United States and the
University.
}}\end{center}

\vfill

\noindent 
Cover:  Events of $\ee \to Z^0 h^0$, simulated with the Large linear collider
detector \hfill \break described in Chapter 15.
Front cover:  $h^0 \to \tau^+\tau^-$, $Z^0 \to b\bar b$.
Back cover:  $h^0 \to b\bar b$, $Z^0 \to \mu^+\mu^-$.

\vfill

\noindent
Typset in \LaTeX\ by S. Jensen.

\vfill
\noindent
Prepared for the Department of Energy under contract number DE--AC03--76SF00515
by Stanford Linear Accelerator Center, Stanford University, Stanford, California.
Printed in the United State of America.  Available from National Technical
Information Services, US Department of Commerce, 5285 Port Royal Road,
Springfield, Virginia 22161.

  \thispagestyle{empty}

  \begin{center}
{\Large American Linear Collider Working Group}
\end{center}
 
\bigskip\bigskip
\begin{center}

T. Abe$^{52}$,
N.~Arkani-Hamed$^{29}$,
D.~Asner$^{30}$,
H.~Baer$^{22}$,
J.~Bagger$^{26}$,    
C.~Balazs$^{23}$,
C.~Baltay$^{59}$,           
T.~Barker$^{16}$,
T.~Barklow$^{52}$,   
J.~Barron$^{16}$
% ...

\end{center}

\newpage

\centerline{$^{1}$ Argonne National Laboratory, Argonne, IL 60439}
\centerline{$^{2}$ Universitat Autonoma de Barcelona, E-08193 Bellaterra,Spain}
\centerline{$^{3}$ University of British Columbia, Vancouver, BC V6T 1Z1, Canada}
\centerline{$^{4}$ Brookhaven National Laboratory, Upton, NY 11973}
\centerline{$^{5}$ Budker INP, RU-630090 Novosibirsk, Russia}
\centerline{$^{6}$ University of California, Davis, CA 95616}
\centerline{$^{7}$ University of California, Riverside, CA 92521}
\centerline{$^{8}$ University of California at San Diego, La Jolla, CA  92093}
\centerline{$^{9}$ University of California, Santa Cruz, CA 95064}
\centerline{$^{10}$ California Institute of Technology, Pasadena, CA 91125}
\centerline{$^{11}$ Carleton University, Ottawa, ON K1S 5B6, Canada}
\centerline{$^{12}$ Universidad Catolica de Chile, Chile}
\centerline{$^{13}$ CERN, CH-1211 Geneva 23, Switzerland}
\centerline{$^{14}$ University of Chicago, Chicago, IL 60637}
\centerline{$^{15}$ Chonbuk National University, Chonju 561-756, Korea}
\centerline{$^{16}$ University of Colorado, Boulder, CO 80309}
\centerline{$^{17}$ Cornell University, Ithaca, NY  14853}
\centerline{$^{18}$ DESY, D-22063 Hamburg, Germany}
\centerline{$^{19}$ DESY, D-15738 Zeuthen, Germany}
\centerline{$^{20}$ Fairfield University, Fairfield, CT 06430}
\centerline{$^{21}$ Fermi National Accelerator Laboratory, Batavia, IL 60510}
\centerline{$^{22}$ Florida State University, Tallahassee, FL 32306}
\centerline{$^{23}$ University of Hawaii, Honolulu, HI 96822}
\centerline{$^{24}$ Indian Institute of Science, Bangalore, 560 012, India}
\centerline{$^{25}$ Indiana University, Bloomington, IN 47405}
\centerline{$^{26}$ Johns Hopkins University, Baltimore, MD 21218}
\centerline{$^{27}$ Kansas State University, Manhattan, KS 66506}
\centerline{$^{28}$ Kyoto University,  Kyoto 606, Japan}
\centerline{$^{29}$ Lawrence Berkeley National Laboratory, Berkeley, CA 94720}
\centerline{$^{30}$ Lawrence Livermore National Laboratory, Livermore, CA 94551}
\centerline{$^{31}$ Universit\"at Leipzig, D-04109 Leipzig, Germany}
\centerline{$^{32}$ Ludwigs-Maximilians-Universit\"at, M\"unchen, Germany}
\centerline{$^{32a}$ Manchester University, Manchester M13~9PL, UK}
\centerline{$^{33}$ Centre de Physique Theorique, CNRS, F-13288 Marseille, France}
\centerline{$^{34}$ University of Massachusetts, Amherst, MA 01003}
\centerline{$^{35}$ Massachussetts Institute of Technology, Cambridge, MA 02139}
\centerline{$^{36}$ Max-Planck-Institut f\"ur Physik, M\"unchen, Germany}
\centerline{$^{37}$ University of Michigan, Ann Arbor MI 48109}
\centerline{$^{38}$ State University of New York, Buffalo, NY 14260}
\centerline{$^{39}$ State University of New York, Stony Brook, NY 11794}
\centerline{$^{40}$ Northern Illinois University, DeKalb, IL 60115}
\centerline{$^{41}$ University of Notre Dame, Notre Dame, IN 46556}
\centerline{$^{42}$ University of Oregon, Eugene, OR 97403}
\centerline{$^{43}$ Oxford University, Oxford OX1 3RH, UK}
\centerline{$^{44}$ Paul Scherrer Institut, CH-5232 Villigen PSI, Switzerland}
\centerline{$^{45}$ University of Pennsylvania, Philadelphia, PA 19104}
\centerline{$^{46}$ Pennsylvania State University, Mont Alto, PA 17237}
\centerline{$^{47}$ Perkins-Elmer Bioscience, Foster City, CA 94404}
\centerline{$^{48}$ Purdue University, West Lafayette, IN 47907}
\centerline{$^{49}$ RWTH Aachen, D-52056 Aachen, Germany}
\centerline{$^{50}$ University of Rochester, Rochester, NY 14627}
\centerline{$^{51}$ Southern Methodist University, Dallas, TX 75275}
\centerline{$^{52}$ Stanford Linear Accelerator Center, Stanford, CA 94309}
\centerline{$^{53}$ Texas A\&M University, College Station, TX 77843}
\centerline{$^{54}$ Tokoku University, Sendai 980, Japan}
\centerline{$^{55}$ University of Wisconsin, Madison, WI  53706}
\centerline{$^{57}$ Uniwersytet Warszawski, 00681 Warsaw, Poland}
\centerline{$^{57}$ Wayne State University, Detroit, MI 48202}
\centerline{$^{58}$ Universit\"at W\"urzburg, W\"urzburg 97074, Germany}
\centerline{$^{59}$ Yale University, New Haven, CT 06520}

\vfill

\noindent
Work supported in part by the US Department of Energy under
contracts DE--AC02--76CH03000,
DE--AC02--98CH10886, DE--AC03--76SF00098, DE--AC03--76SF00515, and
W--7405--ENG--048, and by the National Science Foundation under
contract PHY-9809799.

  \blankpage  \thispagestyle{empty}

\setcounter{chapter}{6}

\setcounter{page}{297} \thispagestyle{empty}
\renewcommand{\thepage}{\arabic{page}}

\emptyheads

\begin{center}\begin{large}{\Huge \sffamily Pathways Beyond the Standard Model} 
\end{large}\end{center}

\blankpage \thispagestyle{empty}

\setcounter{chapter}{8}

\fancyheads

\chapter{Pathways Beyond the Standard Model}
\fancyhead[RO]{Pathways Beyond the Standard Model}

\section{Introduction}

Over the past 30 years or so, high energy physics experiments
have  systematically explored the behavior of the strong,
electromagnetic and weak interactions.   For the strong interactions, 
QCD is generally accepted as the correct 
description, and research on QCD has shifted to 
its application to special regimes such as diffractive and
exclusive processes and the quark-gluon plasma.   For the electromagnetic 
and weak interactions, the progress of the past decade on $W$, $Z$, top, 
and neutrino physics has demonstrated that  their structure is understood
with high precision.

Our current picture of the electroweak interactions requires spontaneous
gauge symmetry breaking.  As yet, there is no direct evidence on the 
means by which the gauge symmetry is broken.  It is remarkable that all 
of the evidence accumulated to date is consistent with the Standard Model
(SM) in which this symmetry breaking is due to a single elementary scalar field,
the Higgs field, 
which generates the masses of the $W$ and $Z$ bosons and the quarks
and leptons.  

However, many features of this simple theory are inadequate.
The Higgs field is an {\em ad hoc} addition to the 
SM.  Its mass and symmetry-breaking expectation value are put in by hand.
The quark and lepton masses are generated by arbitrary couplings to the 
Higgs field.  The existence of three generations of quarks and 
leptons is not explained, nor is the dramatic lack of symmetry
in the masses and mixings of these generations.    

To explain these features, it is necessary to extend the SM.  These 
extensions, in turn, predict new particles and phenomena.  The 
compelling motivation for new experiments at the highest energies is
to discover these phenomena and then to decipher them, so that we can
learn 
the nature of the new laws of physics with which they are associated.

In this document, we are exploring the physics case for a next-generation
$\ee$ linear collider.  To make this case, it is necessary to 
demonstrate that the linear collider can have an important impact on 
our understanding of  these new phenomena.  The argument should
be made broadly  for models of new physics covering the 
whole range of possibilities allowed from  our current knowledge.
It should take into account new information that we will learn from the 
Tevatron and LHC experiments  which will be done before the linear
collider is completed.

Our purpose in this chapter is to give an overview of possibilities 
for new physics beyond the SM.  Our emphasis will be on general orientation
to the pathways that one might follow.
  We will then explain the relevance of the linear collider
measurements to each possible scenario.  We encourage  the reader to
consult  the relevant chapter of the `Sourcebook',
Chapters 3--8, to see
how each quantity we discuss is measured at a linear collider
and why the experimental precision that we expect is justified.

The essay is organized as follows:  In Section 2, we discuss the 
general principle that we use to organize models of new physics.  In 
Sections 3 and 4, we discuss models of new physics in the typical dichotomy
used since the 1980's: on the one hand, models with supersymmetry, on the
other hand, models with new strong interactions at the TeV scale.  In
Section 5, we discuss a new class of models for which the key ingredient 
is the existence of extra spatial dimensions.  It is now understood that these
models stand on the same footing as the more traditional schemes and, in 
fact, address certain of their weaknesses.  Section 6 gives some 
conclusions.

\section{Beyond the Standard Model}

We first discuss some general principles regarding
physics beyond the Standard Model.

From an experimental point of view, it is necessary to study the
interactions of the observed particles at higher energies and with
higher accuracy. This may lead to the discovery of new particles, in
which case we need to study their spectrum and determine their
interactions. Alternatively, it may lead to the observation of
anomalous properties of the observed particles, in which case we could
infer the existence of new particles or phenomena responsible for
these effects.  After this information is obtained in experiments, we
must attempt to reconstruct the structure of the underlying theory.
The linear collider is a crucial complement to the LHC in ensuring
that the experimental information is extensive and precise enough for
this goal to be achieved.

From the theoretical point of view, different ideas lead to models that
provide challenges to this experimental program.  To discuss the range of 
possible models, an organizing principle is needed.  We will organize
our discussion around the major question that we believe most strongly 
motivates new physics at  the TeV scale.  This is the {\em stability crisis}
in the SM explanation for electroweak symmetry breaking.  In technical terms,
this is the problem that the Higgs boson mass is extremely sensitive to 
physics at very high energy scales.  In the SM, the effect of 
quantum fields at the energy scale $M$ is an additive contribution to the 
Higgs boson mass term of order $M^2$.  More physically, this is the 
problem that not only the magnitude but even the sign of the Higgs
boson mass term is not predicted in the SM, so that the SM cannot explain
{\em why} the electroweak gauge symmetry is broken.  From either perspective,
this problem suggests that the SM is a dramatically incomplete picture of
electroweak symmetry breaking.  It is for this reason that we  believe
that new physics must appear at the TeV scale.  We expect that the physics will be more 
exciting than simply the production of some random new particles.  The 
solution of the stability crisis will involve completely new principles 
of physics.  These principles will be reflected in the spectrum and 
properties of the new particles, and in their interactions.
Much as the discovery of the $J/\psi$ convincingly brought together many
different elements of the SM in a coherent picture, so the
discovery and study of these new states will spur us on to the
construction of a new theory that will displace the SM.

We will use the idea of solving the stability crisis 
 to guide our classification of the various models
of new physics. The three approaches to this problem that have received
the most study are supersymmetry, strongly coupled theories, and extra 
dimensions.   The common theme in all three proposed solutions is that
additional particle states and dynamics must be present near the
electroweak scale.  We  briefly describe each approach, 
summarizing in each case the types of new interactions expected and the 
key experimental issues they raise.

Each possible model of new physics must be approached from the viewpoint
expressed at the beginning of this section, that of dissecting 
experimentally the spectrum of new particles and their interactions.
We take particular note of the  important strengths that 
the linear collider
brings to disentangling the physics of these models.
We will see that, in most cases, the linear collider not only contributes
but is {\em essential} to forming this experimental picture.  
Even if none of the specific models we  discuss here is 
actually realized in Nature,  this exercise illustrates
the importance of the linear collider in unraveling the new
world beyond the SM.

%%%%%%%%%%%%%%%%%%%%%%%%%%%%%%
\section{Supersymmetry}

One attempt to cure the stability crisis of the Higgs field
is to introduce a new symmetry---supersymmetry---which relates
fermions and bosons.  To realize this symmetry in Nature, there
must exist supersymmetry partners for each of the known SM particles.
Further, supersymmetry must be broken in the ground state so
that these superpartners are more massive than ordinary particles.  
The Higgs mass terms are then not sensitive to mass scales above the 
superpartner masses.  The Higgs field vacuum expectation value is 
naturally of order 100 GeV if the superpartner masses are also near
this energy scale.

The existence of superpartners implies a rich program for future
accelerators.  The phenomenology of supersymmetry has been studied
in great detail in the literature.  Dozens of papers have been written
on the technical ability of linear collider experiments to discover and study
supersymmetric theories of many different forms.  This material is 
reviewed systematically in Chapter 4 of this book. Different patterns
of supersymmetry breaking masses can yield substantially different
phenomenology at a high-energy collider.
Supersymmetry is not a dot on the theoretical landscape, but rather 
contains a tremendously varied range of possibilities to
be searched for and studied at all available high-energy collider facilities.

In the remainder of this section, we summarize the most important
issues for the study of supersymmetry and the relevant measurements that
can be done at a linear collider.  It is important to keep in mind that
we are likely to be surprised with the spectrum that Nature
ultimately gives us.  The linear collider's ability to cleanly disentangle
the superpartner mass spectrum and couplings would be extremely important
when the surprises occur.  Of course, this is relevant only if the 
linear collider has sufficiently high center-of-mass energy to 
produce the superpartners.  Section~2 of Chapter~4 reviews the 
expectations for the masses of superpartners and gives estimates of
what center-of-mass energies should be required.

\vspace{0.1in}

{\it Mass measurements of accessible sparticles.}
If supersymmetry is
  relevant for electroweak symmetry breaking, then some of the
  superpartners should  be discovered at the LHC.  Furthermore,
  the experiments at the LHC should be able to accurately measure
      some masses or mass differences of the SUSY spectrum.
   This issue is reviewed in Chapter 4, Section 7.  However, 
the systematic measurement of the SUSY spectrum 
   requires a linear collider.

Superpartner masses are measured at a linear collider
in three main ways: from distributions
of the products of an on-shell superpartner decay, from threshold scans,
and from contributions of virtual superpartners to cross sections or 
decay amplitudes.  When
sleptons, charginos, and neutralinos are produced on-shell,
their masses will typically be measured to within about 1\%.  Even  if
the lightest
neutralino LSP is not directly observed, its mass
 should be measurable to within 1\% from these kinematic distributions.
Threshold scans of 
sleptons in $e^+e^-$ collisions and especially in $e^-e^-$ collisions
may yield mass measurements to within one part in a thousand.  Indirect
off-shell mass measurements are more model-dependent but have power
in specific applications. For example, the  $t$-channel
sneutrino contribution to chargino pair production may allow the 
presence of the  sneutrino
to be deduced when its mass is as high as twice the 
center-of-mass energy of the collider.  These techniques are reviewed in 
more detail in Chapter 4, Section 3.
\vspace{0.1in}

{\it Slepton and squark quantum numbers and mixing angles.}
When sparticle mixing can be ignored, the cross sections for pair production
of squarks and sleptons at a linear collider are precisely determined by 
the SM quantum numbers.  This should allow unambiguous checks of the quantum
numbers and spins for sparticles of the first two generations.  In particular,
it is straightforward to distinguish the superpartners of left- and 
right-handed species (\eg,  $\tilde e_L$ from  $\tilde e_R$)  by cross section
measurements with polarized beams.
Third-generation sleptons and squarks are likely to be the most
strongly mixed scalars of supersymmetry, forming mass eigenstates
$\tilde \tau_{1,2}$, $\tilde b_{1,2}$, and $\tilde t_{1,2}$. 
Separation of these eigenstates and accurate 
measurement of their masses are difficult at the Tevatron
and LHC but present no extraordinary problems to a linear collider.
By combining direct
mass measurements with polarization asymmetries for the production
of these sparticles, we can determine the mixing angle needed to form the 
observed mass eigenstates from the left- and right-handed
weak-interaction eigenstates.  The uncertainty in this
determination depends on the parameters of the theory, but it has been
demonstrated for some cases that the error is lower than 1\%.
\vspace{0.1in}

{\it Chargino/neutralino parameters.}
The neutralino and chargino states may be strongly mixed combinations of 
gauge boson and Higgs boson superpartners.  The mass matrix is determined
by four parameters of the underlying Lagrangian:  
$M_1$ (bino mass), $M_2$ (wino mass), $\mu$ 
(supersymmetric higgsino mass) and $\tan\beta$ 
(ratio of Higgs vacuum expectation values).  
Precision measurements of masses, mixing angles, and couplings
 associated with chargino and neutralino 
production can supply  the information to determine these four important
underlying parameters of supersymmetry.  For example, 
measurements of chargino production alone can, in some cases, determine
$\tan\beta$ to better than $10\%$ with only 100 fb${}^{-1}$
of data. The parameters $M_1$, $M_2$, and possibly $\mu$ can be determined
at the percent level in large portions of the accessible supersymmetry
parameter space. 
\vspace{0.1in}

{\it Coupling relations.}
To establish supersymmetry as a principle of Nature, it is important to 
verify some of the symmetry relations that that principle predicts.
An essential consequence of supersymmetry is that the couplings of
sparticles to gauginos are equal to the corresponding couplings of 
particles to gauge bosons.  It has been demonstrated that this 
equality can be tested at a linear collider
to levels better than 1\% for weakly interacting sparticles.  The
precision is sufficiently good that one can even contemplate measuring the tiny
deviations from coupling equivalence that are caused by  supersymmetry-breaking
effects in loop corrections.  This can give an estimate of the masses
of unobserved sparticles with mass well above the collider energy, in the same
way that the current precision measurements predict the mass of the Higgs.
This issue is reviewed in Chapter 4, Section 4.
\vspace{0.1in}

{\it CP violating phases.}
The SM apparently does not have enough CP violation to
account for the baryon asymmetry in the universe.  Supersymmetry has
parameters that may introduce additional sources of CP violation into
the theory.  Testing for the existence of such phases would be 
an important part of a full supersymmetry program.  It has been shown
that the linear collider can determine evidence for additional non-zero 
CP-violating phases in supersymmetric theories if the phases are large
enough ($\phi_i\sim 0.1$), even accounting for the constraints
from electric dipole moment measurements.
\vspace{0.1in}

{\it Lepton number violation.}
Recent data suggest that neutrinos have non-zero masses and mixings.
This implies that non-zero lepton flavor angles
should be present for leptons, in parallel with the CKM angles for 
the quarks.
These rotation angles are difficult to measure using high-energy leptons
because
neutrinos are invisible and are summed over in most observables.  However,
these angles could be detected from superpartner decays, such as
$\tilde \mu^+\tilde\mu^-\to e^+\mu^-\tilde \chi_1^0\tilde\chi_1^0$.
A linear collider can use these measurements to probe the lepton 
flavor angles with greater sensitivity than any  existing experiment
in some parts of parameter space.
\vspace{0.1in}

{\it Complete spectrum.}
The LHC will be a wonderful machine for the discovery of many supersymmetric
sparticles in large regions of parameter space.  The linear collider
can add to the superpartner discoveries at the LHC by detecting states
that are not straightforward to observe in the $pp$ environment.
The discovery abilities of the linear collider begin
to be important at energies above LEPII and become increasingly 
important at energies of 500 GeV and beyond.  
One example of this is slepton studies.
Sleptons with masses above about 300 GeV 
will be difficult to find at the LHC, especially
if they are not produced copiously in the cascade decays of other
strongly-interacting superpartners.  Furthermore, if the left- and
right-sleptons are close in mass to each other they will be difficult
to resolve.  The linear collider produces sleptons directly if 
the CM energy is sufficient.  The two species of sleptons are readily
distinguished using beam
polarization and other observables. 
Another discovery issue arises in the case of a  neutral
wino or higgsino LSP, with a nearly
degenerate charged $\s W^\pm$ just above it in mass. The wino 
case occurs, for example, in anomaly-mediated  and in  $U(1)$-mediated 
supersymmetry breaking.
 In the limit in which all other
superpartners are too massive to be produced at the LHC or LC,
 the linear collider
with energy above 500 GeV and 100 fb${}^{-1}$ is expected to have a 
higher mass reach than the LHC for these states. There are other important
cases, such as R-parity-violating supersymmetry,
 in which the linear collider is needed to discover or resolve
states of the supersymmetry spectrum.
\vspace{0.1in}
 
{\it Supersymmetry and Higgs bosons.} The minimal supersymmetric extension
of the SM (MSSM) predicts that at least one scalar Higgs boson ($h^0$)
must have mass below about 135 GeV.  The mass is controlled at tree-level
by the $Z$-boson mass, and at one loop by the logarithm of superpartner
masses.  The prediction of a light Higgs boson has two virtues: it is a
useful falsifiable test of the MSSM, and fits nicely within the 
upper bound from the current precision EW data.
Over much of the parameter space, the light
MSSM Higgs boson behaves very similarly to the SM Higgs boson.

The other physical scalar Higgs states of the MSSM are $H^0$, $A^0$, and
$H^\pm$.  Unlike the $h^0$ state, these Higgs bosons receive tree-level
masses directly from supersymmetry breaking parameters.  Therefore, it is
not possible to 
rigorously establish upper bounds to their masses.  In large parts of 
parameter space, the masses of these particles are above 300 GeV, and the
only important production processes in $\ee$ annihilation are the 
pair-production reactions $\ee\to H^+H^-, H^0A^0$. Thus, these particles
may not appear at the first-stage linear collider.

If the heavy Higgs boson are not seen directly, 
the effects of the  more complicated Higgs sector 
of the MSSM can
be observed by measuring slight deviations in the couplings of $h^0$ to
fermions and gauge bosons  from those predicted for a SM Higgs boson.
The more massive the heavy Higgs bosons are, the more $h^0$
behaves like the SM Higgs boson.  Nevertheless, inconsistency with 
the SM can be
discerned by precision measurements at the LC over much of the parameter 
space,
even when $m_{A^0}$ is significantly higher than $\sqrt{s}/2$ and out of
reach of direct production. This issue is discussed in Chapter 3, Section 8.
It demonstrates again the importance of
precision Higgs boson measurements to pointing the way to new physics at
higher mass scales. 
\vspace{0.1in}

{\it Probing supersymmetry breaking.}
Finally, precision measurements of supersymmetry masses and mixing angles
serve a purpose beyond simply determining
what Lagrangian applies to the energy region around the weak interaction
scale.  Careful measurements can reveal a pattern characteristic of 
a more fundamental theory.  For example, masses measured
at the weak scale can be evolved using the renormalization
group to a higher scale, where they might be seen to be unified or to 
fit another simple relation.  A pattern that emerged from this study
would point to a specific theory of supersymmetry breaking, indicating
both the mechanism and scale at which it occurs.  This study could also 
support
or refute the hypothesis that our world is derived from a perturbative
grand unified theory with an energy desert, a hypothesis that
does seem to apply to the precisely known gauge couplings measured at $\mz$.
The ability of a linear collider to 
test these tantalizing ideas with precision measurements 
provides a route by which we can climb from the weak scale to 
a more profound theory operating at much higher energies.

\section{New strong interactions at the TeV scale}

A second way to cure the stability crisis of the Higgs field and to 
explain the origin of electroweak symmetry breaking is to introduce a new
set of strong interactions that operate at the TeV scale of energies.
In models of this type, symmetry breaking arises in the weak interactions
in the same way that it arises in well-studied solid-state physics 
systems such as superconductors.  Just as in those systems, the physics
responsible for the symmetry breaking has many other consequences that
lead to observable phenomena at the energy scale of the new interactions.

Two quite distinct implementations of this line of thought have been 
actively pursued. The first follows the possibility that 
the Higgs doublet ({\it i.e.}, the four degrees of freedom which after 
electroweak symmetry breaking become
the Higgs boson and the longitudinal components of the $W^\pm$ and $Z^0$)  
is a bound state that arises from a short-range strongly coupled 
force. Theories that have this behavior are generically called `composite
Higgs' models. These models 
are usually well approximated at low energies by the 
SM, and therefore are consistent with the 
electroweak data.

The second implementation follows the possibility that the new strong
interactions do not generate a Higgs doublet, even as a bound state.
This is possible if  the electroweak symmetry is broken
by the pair-condensation of some new strongly interacting particles.
 The prototype of such theories is `technicolor',
an asymptotically-free gauge interaction that becomes strong at the
TeV scale. The behavior of technicolor theories below the TeV scale 
is typically very different from that of the SM.  In most cases,
there is no Higgs boson 
with an observable coupling to pairs of $Z$ bosons, and the 
new symmetry-breaking interactions generate substantial corrections to
precision electroweak observables.

The linear collider experiments that directly test these two theoretical 
pictures are reviewed in detail in Chapter 5, Sections 3 and 4.
In this section we briefly discuss the two ideas in general terms and
discuss  the relevance of the linear collider for uncovering and studying
these new interactions.

%%%%%%%%
\subsection{Composite Higgs models}

Several ways have been suggested in the literature to form a bound-state
Higgs boson that mimics the properties of the Higgs particle of the SM.
In the top-quark seesaw theory, 
the Higgs boson arises as a bound state of the left-handed top quark
and the right-handed component of a new heavy  vector-like quark.
Although the composite Higgs boson mass 
is typically about 500 GeV, there is agreement with the
precision electroweak data for a range of parameters in which new
contributions from the additional heavy quark compensate the effects
of a heavy Higgs boson.  Depending on the binding interactions, an extended
  composite Higgs sector may form. In this case, mixing among the
CP-even scalar bound states may bring the SM-like Higgs boson down to a mass
below 200 GeV.

Another scenario that  may lead to a composite Higgs boson is 
the SM in extra spatial dimensions, a case that we will discuss
in more detail in the next section.  Here the 
short-range strongly-coupled force is given by the Kaluza-Klein
excited states of 
the $SU(3)_C\times SU(2)_W\times U(1)_Y$ gauge bosons.  The 
 Kaluza-Klein states of the top quark become the
constituents of the Higgs boson. The Higgs boson  
in this scenario has a mass of order 200 GeV.

We now list a number of 
non-standard phenomena that are likely to 
appear in these theories at relatively low energies.  Of course, 
these theories will ultimately be tested by going to the energy scale
of the new interaction and determining its nature as a gauge theory 
or as a field theory of some other type.
\vspace{0.1in}

{\it Deviations in Higgs sector.}
In models in which the Higgs boson appears as a bound state,
it is likely that additional composite scalar states 
will also be present at the TeV scale or below. 
If these states appear, their masses and couplings will provide important
information on the nature of the constituents.  Additional states with
the quantum numbers of the Higgs boson can be produced at a linear collider
in association with a $Z^0$ or singly in $\gamma\gamma$ collisions.
Other states can be studied in pair-production.  In both cases, the
precise measurement of their masses and
branching ratios will provide important information.
In addition, it is possible at a linear collider to recognize
even very small deviations of the properties of the Higgs boson from 
the predictions of the SM.
\vspace{0.1in}
 
{\it Extra fermions.}
The top-quark seesaw model implies the existence of an additional
fermion whose left- and right-handed components have the same charges as 
the  right-handed top quark, $t_R$.  
This quark could have a mass  of many TeV
with little loss in fine-tuning, making it hard to
find directly at any of the next generation colliders, including the LHC.
In this circumstance, however, the improved precision electroweak 
measurements described in Chapter 8 should show a clear deviation from 
the SM in the direction of  positive $\rho$
parameter ($\Delta T > 0$).  This would prove that the SM is incomplete
and give a clue as to the nature of the new physics.
\vspace{0.1in}

{\it Heavy vector bosons.}
Both the top-quark seesaw theory and the extra-dimensional
composite Higgs models imply the existence of heavy vector bosons.
In the top-condensate scenario, the extra heavy vectors could arise
from a topcolor gauge group. In addition, one often requires an 
additional gauge interaction that couples differently to $t_R$ and $b_R$
to explain  why  we see top quark but not bottom quark condensation.
  If a new vector boson couples with some strength
to all three generations, it will appear as a resonance at the LHC, and
its effects will be seen at the LC as a pattern of deviations in all of the
polarized $\ee\to f\bar f$ cross sections.  In both cases, the 
experiments are sensitive to masses of 
 4 TeV and above.   This mass reach
overlaps well with the expectation that the new physics should occur at a 
mass scale of several TeV.   The observation and
characterization of new $Z$ bosons are described in Chapter 5, Section~5.

\subsection{Technicolor theories}

Technicolor theories provide an alternative type of model with new
strong interactions.  These theories do not require a composite Higgs boson.
Instead, they involve
new chiral fermions and a confining gauge interaction 
that becomes strongly-coupled at an energy  
scale of order
1 TeV. The most robust prediction of these theories 
is that there is a vector resonance with 
mass below about 2 TeV that couples with full strength to the 
$J=1$ $W^+W^-$ scattering amplitude.

The general idea of technicolor is severely constrained by the 
precision electroweak measurements, which favor models with a light 
Higgs boson over models where this state is replaced by heavy resonances.
In order to be viable, a technicolor model must provide some new
contributions to the precision electroweak observables that compensate
for the absence of the Higgs boson.  This leads us away from models in 
which the new strong interactions  mimic the behavior of QCD and toward
models with a significantly different behavior.  For such models, it is 
difficult to compute quantitatively and so we must look for qualitative
predictions that can be tested at high-energy colliders.  In this 
situation, the ability of the linear collider to discover new particles
essentially independently of their decay schemes would play an important
role.

We summarize some of the measurements that the linear collider can perform
that are relevant to strongly-coupled theories of this type.  
Our  approach is to identify qualitative features that are likely 
to result from technicolor dynamics.  Because of the uncertainties
in calculating the properties of such strongly-interacting theories,
it is not possible to map out for what parameters
 a given model can be confirmed
or ruled out.   Nevertheless, the linear collider has the opportunity to 
identify key components of technicolor models.
\vspace*{0.1in}

{\it Strong $WW$ scattering.}
As we have noted, the most robust qualitative prediction of technicolor
theories is the presence of a resonance in $WW$ scattering in the vector
($J=1$) channel.   This particle is the analogue of the $\rho$ meson
of QCD.  For masses up to 2 TeV, the `techni-$\rho$' should be seen 
as a mass peak in the $W^+W^-$ invariant mass distribution observed
at the LHC.  In addition, the techni-$\rho$ will appear 
as a resonance in $\ee\to W^+W^-$
for longitudinal $W$ polarizations, for the same reason that in QCD the 
$\rho$ meson appears as a dramatic resonance in $\ee\to\pi^+\pi^-$.
The resonant effect is a very large enhancement of a well-understood
SM process, so the effect should be  unmistakable at the linear collider,
even at $\sqrt{s} = 500$ GeV, well below the resonance. 
As with the case of a $Z^{\prime}$, the two different observations at 
the linear collider and the LHC
can be put together to obtain a clear phenomenological picture
of this new state.  These issues are discussed further in Chapter~5, 
Section~3.
\vspace*{0.1in}

{\it Anomalous gauge couplings.}
If there is no Higgs boson resonance below about 800 GeV, the 
unitarization of the $WW\to WW$ scattering cross-section by new 
strong interactions will lead to a large set of new effective 
interactions that alter the couplings of $W$ and $Z$.  Some of these
terms lead to anomalous contributions to the $WW\gamma$ and $WWZ$ 
vertices.  Through the precision study of $\ee\to W^+W^-$ and 
related reactions,
the 500 GeV linear collider with 500 ${\rm fb}^{-1}$ of integrated luminosity
will detect these anomalous contributions or improve the limits
by  a factor of ten over those that will be set at the LHC.
In the case that there are new strong interactions, the accuracy of
the linear collider measurement is such as to make it 
possible to measure the coefficients of the effective Lagrangian
that results from the new strong interactions. These measurements
 are discussed further in Chapter~5, Section~2.  In addition, many 
technicolor models predict large anomalous contributions to the gauge
interactions of the top quark particularly to the $t\bar t Z$ vertex
function.  The linear collider may provide the only way to measure
this vertex precisely.  The measurement is discussed in Chapter~6, Section~3.
\vspace*{0.1in}

{\it Extra scalars.}
Just as, in QCD, where the strongly coupled quarks lead to octets of
relatively light mesons, technicolor theories often imply the 
existence of a multiplet of pseudoscalar bosons that are relatively
light compared to the TeV scale.  These bosons are composites of the
underlying strongly coupled fermions.  Since these particles have
non-zero electroweak quantum numbers, they are pair-produced in $\ee$
annihilation.  The number of such bosons and their quantum numbers
depend on the precise technicolor theory.  Experimentally, these 
particles look like the particles of an extended Higgs sector, and their
detection and study follow the methods discussed for that case in 
Chapter~2, Section~6.  Particular models may include additional
new particles.  For example, in `topcolor-assisted technicolor', 
there is a second doublet of Higgs bosons, with masses of 200-300~GeV,
associated with top-quark mass generation.

%%%%%%%%%%%%%%%%%%%%%%%%%%%%%%%%%
\section{Extra spatial dimensions}

   It is `apparent'
that the space we live in is three-dimensional, and in fact
precise measurements are consistent with this even down to the small distances
probed by LEP2 and the Tevatron.  But one should not hastily 
conclude that the universe has no more than three dimensions, because
two important loopholes remain.  First,  there could be extra spatial 
dimensions that are not 
accessible to  SM particles such as the photon and the gluon. Second,
there could be  extra spatial dimensions that are compact,
with a size smaller than $10^{-17}$ cm.  In both cases, it is possible
to build models that are in agreement with all current data.

Besides being a logical possibility, the existence of extra spatial dimensions
may explain key features of observed phenomena, ranging from
the weakness of the gravitational interactions 
to the  existence of three generations of quarks and 
leptons.  Most importantly from the viewpoint of the stability problem
of the Higgs field, the assumption that the universe contains more that
three dimensions opens a number of new possibilities for models of 
electroweak symmetry breaking.  In such models, the value of 
the weak-interaction scale results from the fact that some natural
mass scale of gravity in higher dimensions, either the size of the
new dimensions or the intrinsic mass scale of gravity, is of order 
1~TeV.   This, in turn, leads to new observable phenomena in high energy
physics at energies near 1~TeV.  These phenomena, and the possibility 
of their
observation at a linear collider, are discussed in Chapter 5, Section~6.

Once we have opened the possibility of new spatial dimensions, there are
many ways to construct models.  Most of the options can be classified by
two criteria.  First, we must specify
which particles are allowed to propagate in the full space
and which are restricted by some mechanism to live in a three-di\-men\-sion\-al
subspace.  Second, we must specify whether the extra dimensions are 
flat, like the three dimensions we see, or highly curved.  The latter
case is referred to in the literature as a `warped' geometry.  Some ideas
may require additional fields,  beyond the SM fields,
to solve certain  problems (such as flavor violation or  anomaly
cancelation) that can arise from the hypothesized configuration
of particles in the extra-dimensional space.
We now give a brief overview of these possibilities and the role of the
linear collider in each scenario.

\subsection{Flat extra dimensions, containing only gravity}

The first possibility is that all of the particles of the SM---quarks,
leptons, and  Higgs and gauge bosons---are localized on three-dimensional walls
(`3-branes')
in a higher-dimen\-sional space.  Gravity, however, necessarily propagates 
through all of space.  Higher-dimensional gravity can be described in four-dimensional
terms by using a momentum representation in the extra dimensions.  If these
extra dimensions are compact, the corresponding momenta are quantized.
Each possible value of the extra-dimensional momentum gives a distinct
particle in four dimensions.  This particle has mass
$  m_i^2    =    (\vec p_i)^2$, 
where $\vec p_i$ is the  quantized value of the extra-dimensional momenta.
These four-dimensional particles arising from a higher-dimensional field
are called Kaluza-Klein (KK) excitations.  In the later examples, where
we put SM fields also into the higher dimensions, these field will also 
acquire a KK spectrum.

If gravity propagates in the extra dimensions, the exchange of its 
KK excitations will increase the strength of the gravitational force at
distances smaller than the size of the new dimensions.  Then the 
fundamental mass 
scale $M_*$ at which gravity becomes a strong interaction is lower
than the apparent Planck scale of $10^{19}$ GeV.  It is possible that
$M_*$ is as low as 1 TeV if the volume of the extra dimensions 
is sufficiently large.  In that case, there is no stability problem for the Higgs
field.  The Higgs expectation value is naturally of the order of $M_*$.

The KK gravitons can be produced in collider experiments.  In $\ee$
collisions, one would look for $\ee$ annihilation into a photon plus
missing energy.  The cross section for this process has typical
electroweak size as the CM energy approaches $M_*$ and the phase space
for producing the KK gravitons opens up.  The 
expected signals of extra dimensions
are highly sensitive to the number of extra
dimensions.  Nevertheless, if the number of extra dimensions is less than
or equal to six, the signal can be studied at a 
linear collider at CM energies that are a factor of 3--10 below  $M_*$.
 The LHC can also study KK graviton production through processes such as
$q\bar q$ annihilation to a jet plus missing energy.  The sensitivity to
$M_*$ is somewhat greater than that of a 1 TeV linear collider, but it is
not possible to measure the missing mass of the unobserved graviton.

The KK gravitons can also appear through their virtual exchange in 
processes such as $\ee \to f\bar f$,  $e^+e^-\to \gamma\gamma$, and 
 $\ee\to gg$.  The graviton exchange leads to a spin-2 component that
is distinct from the SM expectation.  Although this indirect signal
of KK gravitons is more model-dependent, it is expected that it can be 
seen even at 500 GeV if $M_*$ is less than a few TeV.

\subsection{Warped extra dimensions, containing only gravity}

If the extra dimensions are warped, the KK spectrum of gravitons has 
somewhat different properties.  In the case of flat extra dimension,
the KK particles are closely spaced in mass, but in the case of warped
dimensions, the spacing is of order 1 TeV.  In the simplest model, 
the KK gravitons have masses in a characteristic pattern given by the zeros
of a Bessel function.  The individual states appear as spin-2 resonances
coupling with electroweak strength to $\ee$ and $q\bar q$.  These 
resonances might be seen directly at the LHC or at a linear collider.
If the resonances are very heavy, their effects can be seen from 
additional spin-2 contact contributions to  $e^+e^-\to f\bar f$, even for
masses more than an order of magnitude above the collider CM energy.

%%%%%%%%%
\subsection{Flat extra dimensions, containing SM gauge fields}

It is often assumed that the quarks and leptons are localized on
three-dimensional walls (3-branes) and therefore do not have KK modes,
whereas the gauge bosons propagate in the extra-dimensional space. In this
case, the KK modes of the electroweak gauge bosons contribute at tree
level to the electroweak observables, so that a rather tight lower bound
of about 4 TeV can be imposed on the inverse size of the extra dimensions. The LHC
should be able to see the first gauge boson KK resonance up to about 5
TeV, leaving a small window of available phase space for direct
production of these states.  On the other hand, precision measurements at
a high-energy $e^+e^-$ linear collider can establish a pattern of deviations
from the SM predictions for the reactions $\ee\to f\bar f$ from 
KK resonances well beyond direct production sensitivities.  The
capability of an $e^+e^-$ linear collider in identifying the rise in
cross sections due to KK resonances improves when the center-of-mass
energy is increased.  High luminosity is also important.  For example,
with more than 100 fb${}^{-1}$ of integrated luminosity at a 500 GeV, one
could see the effects of resonance tails for KK masses above 10~TeV 
in models with one extra dimension.

\subsection{Flat extra dimensions, containing all SM particles}

Finally, we consider the case of `universal' extra dimensions, in which
all SM particles are permitted to propagate.
 A distinctive feature of universal extra 
dimensions is that the quantized KK momentum is conserved at each vertex.
Thus, the KK modes of electroweak gauge bosons 
do not contribute to the precision electroweak observables at the 
tree level. As a result,
the current mass bound on the first KK states is as 
low as 300 GeV for one universal extra dimension.
If the KK states do indeed have a mass in the range 300-400 GeV,
we would expect to observe the states at the Tevatron and the LHC. The
linear collider, at a CM energy of 800 GeV,  
would become a KK factory that produces excited states of quarks, leptons,
and gauge bosons.

\section{Surprises}

Our brief discussion of pathways beyond the SM concentrated on three 
very different approaches that have been proposed
to solve  the conundrums of the SM.  
Although some of these ideas are more easily tested than others
at the next-generation colliders, it is important to note that
all three approaches have  many new observable consequences.
In all cases, we expect to see an explosion of new phenomena as 
we head to higher energies.

Though these three approaches are very different, we should not 
delude ourselves into thinking that they cover the full range of 
possibilities.   Letting our imaginations run free, we could envision
models in which quantum field theory itself breaks down at the weak
interaction scale and an even more fundamental description takes over.
Such a possibility would be viable only if it satisfies the constraint
of giving back the predictions of the SM at energies below 100 GeV.
String theory is an example of a framework that resembles the SM at 
low energies but, at the energies of the string scale, is dramatically
different from a simple quantum field theory.  Perhaps there are 
other alternatives to be found.

Exploring physics at shorter distances 
and with higher precision is an endeavor that implies
the possibility of great surprises.  Experiments
at a linear collider will be a necessary and rewarding part of this
program, and will constitute
a major step in our quest to understand how Nature works.

\emptyheads

\begin{center}\begin{large}{\Huge \sffamily Experimental Program Issues}
\end{large}\end{center}

\blankpage \thispagestyle{empty}

\setcounter{chapter}{9}

\fancyheads

\chapter{Scenarios for Linear Collider Running}
\fancyhead[RO]{Scenarios for Linear Collider Running}

In the literature on physics studies at $\ee$ linear colliders, one 
typically finds each process analyzed in isolation with a specific choice 
of energy and polarization.  This naturally raises the question of how the
full program for the linear collider fits together and whether all of the 
important physics topics can actually be scheduled and investigated.  In 
this chapter, we will examine this issue.  We will suggest some simple run 
plans that accomplish the most important goals of the linear collider
program under different physics scenarios.   

Under almost any scenario, one would wish to run the linear collider at
two or more different energies during the course of its program.  Operation
of the collider at energies lower than 500 GeV typically yields lower
luminosity, scaling
roughly as $E_{\rm CM}$.   In this chapter, we will craft scenarios using the 
following guidelines:  We assume that the collider has a single
interaction region that can run at any energy from $\mz$ to 500 GeV,
with instantaneous luminosity strictly proportional to the CM energy.
We plan for a campaign equivalent to 1000 fb$^{-1}$ at 500 GeV, corresponding
to 3--5 years at design luminosity.  We then ask how the collider running
should be allotted among the various possible conditions.  These assumptions
are rather simplistic, but they frame a problem whose solution is instructive.
In Chapter 11,  we describe in a more careful way how a
collider with two interaction regions, sharing luminosity, would be 
configured for a flexible program covering a large dynamic range in CM
energy.

\section{Preliminaries}

In designing a plan for linear collider running, we should consider 
the alternative strategies for energy and for polarization.  In this 
paragraph, we consider these two topics in turn.

There are three different ways to choose the energy of an $\ee$ collider:
\begin{itemize}
\item {\em Sit:}   Choose an energy that is optimal for a particular 
      interesting process, and accumulate integrated luminosity at that
       point.
\item {\em Scan:}  Step through a threshold for pair-production of some
        particle, taking enough data to define the threshold behavior.
\item {\em Span:} Go to the highest available energy, and take a large
       sample of data there.
\end{itemize}

\noindent
In the application of $\ee$ colliders to the $J/\psi$ and $\Upsilon$ systems,
and to the $Z^0$, the $\ee$ annihilation cross section contained narrow 
structures that put great importance on the exact choice of the beam 
energy.  For most of the important processes considered for study at the 
next-generation linear collider, the choice of energy  should be less
of an issue, since 
the Higgs boson, the top quark, supersymmetric particles, \etc,  will be
 studied mainly in  continuum production of a pair of particles.  These 
processes have cross sections that peak within 50--100 GeV of the threshold
and then fall as $E_{{\rm CM}}^{-2}$.  This dependence is somewhat compensated by 
the higher collider luminosity  at higher energy.   Since the signatures
of different particles seen in $\ee$ annihilation are distinctive,
many different reactions can be studied at a single energy.
   
As an example, consider the measurement of Higgs boson branching 
ratios.  For this study, the Higgs boson is produced 
in the reaction $\ee\to Z^0 h^0$.  For a 
Higgs boson of mass 120 GeV, the peak of the cross section is at 
250 GeV.  However, taking into account the increase of luminosity with 
energy, the penalty in the total number of Higgs bosons in working at 
500 GeV instead of at the peak of the cross section is only a factor of 2.
At higher energy, more reactions become accessible, and more effort must
be made to isolate the Higgs sample.  On the other hand, the Higgs 
production process has a distinctive signature, the monoenergetic $Z^0$.
As the energy increases, the kinematics become more distinctive as
the Higgs and the $Z^0$ are boosted into opposite hemispheres.
We conclude that LC experimenters will continue to accumulate statistics
for the Higgs branching ratio study as they move to higher energies.
Thus, though concentration on this process would favor a sit at an energy
below 300 GeV, one could well adopt a span strategy if other physics
required it.  This example illustrates that it is important, in future
studies of linear collider measurements, to evaluate explicitly how the
quality of the measurement depends on CM energy.

Only a few reactions among those anticipated for the LC require a detailed
scan of some energy region.  These include the measurement of the top quark
mass by a threshold scan, the precision measurement of supersymmetric
particle masses (to the parts per mil rather than the percent level), and,
in the precision electroweak program of Chapter 8, the measurement of the 
$W$ mass to 6 MeV.  The top quark mass measurement actually becomes limited
by theory errors after about 10 fb$^{-1}$ of data, though a longer run 
would be justified to obtain a precision measurement of the top quark width
and the decay form factors.  Other threshold scans require similarly small
increments of luminosity, except for the cases of sleptons, where the 
threshold turns on very slowly, as $\beta^3$, and the $W$, where extreme
precision is required.

As for the choice of beam polarization in LC running, it is important to 
understand how polarization will be implemented.  The choice of a polarized
or unpolarized electron source is not a limiting factor for the electron
currents in the machine.  So there is no penalty in choosing a polarization 
that is as large as possible---80\%, with current technology.  Polarized
electrons are created by shining circularly polarized light on an 
appropriate cathode.  In the SLD polarization program at the $Z^0$, the
polarized light was created by passing a linearly polarized laser beam
through a Pockels cell, a device that is effectively a quarter-wave plate
whose sign is determined by an applied voltage.  The signal applied to the
cell changed sign randomly at the 120 Hz repetition rate of the machine.
This random sign was supplied to the experimenters and used to determine
the initial-state polarization in detected events.  We anticipate that the 
beam polarization will be created in a similar way at the LC.  Thus, 
there will be no `unpolarized' running.  The normal running condition will be
a half-and-half  mixture of left- and right-handed electron polarization,
switching randomly
at the repetition rate for bunch trains. In this arrangement, it 
is straightforward to measure polarization-averaged cross sections.
The rapid switching allows
polarization asymmetries to be measured with many systematic errors 
cancelling.  

For certain processes, it is advantageous to take the bulk of the data in 
a single state of beam polarization.  For example, the supersymmetric 
partners of the right-handed sleptons are most easily studied with a 
right-hand polarized electron beam, while $WW$ pair production and fusion
processes such as 
$W^+W^- \to t\bar t$ receive most or all of their cross section 
from the left-handed electron beam.  In contrast, 
$e^+e^- \rightarrow Z^0h^0$ has only a weak polarization dependence.
It is possible that our knowledge of 
physics at the time of the LC running will single out one such process
as being of great importance and call for a run with an unequal 
(90\%/10\%) distribution
of beam polarizations.  As in the case of the energy choice, this is a 
shallow optimum, winning back, in the best case, less than a factor of
2 in luminosity.

\section{Illustrative scenarios}

With these considerations in mind, we now propose some sample run plans
appropriate to different physics scenerios. 
  For each plan, we quote the luminosity sample to be
obtained at each energy and, in parentheses, the corresponding sample
scaled to 500 GeV.  These latter values are constrained to add up to 
1000 fb$^{-1}$.  

In most cases, the luminosity assigned below to 500 GeV would be
accumulated at the highest machine energy if higher energies were
available.  Many physics issues, including the measurement of the 
Higgs coupling to $t\bar t$ and the Higgs self-coupling in addition to
studies of new heavy particles, benefit greatly from CM energies above
500 GeV.  The integrated luminosities given are totals, which might be
accumulated in any order.  In the scenarios presented here, we omit,
for simplicity, the possibility of positron polarization and $\gamma\gamma$
or $e^-e^-$ running.  These options are discussed in the later chapters
of this section.  In considering any of these options, it is important to
keep in mind that these options entail trade-offs against $e^+e^-$ integrated
luminosity.

\subsection{A Higgs boson, but no other new physics, is seen at the LHC}

In this case, we would want to apply a substantial amount of luminosity to
a precision study of the branching ratios of the known Higgs boson.  It 
will also be important to search for Higgs bosons not seen at
the LHC, to search for new particles with electroweak couplings that might 
have been missed at the LHC, and to measure the $W$ and top gauge couplings
to look for the virtual influence of new particles. Thus:

\begin{itemize}
\item 300 GeV:  250 fb$^{-1}$  (420 fb$^{-1}$) sit
\item 350 GeV:  100 fb$^{-1}$  (140 fb$^{-1}$) top threshold scan 
\item 500 GeV:  440 fb$^{-1}$  (440 fb$^{-1}$) span
\end{itemize} 

\noindent
This run plan gives a data sample for the Higgs boson branching ratio
measurement equivalent to 600~fb$^{-1}$ at
350 GeV.

\subsection{No Higgs boson or other new particles are seen at the LHC}

In this case, we would want to apply the largest
amount of luminosity to the highest available energy. The 
issues for this study would be the 
 search for additional Higgs bosons not seen at
the LHC and the  search for new particles.  The measurement of the 
$W$ and top gauge couplings would be of essential importance.
Because the 
absence of a light Higgs conflicts with the precision electroweak fits
within the SM,
it will also be crucial in this case to include running at the $Z^0$ and 
the $WW$ threshold.

\begin{itemize}
\item 90 GeV:    50 fb$^{-1}$  (280 fb$^{-1}$) sit
\item 160 GeV:   70 fb$^{-1}$  (220 fb$^{-1}$) $W$ threshold scan 
\item 350 GeV:   50 fb$^{-1}$   (70  fb$^{-1}$) top threshold scan
\item 500 GeV:  430 fb$^{-1}$  (430 fb$^{-1}$) span
\end{itemize} 

\subsection{Light Higgs and superpartners are seen at the LHC}

In this case, it is necessary to compromise between the optimal energies
to study each of the new states, the optimal energy for the Higgs study---since
a light Higgs must also appear in supersymmetric models---and searches
for new superparticles, such as the extended Higgs particles and the 
heavier charginos and neutralinos, that could have been missed at the LHC.
The program will begin with extended running at 500 GeV, and perhaps
also at a lower energy, to determine the superpartner masses to
percent-level accuracy.  This could be followed by detailed
threshold scans.

Martyn and Blair~\cite{Martyn-Blair} have studied a particular scenario
in which the lightest neutralino has a mass of 70 GeV, the lighter charginos
and sleptons lie at about 130 GeV, and the heavier charginos and neutralinos
are at about 
 350 GeV.  Converting their suggested program to our rules, we have
for this case:

\begin{itemize}
\item 320 GeV:  160 fb$^{-1}$  (250 fb$^{-1}$) sit
\item 500 GeV:  245 fb$^{-1}$  (245 fb$^{-1}$) span
\item 255 GeV:   20 fb$^{-1}$  (40 fb$^{-1}$) chargino threshold scan
\item 265 GeV:  100 fb$^{-1}$  (190 fb$^{-1}$) slepton 
                                    ($\ell_R^-\ell_R^+$) threshold scan
\item 310 GeV:   20 fb$^{-1}$  (30 fb$^{-1}$) slepton 
  ($\ell_L^-\ell_R^+$) threshold scan
\item 350 GeV:   20 fb$^{-1}$  (30 fb$^{-1}$)  top threshold scan
\item 450 GeV:   100 fb$^{-1}$  (110 fb$^{-1}$) neutralino 
                               ($\chi_2^0\chi_3^0$)          threshold scan
\item 470 GeV:  100 fb$^{-1}$  (105 fb$^{-1}$)  chargino 
              ($\chi_1^-\chi_2^+$)          threshold scan
\end{itemize}

\noindent
The threshold scans would be done with the dominant beam polarization chosen,
respectively, right, left, equal, left, left.
The threshold with $\beta^1$ cross sections are given small amounts of running
time; thresholds with $\beta^3$ cross sections or cross sections that are
intrinsically small are given 100  fb$^{-1}$.  The running time at the 
top threshold is more than sufficient
 to push the determination of $m_t$ to the 
systematics limit.  While running at each threshold, pair production of
all lighter species can also be studied.  In particular, the total 
statistics for the Higgs branching ratio measurement is equivalent to about
700  fb$^{-1}$ at 350 GeV.

 \emptyheads
\blankpage \thispagestyle{empty}
\fancyheads

\setcounter{chapter}{10}
 \chapter{Interaction Regions}
\fancyhead[RO]{Interaction Regions}

\section{Introduction}

The Standard Model has received considerable experimental attention in
the past two decades, and much is known about its electroweak sector and
about its flavor sector.  Recent precision experiments have the
sensitivity to look beyond the SM for new physics.  However, the
mechanism for symmetry breaking in the SM is still unknown, and many
questions, such as the existence of SUSY, still are answered only by
speculation.  A future linear collider will provide the tools with which
we may probe the mechanism of symmetry breaking and address the questions of
new physics beyond the SM.  We seek the best configuration of a linear
collider facility that maximizes the potential for answering these
questions.

The number of interaction regions is a very important issue, affecting
the project cost, the physical footprint of the collider complex, the
number of detectors that can be accommodated, the breadth of the physics
program, and almost certainly the amount of enthusiasm and support the
linear collider would receive in the world's high energy physics
community.  In this section we look at the nature and number of
interaction regions to accompany the accelerator complex of a linear
collider.  The baseline configurations for TESLA and the NLC are briefly
discussed here.  This section gives only a brief overview of the
technical designs.  One must go to the relevant reports and documents to
get more technical details.

Both the TESLA and the NLC designs for the IRs allow for two regions.
The TESLA philosophy in its baseline design differs somewhat from that
of the NLC.  The baseline design for TESLA includes only one IR,
with real estate available for a second IR and a second beam
delivery system, if and when the funds become available.  The NLC
baseline design contains two IRs, as described below.

The arguments favoring the two-IR collider configuration come first from
the physics program.  The rich program of particle physics could best be
investigated by two active IRs with two or more detectors.  However, one
must consider the tradeoff between the increased breadth of the physics
program and the increased costs incurred.  One of the ``costs''
encountered is the unavoidable sharing of the available luminosity
between the two IRs.  Strategies for simultaneous running in the NLC are
briefly discussed.

However, it should be pointed out that the strongest motivation for two
IRs may come from external factors.  The future linear collider will
surely be an international facility.  In order for there to be
international participation in the financing of the collider, it would
be wise to incorporate two IRs to facilitate broad participation in the
detectors and the experimental program.  This philosophy on
international participation in the linear collider is surely part of the
strategy for incorporating two IRs in the TESLA and NLC designs.

\section {The two interaction region design at TESLA}

TESLA has provision
for two IRs, one which is in the baseline design, and a second which is
not currently in the baseline, but may be added.  The TESLA
linear collider cannot serve two IRs with luminosity simultaneously.  It
is possible, however, to switch the beam between the two experimental
stations.  The primary IR will receive beams at a zero crossing angle,
while the secondary IR will have a crossing angle of 34 mrad.  If the
secondary IR is run in the $e^+e^-$ collider mode (with crab crossing),
it is anticipated to have the same luminosity as the primary IR.  The
crossing angle also makes the secondary IR suitable for $\gamma \gamma$
and $e\gamma$ collider modes of operation using backscattered laser
beams, as described in Chapter 13.  Electron-electron collisions are
possible at one or both IRs, by reversing magnet polarities and
providing a second polarized electron source.  This option is discussed
in Chapter 14.
The layout of the two IRs and their technical
parameters can be found in the TESLA TDR \cite{TESLATDR}.

\section {The dual-energy interaction region design at the NLC}

To allow for a collider design for the desired physics program that
extends from the $Z$-pole to many TeV, the NLC group has introduced a
dual-energy IR design \cite{IR-NLC}.  The first IR is in a direct line with the main
linacs that accelerate the beams.  The second IR is reached by bending
the beam away from this direct line.  Both IRs have crossing angles, as
described below.  The IRs would be designed to operate in different
energy ranges, the first from 250 to 1000 GeV, the second from 90 to 500
GeV.

There are two motivations for this choice.  First, by having one of the
two IRs in a direct line with the main linacs that accelerate the beams,
this IR can operate at multi-TeV energies in subsequent machine energy
upgrades.  This layout eliminates the bending where incoherent synchrotron
radiation would dilute the beam emittances.  Second, Final Focus beamlines
are naturally optimized to operate over roughly a factor of four to five in
beam energy.  At the high end of the range, the luminosity decays
rapidly due to increasing synchrotron radiation.  At lower energies,
the luminosity scales proportionally to the collision energy until a limit
is reached at roughly 25\% of the maximum energy.  Below this limit,
the luminosity decays as the square of the collision energy due to increasing
aberrations and limited vacuum and masking apertures.  At either end, a smoother
dependence of luminosity on energy can be retained by realigning the Final Focus
components to change the total bending.  The choices we have indicated,
with two Final Focus systems of fixed configuration, give the NLC
overlapping coverage of the energy region that is thought to be initially
of interest.

\begin{figure}[htbp]
\begin{center}
\epsfig{file=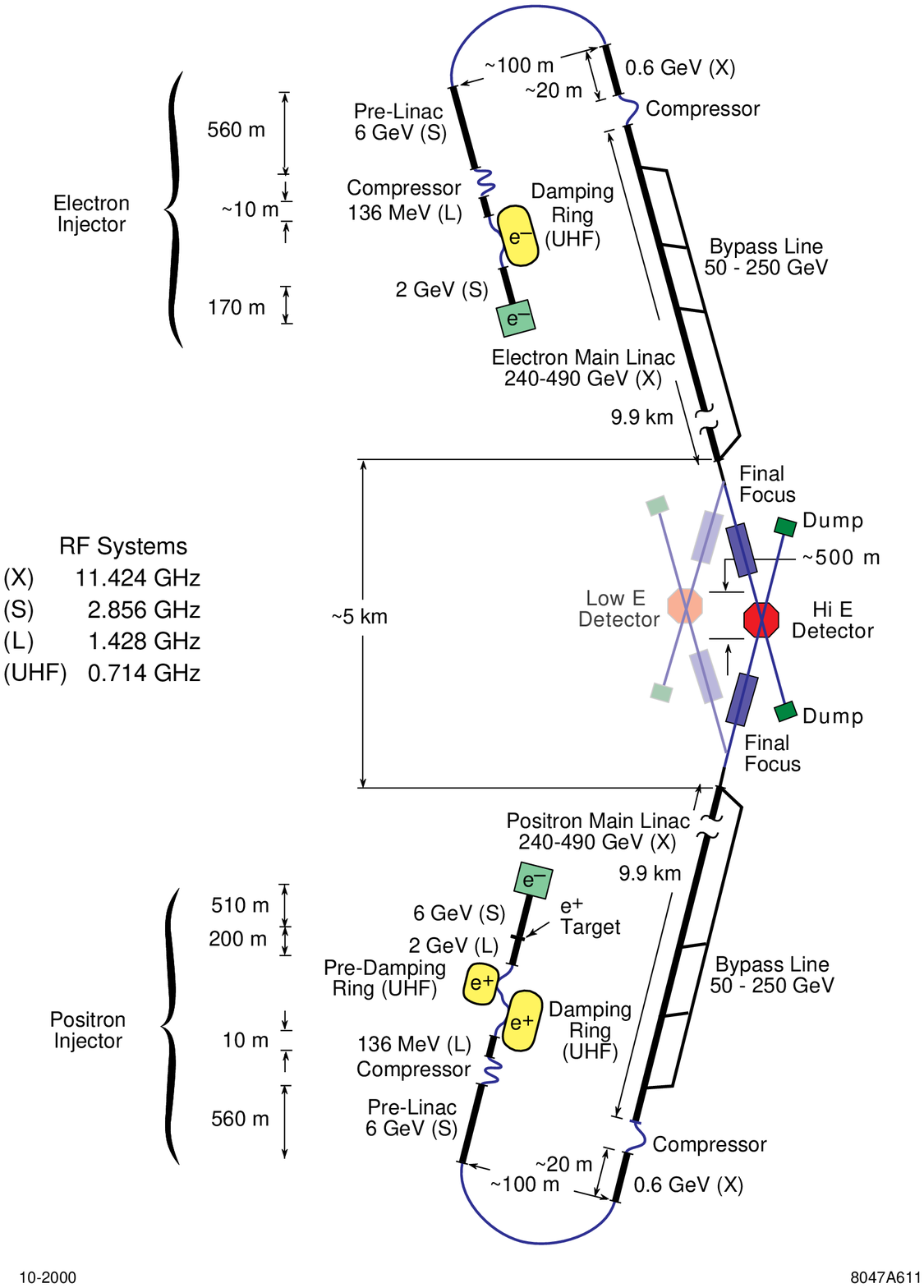,width= 5.50 in}  %%  4.0 in}
\caption{\label{fig:nlcscheme}
Schematic of the non-zero crossing angle of the two
linacs and the Dual Energy IR layout.}
\end{center}
\end{figure}

\begin{figure}[htbp]
\begin{center}
\epsfig{file=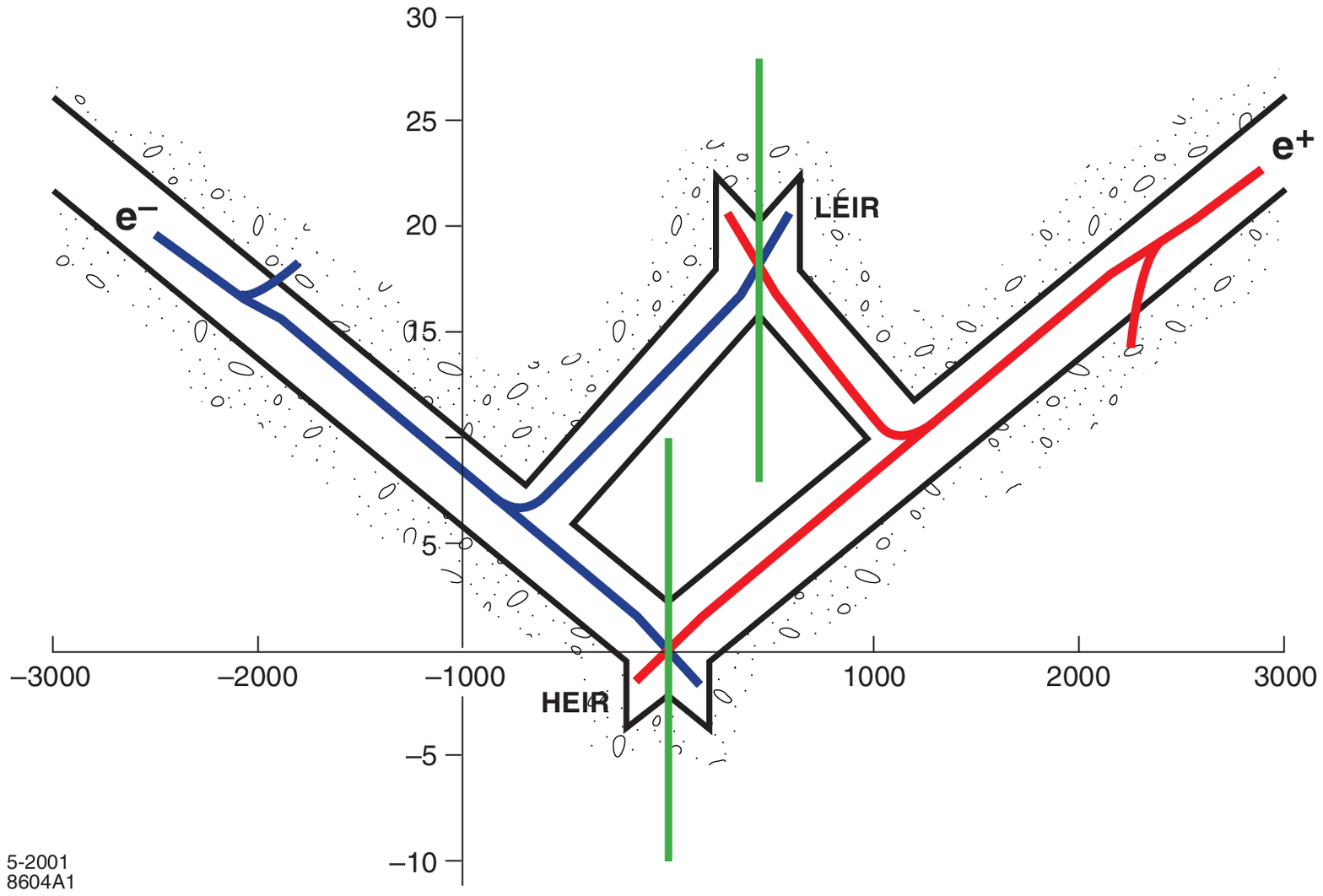,width=5in}  %%  3.5 in}
\caption[*]{\label{fig:TwoIRscheme}
Schematic of the accelerator tunnels leading to the two
interaction regions.  The IRs are separated laterally by 25~m and
longitudinally by 440~m.  The crossing angles at the HEIR and
LEIR are 20~mrad and 30~mrad, respectively.  Note that the
figure is extremely compressed in the horizontal direction; the
detectors occupy the volume of the vertical rectangles that intersect
the two beamlines at their crossing points.}
\end{center}
\end{figure}

Because the straight-ahead IR could support multi-TeV beam collisions,
we refer to this as the `high-energy' IR (HEIR).  The bending required
to reach the second IR limits the maximum energy attainable.  Thus, we
refer to this as the `low-energy' IR (LEIR).  Schematic plans of the NLC
machine and the two-IR layout are shown in Figs.~\ref{fig:nlcscheme}
and~\ref{fig:TwoIRscheme}.

With this starting point, the collider layout is determined by the
length of the beam delivery systems, the required transverse separation
of the IRs, and the desired crossing angle in the interaction regions.
Given the new Final Focus optics design which utilizes local chromatic
correction, the Final Focus can be relatively short.  The present NLC
Final Focus design is 700 meters long.  This length is sufficient up to 5 TeV 
in the center of mass.  In addition to the Final Focus optics
itself, there are diagnostic regions and beam collimation regions
upstream of the IP.  Depending on the operating mode, these regions
could likely be shared.  In the present NLC design, these regions are
roughly 1300 meters long for a total beam delivery system length of 2 km
per side.  This length could be reduced; however it is relatively
inexpensive and provides a conservative solution to the beam optics and
the beam collimation problems.

To attain reasonable transfer efficiency of the rf to the beam in a
normal conducting linear collider, the bunches must be spaced together very
closely.  In this case, both IRs must have a non-zero crossing angle
to prevent interactions between bunches at satellite crossings.
Typical values for the crossing angle could range from 6 mrad to 40
mrad.  The larger angles result in easier beam extraction and IR
integration but lead to more difficult tolerances.  Simplifying the beam
extraction is important if one believes that it is important to measure
the beam energy spread and polarization after collision at the IP.  The
crossing angles allow for these measurements in the NLC but not at the
primary IR at TESLA.

Without consideration of the extraction line, the minimum crossing angle
is set by the `multi-bunch kink' instability.  At CM energies below 1.5
TeV, the minimum angle in a normal conducting design is roughly 2 mrad.
However, studies of the CLIC 3 TeV IR suggest that a minimum crossing angle
of 15 mrad is necessary at multi-TeV energies.  For these reasons, a
crossing angle of 20 mrad at the HEIR and between 20--40 mrad at the
LEIR is suggested.

The IR halls have been sized assuming that one would house the NLC
L or SD  Detector and that one would house the P Detector.
Table~\ref{tab:irparam} gives a list of the hall parameters.  The hall
length (transverse to the beam) is large enough to allow assembly of the
detector while a concrete wall shields the interaction point.  The wall
would also serve as radiation shielding if the detector is not deemed to
be `self-shielding'.  If the detector were built in place on the beam
line, and could be self-shielding, the length could be reduced by
roughly a factor of three.  The hall width (parallel to the
beamline) is set by the constraint that the doors open just enough to
allow servicing of the inner detectors.

The baseline design assumes that the two IR halls are physically
separated so that activities and mechanical equipment operating in one
hall are seismically isolated from the other hall.  For example, the
LIGO facility has used 100 m as a minimum separation between rotating
machinery and sensitive detectors.  While the active detection
and compensation of culturally induced ground vibration is a key element
of the NLC R\&D program, passive compliance with vibration criteria is
the ideal.  In principle each of the IR halls could be designed to
accommodate two detectors that share the beamline in a push-pull manner,
thus increasing experimental opportunities, or the overall NLC layout
could be changed to support only one push-pull IR at a considerable cost
savings.  In any push-pull scheme, major installation activities might
need to be curtailed if they introduced uncompensated vibration of the
final magnets producing data for the detector currently on the beam
line.

\begin{table}[h]
\begin{center}
\begin{tabular}{|l|c|c|} \hline\hline
Parameter & Small Detector & Large Detector \\ \hline
Detector footprint &  12 $\times$ 11 m   &  20 $\times$ 20 m  \\ \hline
Pit length & 40 m & 62 m \\ \hline
Pit width & 20 m  & 30 m\\ \hline
Pit depth below beamline & 5 m & 7 m \\ \hline
Door height & 10 m & 13 m \\ \hline
Door width & 10 m & 13 m \\ \hline
Barrel weight & 2000 MT & 7300 MT \\ \hline
Door weight & 500 MT & 1900 MT \\ \hline
Total weight & 3100 MT & 11100 MT \\ \hline\hline
\end{tabular}
\caption{\label{tab:irparam}
 The Baseline Interaction Region Parameters}
\end{center}
\end{table}

All of these features are illustrated in the schematic designs shown in
Figs.~\ref{fig:nlcscheme} and ~\ref{fig:TwoIRscheme}.  The main linacs
are aligned to provide the 20 mrad crossing angle at the HEIR.  The
LEIR beamline is bent from the straight-ahead beams.  The transverse
separation between the two IR collision points is currently set at
roughly 25 meters.  However, roughly 440 meters longitudinal separation
of the two IR halls has been provided for increased vibration isolation.
In addition, bypass lines are installed along the side of the linac so
that lower-energy beams can be transported to the Final Focus without
passing through the downstream accelerator structures.

\subsection {The low-energy interaction region at the NLC}

\begin{figure}[t!]
\begin{center}
\epsfig{file=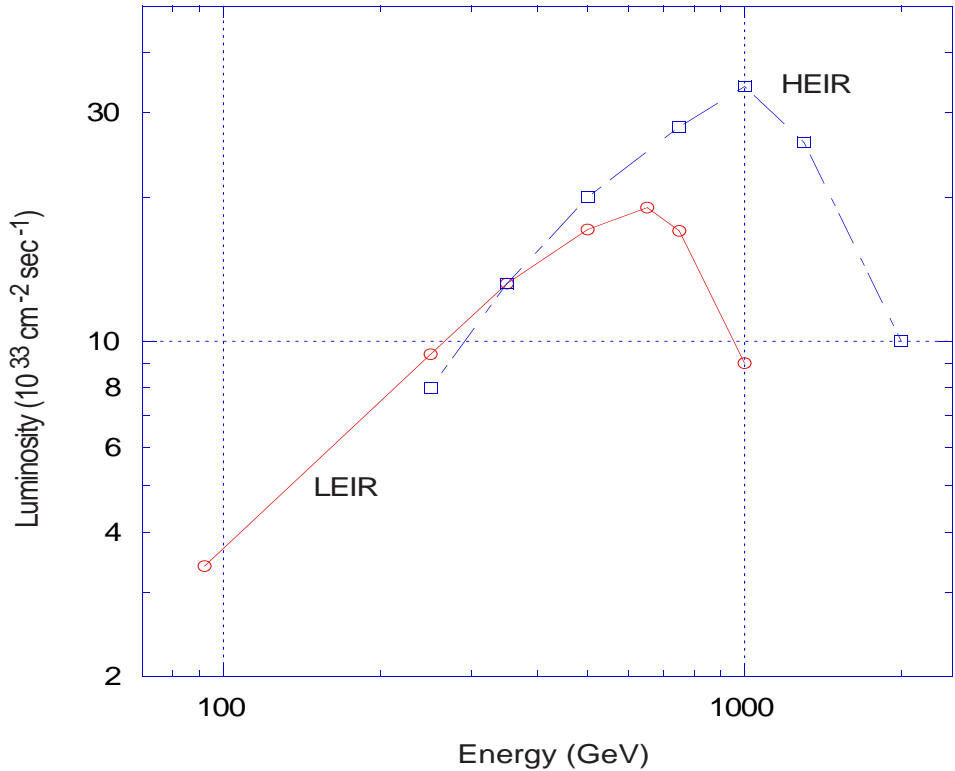,width=5 in,height=4in}  %% 3.0 in}
\caption[*]{\label{fig:nlclum}
The baseline luminosity versus CM energy for the
NLC LEIR and HEIR.  The two IRs have been designed to have
comparable performance in the region between 250~GeV and 500 GeV,
however, the NLC HEIR beam delivery system has been optimized
for a maximum energy of 500 GeV, the HEIR for 1 TeV.}
\end{center}
\end{figure}

The experimental program in the LEIR is determined by the
range of accessible center-of-mass energies and the available
luminosity.  The amount of luminosity that should be dedicated to a
particular $\sqrt{s}$ will depend on the physics that is revealed by the
Tevatron and the LHC.  This need for flexibility imposes the requirement
that the LEIR have high performance at least over the range $m_{Z} \leq
\sqrt{s} \leq 2m_{t}$.  Figure~\ref{fig:nlclum} shows the luminosity for
the baseline design of the LEIR versus the center-of-mass energy.
In the following, we outline the basic LEIR physics program as a
function of increasing beam energy.

The lowest operating energy of the LEIR is determined by the requirement
that high-statistics studies at the $Z$-pole be possible.  The goal of a
next-generation $Z$-pole experiment would be a significant reduction in
the experimental errors in key electroweak parameters, as explained in
Chapter 8. The success of this program relies on the availability of
longitudinally polarized beams.  Polarized electron beams will be
available in the initial configuration.  It would be desirable
eventually to have positron polarization as well.  Issues and
technologies for positron polarization are discussed in Chapter 12.  One
feature pertaining to beam polarization in the LEIR is the need to
account for the spin precession in the bends in the beam transport
system.  Another issue is the desire to account for the depolarization
that arises during collision.  For this reason, a crossing angle is
desirable, since it eases the polarization measurement after the IP.

Precise determination of the electroweak parameters could be
particularly valuable in understanding the SM and physics beyond,
particularly at a time when the Higgs boson mass is experimentally
determined.  In the event that only a single Higgs boson is observed
with no other direct evidence of new physics from the LHC programs, the
precision electroweak measurements will be a crucial aspect of the NLC
program.  A benchmark for such a program would be to accumulate a sample
of $10^9$ $ Z^0$ decays.

The $W$-pair threshold occurs near $\sqrt{s} = 160$ GeV with the maximum
production cross section at $\sqrt{s} \sim 200$ GeV.   In the event that a significantly 
improved measurement of the $W$ mass is required, it will be necessary to have
dedicated running at the $W$-pair threshold.  Studies have shown that an
error on the $W$ mass of 6 MeV would be obtainable with 100 fb$^{-1}$.
Given the otherwise very limited physics program in this energy range,
the need for high instantaneous luminosity is evident.\footnote{Although
investigation of $W$-boson properties will be an important goal of any
NLC program, many of these studies, {\em e.g.}, the determination of
Triple Gauge Boson couplings, are best performed at the highest
achievable center-of-mass energy.  This issue is discussed in Chapter 5,
Section 2.}

Beyond the $W$-pair threshold, it is highly likely that next benchmark
center-of-mass energy will be the production cross section peak for a light Higgs
boson.  Precise measurements of the Higgs mass, width, spin-parity, and
branching fractions are essential to help understand the role this
object would play in electroweak symmetry breaking.  The associated
production process $ e^+e^- \rightarrow {Z}^0{h}^0$, with ${Z}^0
\rightarrow \ell^+\ell^-$ and $\ell$ an electron or muon, provides a
model-independent tag of Higgs production.  The Higgs signal is easily
identifiable in the dilepton recoil mass distribution.  The maximum
cross section for associated production occurs at roughly $m_{Z} +
\sqrt{2} m_{\rm h}$.  In minimal SUSY, the mass of the lightest CP-even
scalar is required to satisfy $m_{h} \lsim 135$ GeV.  The precision electroweak fit 
to the SM calls for a Higgs boson with mass below 200 GeV.  It is therefore
essential that the LEIR design be capable of delivering high luminosity
in the range $220 \lsim \sqrt{s} \lsim 340$ GeV.  The study of a light
Higgs boson will also benefit from control of the beam polarization; for
example, for the measurement of the $hWW$ coupling, one can exploit the
large difference in the $\nu \bar{\nu} h^{0} $ production
cross section for $e^-_L$ and $e^-_R$ beams.  For some processes,
positron polarization is also desirable.  In many scenarios, the
precision study of a light Higgs boson would be the principal focus of
the LEIR program.

The $t \bar t$ threshold occurs near 350 GeV.  The low-energy IR 
would be the natural facility to focus on this important topic.  The
threshold onset is a difficult process to study experimentally because
of the resolution smearing caused by the natural energy spread from
bremsstrahlung in the initial state, and from energy spread in the
linear collider.  The amount of dedicated running at the $t \bar t$
threshold will be dictated by the Higgs physics program.  If a light
Higgs is present, $m_{H} \lsim 180$ GeV, it may be desirable to run
below the $t \bar{t}$ threshold to control physics backgrounds and to
optimize the Higgs production rate.  For the case where the physics of
electroweak symmetry breaking has conspired to produce a heavy Higgs
boson that somehow satisfies the precision constraints, the study of the top
quark properties will assume a central importance.  The integrated
luminosity requirements for the LEIR at or above the $t \bar{t}$
threshold in such a scenario will be the order of 100 fb$^{-1}$
necessitating instantaneous luminosities of at least $5 \times 10^{33}$
cm$^{-2}$s$^{-1}$.

Other physics options for the low-energy IR have been considered
extensively.  The region would serve well as the location for a `second
generation' detector for $\gamma \gamma$ collisions.   Similarly, an $e^-e^-$
program might be done in the LEIR, should the physics motivations lead
in this direction.

In summary, a low-energy IR has many uses and advantages in an
NLC program.  It would provide considerable flexibility in the physics
program, and would preserve many physics opportunities in scenarios in
which the NLC is upgraded to multi-TeV operations for high-energy
studies in the other IR region.

\subsection {The high-energy interaction region at the NLC}

The design of the NLC allows for an IR region capable of upgrading to
multi-TeV operations in an energy-upgraded NLC.  To assure this
possibility, the beam delivery systems are aligned in a straight-ahead
configuration relative to their respective linacs, with very little
bending of the incoming beams between the linear accelerator structure
and the IR.  To preserve the non-zero crossing angle required at
the point of collisions, the two halves of the collider structure are
not parallel but rather cross at an angle at the collision point.
Figure~\ref{fig:nlclum} shows the luminosity versus CM energy for the
baseline design of the HEIR.

The HEIR physics program is intimately related to the scenario that is
realized in Nature for electroweak symmetry breaking.  In the event that
supersymmetry is discovered, the focus of the HEIR program will be the
measurement of sparticle properties.  It is unlikely that the
full SUSY spectrum will be accessible at $\sqrt{s} = 500$ GeV;
therefore, the energy reach of the HEIR should be upgradable to
the multi-TeV region.  Symmetry-breaking arising from some new strong
dynamics would also be likely to put a premium on the energy reach.  It is
clear that in comparison to the LEIR, the physics requirements for the
HEIR are, to first order, straightforward:  the highest possible
luminosity at the highest possible energy.

The energy span of the HEIR runs from 250 GeV to 500 GeV in the initial
phase.  Therefore the physics program can in principle include
everything from 250 GeV on up, a region which overlaps in energy with
the LEIR.  Studies of $W$-pairs, low-lying SUSY states, and the $t\bar
t$ threshold could occur in the HEIR.  Although, in the case of a light
Higgs boson, much of the precision Higgs physics could be
performed at the LEIR, there is Higgs physics unique to the HEIR.  For a
light Higgs boson consistent with the current theoretical and
experimental constraints, the maximum cross section for the rare process
$ e^+e^- \rightarrow {Z}^{0}{h}^{0} {h}^{0}$ occurs at $\sqrt{s} \sim$
500 GeV.  This process is of great interest, since it enables
measurement of the Higgs self-coupling which in turn can be related to
the shape of the Higgs potential.  The $W$-fusion process, $e^+e^-
\rightarrow \nu \bar{\nu} {h}^{0}$, which is sensitive the $hWW$ vertex,
has a cross section that increases with center-of-mass energy.  The
measurement of the Higgs self-coupling sets a benchmark for the
accelerator performance.  Depending on the exact mass, a measurement of
this quantity requires integrated luminosities the order of 1000
fb$^{-1}$, which corresponds to 3--4 years at design luminosity.

Supersymmetry is a primary candidate for physics
beyond the SM.  Almost all versions of SUSY models result
in low-lying states that would appear in $e^+e^-$ annihilations below
500 GeV.  Although the discovery phase for SUSY is likely to occur at
the Tevatron or LHC, the NLC will play a key role in the detailed study
of the sparticle spectrum and subsequent delineation of the soft
SUSY-breaking Lagrangian.

To exploit fully the physics potential of the NLC, a number of special
operating conditions may be necessary for the HEIR.  For example, in
EWSB models with extended Higgs sectors, of which SUSY is the most
widely studied, a $\gamma \gamma$ mode of operation for the HEIR may be
crucial.  For example, the $\gamma \gamma$ mode enables production of a
single Higgs boson; for the case of a nominal 500 GeV center-of-mass,
this would effectively increase the mass reach from 250 GeV to 400 GeV
for production of heavy neutral Higgs particles.  Operation with
transversely polarized photon beams allows separate production of the
CP-even and CP-odd states.  Control of the electron and positron beam 
polarization will also be
extremely useful.  For Higgs physics it can be used to increase the
nominal production cross section for the self-coupling measurement.
Beam polarization will also be useful in unraveling gaugino and slepton
mixing.  The need for an $e^-e^-$ operating mode may be necessary to
decipher selectron production.

It is likely and perhaps desirable that there be a staged evolution of
the HEIR center-of-mass energy.  Although the goal of the initial phase
of the NLC is 500 GeV for the HEIR, it may be possible to start physics
earlier at a lower collision energy.  An intermediate commissioning
stage with $\sqrt{s} \sim 250$ GeV and modest luminosity could
potentially be very relevant and exciting, especially if direct evidence
from the LHC indicated the production of a light Higgs boson or a threshold
for supersymmetric states.  Another obvious commissioning stage could be
the $t \bar t$ threshold at 350 GeV.  Even at 10\% of design luminosity,
the physics program promises to be rich.  For example, dedicating 10
fb$^{-1}$ to a scan of the $t \bar t$ threshold would already lead to a
top quark mass measurement with a 200 MeV error, as discussed in Chapter
6, Section 2.

\subsection{Alternative interaction region scenarios}

The baseline scenario that we have assumed considers two interaction
regions---a high-energy region limited only by the available
accelerating structures and a second region that is limited in energy
or by the support of $\gamma \gamma$ or other options.  It is
appropriate to discuss alternative scenarios and the interplay between
the physics programs of the high- and low-energy interaction regions.
The issue is complicated by the diversity of physics scenarios that may
arise.  An additional consideration is the possible staging of the maximum
center-of-mass energy.  The possibilities can be broadly classified into types:
\begin{itemize}
\item[a)] Single interaction region with one detector;
\item[b)] Single interaction region with two detectors;
\item[c)] Two interaction regions, high-energy and low-energy;
\item[d)] Two high-energy interaction regions.
\end{itemize}

For scenario (a), there is an obvious cost advantage; however, the NLC
physics program could be unduly compromised.  The physics program would
be tightly coupled to the available center-of-mass energy.  Depending on
the details of the actual physics scenario, it may not be possible to
simultaneously satisfy the various needs of a diverse user community.
The resolution of mutually exclusive requirements for
luminosity and choice of the center-of-mass energy may not be
straightforward.  

It is difficult to identify the merits of scenario (b), given the
limitations of a single IR outlined above for scenario (a).  Given that
the total luminosity accumulated by both experiments will be comparable
to that for a single experiment, this scenario would only be of interest
if the two detectors were of sufficiently  different capabilities or there
were very strong sociological arguments for a second collaboration.  One
possible scenario where differences between detectors could arise is if
there were a need to have a dedicated $\gamma \gamma$ collider program.
In such a scenario, it would be more natural to consider a push-pull
capability for one of the IRs in a two-IR facility.  The two IR regions
allow for a push-pull configuration in a least one of the two regions.
The footprint of the push-pull IR hall must not infringe on the beamline
of the adjacent region.  In addition, access to the detector captured
between the two beamlines must be possible, and adequate shielding must
be provided to permit work in the IR hall when beams are alive in the
machine.  Scenarios for staging two detectors would have to be
considered and understood.  These are complicated issues that would
involve assumptions that might not be appropriate at a future date.
Nevertheless, provision for staging two detectors in a push-pull
configuration would be a low-cost and effective means to keep open
future possibilities for a unique and special-purpose detector.

The scenario that has been chosen as the baseline is (c); there are a
number of considerations in its favor.  It makes it possible to have
parallel physics programs running simultaneously, a clearly desirable
feature.  The upgrade path for the HEIR is less complex.  It provides
for a lower-energy IR that can be dedicated to precision studies of the
Higgs boson, $Z$-pole or $t \bar{t}$ system.  Moreover, in this scenario
both the HEIR and LEIR will cover the preferred energy range for the
study of a light Higgs.  The two-IR design adds a degree of flexibility
that enables the NLC to address essentially any physics scenario that
could arise.

The scenario (c) affords a natural context for energy staging.  As
mentioned in Section 2.2, staging the HEIR energy at the beginning of
the NLC program would make it possible to perform an initial
investigation of the region above 250 GeV.  Commissioning of the LEIR
program might follow the completion of the full complement of
accelerating structures required to reach 500 GeV though, with a bypass
line, this might alternatively begin before the accelerator is complete.
Many of the high-luminosity measurements foreseen for the LEIR would
benefit from longitudinally polarized positron beams, which are not
likely to be available at the initial stages of running.

Given the need to have minimal bending in the beam delivery system in
order to preserve beam emittances, scenario (d), which has two 
high-energy IRs of similar performance, becomes technically challenging and
more costly.  Given the interest exhibited by many members of the
physics community in the low-energy potential of the NLC, and the need to
perform high-statistics studies of the $Z$-pole in a number of physics
scenarios that could arise, it would seem prudent to have at least one
IR capable of delivering that physics.

\subsection {Simultaneous operation}

The NLC design has in it the capability for simultaneous operations in
the two IRs.  In the baseline design, the accelerator delivers bunch
trains at a rate of 120 Hz.  With pulsed magnets, the beams can be sent
alternately to two IRs, resulting in an even split of 60--60 Hz.  Uneven
splitting of the 120 pulses per second is technically more challenging,
and is not envisioned as an option.

A higher pulse rate in the NLC is possible, but is not in the baseline
design.  It appears technically feasible, for example, to operate at 180
Hz.  This would require modifications to the damping rings and
additional cooling for the klystrons and modulators in some regions of
the accelerator.  But these changes would allow operation, for example,
with 60 Hz of low-energy beams in the LEIR and 120 Hz of beams in the
HEIR.  This mode of operation would clearly enhance the experimental
program and augment the total luminosity delivered to the experimenters.

\setcounter{chapter}{11}

\chapter{Positron Polarization}
\fancyhead[RO]{Positron Polarization}
\section{Introduction}

The baseline designs for NLC and TESLA include a polarized electron
beam, but the positron beam is unpolarized.  In this chapter, we
investigate the physics merits of positron polarization and summarize
the status of proposed polarized positron source 
designs.  These questions have also been discussed 
in~\cite{Moortgat-Pick:2001kg}.

The importance of electron beam polarization has been demonstrated in
 $Z^0$ produc\-tion at the Stanford Linear Collider (SLC), where $75\%$ 
electron
 polarization was achieved.  This level of electron polarization provided
 an effective luminosity increase of approximately a factor of 25 for
 many $Z$-pole asymmetry observables.  In particular, it allowed the SLD
 experiment to make the world's best measurement of the weak mixing
 angle, which is a key ingredient for indirect predictions of the SM
 Higgs mass.  The electron polarization at SLC also provided a powerful
 tool for bottom quark studies, providing a means for $b$ and $\bar b$
 tagging from the large polarized forward-backward asymmetry,  and for
 studies of parity violation in the $Zb\overline{b}$ vertex.  At a 500
 GeV linear collider, electron polarization will increase sensitivity to
 form-factor studies of $W^+W^-$ and $t\overline{t}$ states, control the
 level of $W^+W^-$ backgrounds in new physics searches, provide direct
 coupling to specific SUSY chiral states, and enhance
 sensitivity to new physics that would show up in the spin-zero channel.

 But what will positron polarization add?  First, the presence of
 appreciable positron polarization is equivalent to a boost in the
 effective electron polarization.  Measured asymmetries that are
 proportional to the polarization will increase; fractional errors in
 these quantities will accordingly decrease.  Second, cross sections for
 many processes will grow.  Any process mediated by gauge bosons in the
 $s$-channel naturally wastes half the incident positrons.  Left-handed
 electrons, for example, only annihilate on right-handed positrons.  The
 same is true for $t$-channel exchanges with unique handedness in their
 couplings, such as neutrino exchange in $W$-pair production.  By
 polarizing the positrons and coordinating their polarization with that
 of the electrons, the cross sections for these processes can double (in
 the limit of 100\% polarization).  
 Finally, polarimetry will benefit
 from positron polarization.  As the effective polarization increases,
 its error decreases, allowing measurements with very small systematic
 errors.  Such small errors are needed for high-precision work at the
 $Z$ pole and will benefit studies of production asymmetries for
 $W^+W^-$.  And, by using measurements of rates with all four helicity
 states (RL,LR,RR,LL) the beam polarizations can be inferred directly
 without additional polarimetry. 

  What positron polarization can bring,
 poor yields of polarized positrons can take away, so the yield of any
 source of polarized positrons is very important.  Several schemes have
 been advanced for polarizing positrons.  All are ambitious, large
 systems which are mostly untested.  R$\&$D is required before decisions
 are made about how and when to include positron polarization in linear
 collider design.

 \section{The physics perspective}

 \subsection{The structure of electroweak interactions at high energies}
 \label{sec:one}

 The primary purpose of a linear collider will be to study the mechanism
 of electroweak symmetry breaking (EWSB).  Beam polarization at a
 high-energy linear collider can play an important role in this endeavor
 because:  (1) the electrons and positrons in the beams are essentially
 chirality eigenstates; (2) gauge boson interactions couple $\eml\epr$ or
 $\emr\epl$ but not $\eml\epl$ or $\emr\epr$; and (3) the $SU(2)_L$
 interaction involves only left-handed fermions in doublets, whereas
 right-handed fermions undergo only hypercharge $U(1)_Y$ interactions.
 At typical LC energies, where masses are small compared to $\sqrt{s}$,
 one can replace the exchange of $\gamma$ and $Z$ bosons with the $B$ and
 $W^3$ bosons associated with the unbroken $U(1)_Y$ and $SU(2)_L$.

 As a concrete application of these points, consider $e^+e^-\to W^+W^-$
 production, which is a background to many new physics searches.  There
 are three tree--level Feynman diagrams for this process, one involving
 the $t$-channel exchange of $\nu_e$ and the others involving the
 $s$-channel exchange of $\gamma$ and $Z$.  The polarization choice
 $e^+e^-_R$ will eliminate the first contribution, since $W$ bosons have
 only left-handed interactions.  Decomposing the $s$-channel diagrams into
 a $W^3$ and a $B$ contribution, the $W^3$ diagram is also eliminated
 using $e^-_R$ polarization for the same reason.  The only remaining
 diagram now vanishes for symmetry reasons---the $B$ and $W$ bosons
 involve different interactions and do not couple to each other.  In
 reality, there is a small but non-vanishing component to $W^+W^-$
 production, because of EWSB.  The polarization choice $e^+_R$ would
 eliminate this background at tree-level.  Of course, it also important
 to consider the behavior of the signal process under the same choices of
 polarization and the fact that 100\% beam polarization is difficult in
 practice.

 In the example above, note how the polarization of only one beam had a
 dramatic effect.  Once the electron polarization was chosen, only
 certain positron polarizations contributed.  One can imagine also the
 case where the desired effect is to enhance the $W^+W^-$ signal.  Then,
 by judiciously choosing the polarization combination $e^-_L e^+_R$, the
 production rate is enhanced by a factor of four relative to the unpolarized
 case, and a factor of two beyond what is possible with only electron
 polarization.  When either searching for rare processes or attempting
 precision measurements, such enhancements of signal and depletions of
 background can be quite important.

We use the convention that the sign of polarization is {\em positive} for
{\em right-handed} polarization, both for electrons and for positrons.
Then, for example, 
 for the case of single gauge boson production, the production cross section
 is proportional to
 \begin{eqnarray}
 (1-\pem)(1+\pep)c_L^2 + (1+\pem)(1-\pep)c_R^2,
 \label{eq:poldep}
 \end{eqnarray}
 where $c_L$ and $c_R$ are chiral couplings.  Equation~(\ref{eq:poldep}) is at
 the heart of the forward-backward asymmetry that arises when $c_L\ne c_R$.
 If two measurements of the cross section are made with a different {\it sign}
 for the polarizations $\pem$ and $\pep$, then the difference of the two measurements
 normalized to the sum is:
 \begin{eqnarray}
 {{N_L-N_R}\over{N_L+N_R}}\quad  = \quad {\cal P}_{\rm eff}{c_L^2 - c_R^2
 \over c_L^2 + c_R^2}\quad  \equiv \quad
 {\cal P}_{\rm eff}A_{LR},
 \end{eqnarray}
 where
 \beq
 \peff = {{\pem - \pep} \over {1-\pem\pep}} \ .
 \eeq{definepeff}
 In $Z$ boson production, $A_{LR}$ depends on the difference between
 $1/4$ and $\sin^2\theta_W$.  Since the error in an asymmetry $A$ for a
 fixed number of events $N=N_L+N_R$ is given by $\delta A = \sqrt{
 (1-A^2) /N}$, increasing ${\cal P}_{\rm eff}$ makes measurable
 asymmetries larger and reduces the error in the measured asymmetry
 significantly if $A^2$ is comparable to 1. When only partial electron
 polarization is possible, a small positron polarization can
 substantially increase $\peff$, while also decreasing systematic errors.
 These asymmetry improvements utilizing polarized positrons are exploited
 in the Giga-Z mode for a linear collider.  With Giga-Z, polarized
 positrons are needed to take full advantage of the large statistics
 possible at a linear collider---50 times more data than the integrated
 LEP-I data sample and 2000 times more data than SLD's sample.  With a
 Giga-Z data sample, one expects to achieve a factor of 20 improvement over
 SLD's $A_{LR}$ and $A_b$ measurements.  These improved measurements can
 be used to perform exquisite tests of the Standard Model.  Together with
 a precise measurement of the top quark mass (to 100 MeV from a threshold
 scan at a linear collider), the $A_{LR}$ measurement can be used to
 predict the Standard Model Higgs mass to 7\%.  The Giga-Z program is
 discussed in more detail in Chapter 8.

 Equation~(\ref{eq:poldep}) is also applicable to other situations.  In
 general, as long as a process has a helicity structure similar to that
 of $s$-channel gauge boson production, the rate is \begin{eqnarray}
 (1-\pem\pep)\sigma_{\rm unpol} \left(1 + {\cal P}_{\rm eff}{c_L^2 -
 c_R^2 \over c_L^2 + c_R^2}\right), \end{eqnarray} where $\sigma_{\rm
 unpol}$ is the unpolarized cross section.  Notice that polarization can
 increase the cross section by at most a factor of four, as can occur for
 $W^+W^-$ production where $c_R\simeq 0$.

 \subsection{Standard  Model-like Higgs boson}
 \label{sec:twoa}

 One process of particular interest for a LC is Higgs boson production.
 The primary modes at a LC are associated production with a $Z$ boson
 ($Zh$) and vector boson fusion ($\nu\bar\nu h$).  The $Zh$ process is
 particularly simple, since the direct coupling of the Higgs boson to
 electrons is negligible.  Polarization effects appear only at the
 initial $e^+e^-Z$ vertex.  The $Z$ process allows for the discovery and
 study of a Higgs boson with substantial couplings to the $Z$ boson
 independently of the Higgs boson decay mode, using the $Z$ recoil
 method.  Therefore, the relative size of signal and background is of
 great interest.

 \begin{table}[!ht]
 \begin{center}
 \begin{tabular}{|c|c||c|c||c|c||c|c|}    \hline\hline
 \multicolumn{2}{|c||}{ } & \multicolumn{2}{c||}{${\sigma(Zh)}$} &
  \multicolumn{2}{c||}{${\sigma(ZZ)}$} &
 \multicolumn{2}{c|}{${\sigma(W^+W^-)}$} \\
 \multicolumn{2}{|c||}{ } & \multicolumn{2}{c||}{$c_L^2=.58~~c_R^2=.42$}
  & \multicolumn{2}{c||}{$c_L^2=.65~~c_R^2=.35$} &
 \multicolumn{2}{c|}{$c_L^2\simeq 1~~c_R^2\simeq 0$} \\ \cline{3-8}
 \cline{3-8}
 \multicolumn{2}{|c||}{ } &    $E=1$    & $E=.8$ & $E=1$  & $E=.8$ & $E=1$
   & $E=.8$  \\
 $\pem$  & $\pep$ &    $P=1$    & $P=.6$ & $P=1$  & $P=.6$ & $P=1$  & $P=.6$
  \\ \hline\hline
   0     &  0     &       1       &     1        &  1       &     1
    &   1      &       1       \\ \hline
 $+E$    &  0     &      0.84     &    0.87      &   0.69   &    0.75
     &      0     &    0.2    \\
 $-E$    &  0     &      1.16     &    1.13      &   1.31   &    1.25
     &      2     &    1.8    \\
 $+E$    &  $-P$  &      1.68     &    1.26      &   1.37   &    1.05
     &     0     &     0.08   \\
 $-E$    &  $+P$  &      2.32     &    1.70      &   2.62   &    1.91
     &      4     &    2.88   \\ \hline\hline
 \end{tabular}
 \end{center}
 \caption{Behavior of various Standard Model cross sections relevant
 for Higgs boson studies as a function
 of polarization for full and partial electron and positron polarization.
 The numbers listed are normalized to the unpolarized cross section.
 \label{tab:one}}
 \end{table}

 At tree-level, the $Zh$ cross section depends on polarization as
 indicated in Eq.  \leqn{eq:poldep} with the couplings $c_L= - \half +
 \sstw$, $c_R = \sstw$.  Numerically, the two squared coupling factors
 appear with the relative weights (normalized to unity) $0.58$ to $0.42$.
 Table~\ref{tab:one} shows the relative behavior of the $Zh$ cross
 section for full (100\%) and partial electron (80\%) and positron (60\%)
 polarization.  Even for partial polarization, a
 substantial increase to the production cross section occurs over the
 unpolarized case.  Other Higgs boson production processes, such as
 $e^+e^-\to HA$ in the MSSM or $e^+e^-\to Zhh$ in the SM or MSSM
 (relevant for measuring the Higgs self-coupling), proceed through the
 $Z$ resonance and have the same chiral structure.

 Significant backgrounds to the $Zh$ search can arise from $W^+W^-$ and
 $ZZ$ production.  The polarization dependence of these processes is also
 shown in Table~\ref{tab:one}.  The physics of the $W^+W^-$ background
 was discussed previously.  It is relevant to note from
 Table~\ref{tab:one} that without full polarization---which may be
 difficult to obtain in practice---the $W^+W^-$ background cannot be
 fully eliminated.  On the other hand, the partial polarization of both
 beams can approximately recover the benefits of full polarization, since
the effective polarization $\peff$ is close to 1.
 Another potential background, $ZZ$ production, has a similar behavior as
 the signal $Zh$, except that an additional $Z$ must be attached to the
 incoming $e^+e^-$.  Therefore, the relative weight of the different
 polarization pieces goes as the square of those for $Zh$ production.
 For the case of partial polarization of both beams and
 ($\pem=+80\%,\pep=-60\%$), where the $W^+W^-$ background is
 substantially decreased, there is a small increase in
 $\sigma(Zh)/\sigma(ZZ)$.  The efficacy of polarization will depend on
 the most significant background.  Note that for a Higgs boson mass that
 is significantly different from $m_Z$, propagator effects and
 non-resonant diagrams need to be included, but the results should not be
 significantly different from those shown here.

 The other Higgs production process of interest is $WW$ fusion, which has a
 similar behavior to the $WW$ background.  When operating at energies
 where $Zh$ and $WW$ fusion are comparable, polarization can be used to
 dial off the fusion contribution.  This may be important for
 the study of inclusive Higgs production using the recoil technique.

  \

 \subsection{Supersymmetric particle production}
 \label{sec:thra}

 The production and study of new particles with electroweak quantum
 numbers should be the forte of a linear collider, where the major
 backgrounds are also electroweak in strength.  Supersymmetry is a
 concrete example of physics beyond the SM that predicts a spectrum of
 new electroweak states related to the SM ones by a spin transformation.
 We now discuss some aspects of supersymmetry measurements affected by
 beam polarization.  For further discussion of supersymmetry mass and
 coupling measurements, see Chapter 4.

 \subsubsection{Slepton and squark production}

 One of the simplest sparticle production processes to consider is $\widetilde \mu$
 pair production, where the interaction eigenstates $\smur$ and $\smul$
 are expected to be nearly mass eigenstates.  Gauge bosons couple to the
 combinations $\smur\smur^*$ and $\smul\smul^*$.  $\smur$ has only
 couplings to the hypercharge boson $B$. The initial $e^+e^-$ state
 has different hypercharge depending on the electron polarization:
   $e^-_L$ has $Y=-1/2$, whereas $e^-_R$
 has $Y=-1$.  The production cross section depends on $Y^2$ and thus is
 four times larger for $e_R^-$ than for $e_L^-$.  Furthermore, the choice
 $e_R^-$ significantly reduces the background from $W^+W^-$ production,
 which comes both from decays to $\mu^+\nu_\mu\mu^-\bar\nu_\mu$ and from
 feed-down from decays to $\tau$.  Since $e_R^-e_R^+$ components do not
 contribute to the signal, left-polarizing the positron beam doubles the
 signal rate.  $\smul$ pair production depends on both $B$ and $W^3$
 ($\gamma$ and $Z$) components.  Switching the electron polarization will
 emphasize different combinations.  In all, a judicious choice of the
 positron polarization will make more efficient use of the beam, increase
 the cross section, and suppress the backgrounds.

 For third-generation sparticles such as $\stau$ and $\st$, there may be
 significant mixing between the mass and interaction eigenstates, leading
 to new observables.  As for the $\smu$ case, the production cross
 section itself is sensitive to the electron polarization.  However,
 increased sensitivity to the mixing may be obtained from a measurement
 of the left-right asymmetry.  For $\st$ production, the addition of 60\%
 polarization in the positron beam increases the accuracy of the mixing
 angle measurement by 25\%, while decreasing systematic
 errors~\cite{Bartl:2000wj}.  Of course, the former effect can be
 achieved with only $e^-$ polarization by increasing the integrated
 luminosity.

 Selectron  production may benefit more from positron
 polarization because of the $e^+e^-$ initial state at a LC.  The
 exchange of neutralinos $\widetilde \chi^0$ in the $t$-channel
 introduces more structure beyond the $s$-channel exchange of $\gamma$
 and $Z$.  The processes $e_L^- e_L^+\to \sel\ser^*$ {and} $e_R^-
 e_R^+\to \ser\sel^*$ proceed through $\nino$ exchange only.  Considering
 the case that $\sel$ and $\ser$ are close in mass, the polarization of
 both beams can play an essential role in disentangling the different
 interaction states.  For example, $e_L^- e_L^+$ polarization will only
 produce the negatively-charged $\sel$ and the positively-charged
 $\ser^*$.  Switching the polarization of both beams will produce only
 negatively-charged $\ser$ and positively-charged $\sel^*$.  Since the
 endpoints of the lepton spectrum can be used to reconstruct the
 selectron and neutralino masses, the electrons and positrons yield
 separate information about $\sel$ and $\ser$.  Without the positron
 polarization, one would always have contamination from $\sel\sel^*$ and
 $\ser\ser^*$ production.  Conversely, the observation of the switch from
 one species to another with the change in positron polarization would
 give more weight to the SUSY interpretation of the events.  The study of
 $t$-channel exchange in selectron production is an important method for
 studying neutralino mixing, since the components of the neutralinos that
 are Higgsino-like do not contribute.  Therefore, it is valuable to be
 able to isolate the $t$-channel exchanges experimentally by using
 polarization.

 \subsubsection{Chargino and neutralino production}

 The study of chargino pair production $e^+e^-\to\cinom\cinop$ gives
 access to the parameters $M_2$, $\mu$, $\tan\beta$, $m_{\widetilde
 \nu_e}$.  It is conservative to assume that only the lightest chargino
 is kinematically accessible.  In this case, studies have considered the
 case of extracting the SUSY parameters from the measurement of cross
 sections for full $\eml\epr$ ($\sigma_L$), $\emr\epl$ ($\sigma_R$) and
 transverse ($\sigma_T$) polarizations \cite{pChoi:2000hb}.  By analyzing
 $\sigma_R$ and $\sigma_L$, the two mixing parameters of the chargino
 sector can be determined up to  at most a four-fold ambiguity,
 provided that the electron sneutrino mass is known and one assumes the
 supersymmetric relation between couplings in the interaction Lagrangian.
 The addition of transverse polarization allows the ambiguity to be
 resolved and gives a handle on the sneutrino mass.  The role of {\it
 transverse} polarization is to allow interference between two different
 helicity states so that a product of two mixing factors appears in a
 physical observable instead of sums of squares of individual mixing
 factors, resolving the sign ambiguity.  Given the
 measurement of the chargino mass and the mixing parameters, the
 Lagrangian parameters $M_2, \mu, \tan\beta$ can be determined up to 
 two-fold ambiguity in modulus and a 2$\pi$ ambiguity in the phase
 combination ${\rm arg}(m_2) + {\rm arg}(\mu)$.  Such studies need to be
 redone with more detail, considering partial beam polarization,
 backgrounds, cuts, and the likely absence of transverse polarization,
 but there is promise that SUSY parameters can be extracted from real
 data.

 Other investigations have considered the consequences of partial
 longitudinal polarization at a purely theoretical level, focusing on the
 case $|\pem|=.85, |\pep|=.60$, and studying production cross sections
 near threshold~\cite{Moortgat-Pick:2000uz}.  Comparing a gaugino-like
 and Higgsino-like chargino, the total cross sections including the decay
 $\cinom\to e^-\bar\nu \widetilde\chi^0_1$ are calculated as a function 
 of electron and positron
 polarization.  For an unpolarized positron beam, 
 the cross sections from $\eml$ are larger than those from $\emr$  for
 both the gaugino and Higgsino cases.
 However, the addition of positron polarization gives access to more detailed
 information.  For example, one has the relation that $\sigma (e^-_Re^+_L)$
is less than the unpolarized cross section 
 for gaugino-like charginos,
 and greater  for Higgsino-like charginos.  The sensitivity
 of the forward-backward asymmetry $A_{FB}$ to polarization, and how this
 effect can be used to bound the sneutrino mass, has also been
 discussed \cite{Moortgat-Pick:1999ck}.  Similar considerations can be
 applied to the case of $\nino\nino$ production.  These analyses would
 benefit from more detailed studies, including backgrounds and addressing
 the issue of measuring branching ratios.

 \subsection{Some other new physics}

 Contact interactions can arise from many sources of new physics, such as
 compositeness, a heavy $Z^{\prime}$, leptoquarks, KK excitations, \etc\
 The low-energy effect of such physics can be parameterized in an
 effective Lagrangian as
 \begin{eqnarray*}
 {\cal L}_{\rm eff} 
= {\tilde g^2 \over \Lambda_{\alpha\beta}}\eta_{\alpha\beta}
 (\bar e_\alpha \gamma_\mu e_\alpha )
      (\bar f_\beta \gamma_\mu f_\beta), f\ne e,t.
 \end{eqnarray*}
 The chiral components are extracted by varying $P_{\rm eff}=\pm P$ (this
 is just $A_{LR}$).  Positron polarization increases the reach on
 $\Lambda_{\alpha\beta}$ by $20-40\%$ depending on the nature of the
 couplings \cite{Babich:2000kx}.

 Low-energy signatures of string theory may include
 spin-zero resonances with non-negligible couplings to
 the electron and sizable amplitudes \cite{Cullen:2000ef}, \ie,
   $\A(e^-_R e^+_R \to \gamma_{03}^*)  = \sqrt{2} e M_S$ and
   $\A(e^-_L e^+_L \to \gamma_{04}^*)  = \sqrt{2} e M_S$.
 With positron polarization, the SM backgrounds to these processes should be
 negligible.

 \subsection{Transverse polarization}

 Finally, we should comment on transverse polarization, which has been
 considered in some chargino studies.  Transversely polarized beams are
 linear combinations of different helicities with equal weight.  
 Transverse polarization
 can introduce an azimuthal dependence into production cross sections,
  proportional to the degree of
 polarization.  However, all such effects
 in the SM are negligible upon azimuthal averaging for an $e^-e^+$
 collider, because of the small electron mass and Yukawa 
coupling~\cite{Hikasa:1986qi}.  Thus, transverse polarization can be used as a
 probe of physics beyond the SM, when small amplitudes from new physics
 interfere with larger SM ones.  Without the positron polarization,
 however, there is no visible effect.

 \section{Experimental issues}
 \subsection{Polarimetry}

 The baseline NLC design includes a laser-backscattering Compton
 polarimeter to measure the electron beam polarization with an expected
 accuracy of $1\%$ or better~\cite{Woods,Rowson}.  For the Giga-Z physics
 program, an accuracy of 0.25$\%$ should be achievable in an optimized
 setup, which is a factor two improvement over SLD's Compton polarimeter.
 Above the $W$-pair threshold, the SM asymmetry in forward $W$ pairs can also
 be used~\cite{Woods}.  Sub-$1\%$ polarimetry using this technique will
 require reduction of the background to the $W$-pair sample below $1\%$.

 If the positron beam can also
 be polarized,
 significant improvements in polarimetry are possible.  At Giga-Z, the
 polarimetry error can be improved to $0.1\%$ using the `Blondel scheme'.
 In this method, one
 measures the three independent asymmetries~\cite{Blondel,Gambino}:
 \begin{eqnarray}
 A_1 & = & {{N_{LL}-N_{RR}} \over
	    {N_{LL}+N_{RR}}} \nonumber \\
 A_2 & = & {{N_{RR}-N_{LR}} \over
	    {N_{RR}+N_{LR}}} \nonumber \\
 A_3 & = & {{N_{LR}-N_{RL}} \over
	    {N_{LR}+N_{RL}}} = \peff \alrm,
 \end{eqnarray}
 where $\peff$ is given by Eq.  \leqn{definepeff}.  From these three
 measurements, one can determine \alr\ (and hence the weak mixing angle)
 along with $\pem$ and $\pep$.  It should be noted that $\peff$ is
 typically substantially higher than either $\pem$ or $\pep$ and has a
 smaller uncertainty.  For example, if $\pem = 80\%$ and $\pep= -60\%$,
 then $\peff = 94.6\%$, and the error on $\peff$ is proportional to the
 difference from 100\%.  With a Giga-Z sample using these
 polarization values, $\alrm$ can be determined to an accuracy of
 $10^{-4}$ and the beam polarizations to an accuracy of $10^{-3}$.  These
 estimates are derived in Chapter 8, Section 1. An advantage of the
 Blondel scheme for polarimetry is that the luminosity-weighted
 polarization, \pelum, is directly measured.  A Compton polarimeter
 measures the average beam polarization and small corrections may be
 needed to extract \pelum.  It should be noted that a Compton polarimeter
 is still needed to measure the difference between the right-handed and
 left-handed beam polarizations.  One also needs to understand the
 relative luminosities for the four beam polarization states (at the level
 $10^{-4}$ for Giga-Z).

Away from the $Z$-pole, the Blondel scheme with polarized positrons
can also be applied to  $W$-pair events.  Using $W$ pairs
when both beams are polarized, an error on the beam polarizations of
0.1$\%$ should be achievable.  The large $W$-pair physics asymmetry can be
fit together with the beam polarizations, without sensitivity to
backgrounds or assumptions about the polarization asymmetry in $W$
interactions.

\subsection{Frequency of spin flips}

Depending on the method for producing polarized positrons, it may be
difficult to achieve fast reversals of the positron helicity.  For the
polarized electron source, helicity reversals are easily done at the
train frequency (120~Hz for NLC or 5~Hz for TESLA) using an
electro-optic Pockels cell in the polarized source laser system.  At
SLC, the 120 Hz random helicity was very useful in controlling possible
small left-right asymmetries in luminosity.  Helicity reversals that are
fast compared to any time constants for machine feedbacks are desirable.
If fast helicity reversals are not possible, then relative integrated
luminosities for the different polarization states need to be measured
to better than $10^{-4}$ for Giga-Z.  This should be achievable using
forward detectors for Bhabha and radiative Bhabha events.

\subsection{Run time strategy for LL, LR, RL, RR}

One of the advantages of polarizing the positron beam is the increase in
event rate by running in the (higher cross section) LR or RL
polarization states.  However, to take advantage of the Blondel
technique for polarimetry and $\alrm$ measurements, it is necessary also to
accumulate data in the LL and RR states.  However, it has been
shown that only 10$\%$ of the running time has to be spent in the 
lower-event rate LL and RR states to achieve adequate statistics for the
asymmetry measurements \cite{monig}.  One anticipates equal run times
for the LR and RL configurations, even though some physics analyses may
benefit most from selecting one of these configurations for enhancing or
suppressing $W$ pairs or to enhance a cross section for a new process.
Of course, some new physics searches will benefit from choosing those
configurations that are suppressed in the SM.

\section{Sources of polarized positrons}

Several techniques have been suggested for producing polarized positrons
for a linear collider.  Present designs are largely conceptual, and much
work remains before they can be realized.

In 1979, Mikhailichenko and Balakin~\cite{undulator} proposed generating
circularly polarized photons by running a high-energy electron beam
through a helical undulator.  These photons are directed onto a thin
target, where they produce $e^+e^-$ pairs.  Selecting positrons near the
high-energy end of the spectrum gives a sample with appreciable
polarization.  Okugi \etal~\cite{omori} have proposed generating
polarized photons by colliding intense circularly polarized laser pulses
with few-GeV electron beams.  Variations on this theme have been
proposed in an attempt to mitigate the rather extreme requirements on
laser power by using an optical cavity to concentrate and store multiple
laser pulses \cite{Frisch,poty}.  Finally, Potylitsin~\cite{ebeam} has
proposed directing a 50 MeV beam of polarized electrons onto a thin
target.

\subsection{Helical undulator}

In the baseline TESLA design, unpolarized positrons are generated by
photons produced when the full-energy electron beam is passed through a
100~m long wiggler prior to collision.  The photon beam is directed to a
thin, rotating target where $e^+e^-$ pairs are produced, and the
positrons are subsequently captured, accelerated, and damped.  This
novel approach reduces the power dissipated in the positron target to
manageable levels and significantly reduces radiation in the target
area.

Replacing the wiggler with a helical undulator would in principle allow
polarized positrons to be produced.  The magnetic field created by a
helical undulator has two transverse components that vary sinusoidally
down the length of the device, the vertical component shifted in phase
by $90^{\circ}$ from the horizontal.  Such a field is created by two
interleaved helical coils of the same handedness, driven by equal and
opposite currents.  Typical fields are of order 1 T; the period of the
sinusoidal field variation is about 1 cm.  The resulting electron
trajectory for a 150 GeV beam is a helix whose axis coincides with that
of the undulator; the radius of curvature is measured in nanometers!
The undulator coils must be quite compact, with an internal radius of
several millimeters and an outer radius of about 1
centimeter~\cite{Flottmann}.

Efficient positron production requires photon energies of about 20 MeV,
which in turn necessitates electron beam energies of approximately
150--200 GeV.  The photons produced within $\theta \approx 1/ \gamma$
have high average polarization.  Collimators which are arranged to
absorb the radiation at larger angles remove about $80\%$ of the flux.
To compensate this loss, the undulator length must be about 200 meters,
somewhat longer than that of the wigglers used in the TESLA positron
source.  The undulator requires a very low-emittance electron beam,
which probably prevents reuse of the electron beam after it has been
used for high-energy collisions.  It is possible that one could direct
the primary high-energy electron beam through the undulator prior to
collision.  A drift space of about 200 meters between the undulator and
the target is required to achieve the required photon beam size.

The highly polarized photons produced in the undulator are directed
against a 0.4 $X_0$ target, where pair production can occur.  Positrons
produced with energies above 15 MeV are highly polarized.  With this
energy cut, roughly 0.025 $e^+$/incident photon is collected and $60\%$
polarization is obtained~\cite{Flottmann}.  Collection of the positrons
requires solenoidal magnets, rf acceleration, and a predamping ring to
handle the enlarged phase space.  On paper, the scheme can generate the
needed positron bunch currents.

The undulator scheme makes excellent use of the high-energy electron
beam as the source of polarized photons.  The low emittance requirements
probably preclude the use of the post-collision beam.  Whether the
primary, pre-collision beam should be run through the undulator, or a
dedicated beam should be generated for the sole purpose of positron
production is a choice still being debated.  A helical undulator
generates positrons of a single helicity, so other means must be
developed to flip the spin, and preferably to do so rapidly.  Many of the
photons could be absorbed in the undulator coil, so a workable design
must accommodate many kilowatts of power dissipation.

\subsection{Backscattered laser}

A second method for producing highly polarized photons with enough
energy to produce electron-positron pairs on a thin target involves
backscattering an intense circularly polarized laser beam on a 
high-energy electron beam.  The highest energy photons are strongly polarized
and have helicity opposite to that of the incident laser light.  As
above, positrons are produced when these photons intercept a thin
target.  The highest-energy positrons are strongly polarized.

Omori and his collaborators have made a conceptual design of a 
laser-backscattering polarized positron source suitable for
NLC/JLC~\cite{omori2}.  They arrange for multiple collisions between
polarized laser pulses from 50 CO$_2$ lasers and a high-current 5.8 GeV
electron beam.  The laser system must provide 250 kW of average optical
power, which is regarded as extremely ambitious.  Positron production is
accomplished just as in the helical undulator scheme above.  Simulations
indicate that $9.4\%$ of the incident photons produce a positron above
20 MeV, $26\%$ of which are accepted into the pre-damping ring, with an
average polarization of $60\%$~\cite{omori2}.

This scheme makes production of polarized positrons independent of the
high-energy electron beam, hence independent of its energy, but does so
at the very considerable expense of a dedicated high-current linac and a
very complex laser system.  The estimated power required by those
systems is roughly $10\%$ of that required for the whole collider
facility.

\section{Conclusions}

A polarized positron beam at a LC would be a powerful tool for enhancing
signal-to-background, increasing the effective luminosity, improving
asymmetry measurements with increased statistical precision and reduced
systematic errors, and improving sensitivity to non-standard couplings.
  Suppression of $W$-pair backgrounds can be improved
by a factor 3 with 60$\%$ positron polarization.  By limiting the
running time allotted for LL and RR modes to 10\%, the effective
luminosity for annihilation processes 
can be enhanced by 50\%.  For asymmetry measurements, the
effective polarization is substantially increased (\eg, from 80\% to
95\%) and the systematic precision is improved by a factor 3. With these
features, a polarized positron beam may provide critical information for
clarifying the interpretation of new physics signals.  Polarized
positrons are needed to realize the full potential for precision
measurements, especially those anticipated for Giga-Z running at the
$Z$-pole.

Designs of polarized positron sources have not reached maturity. 
Several approaches have been proposed, the most promising of
which uses a helical undulator, but to date no real engineering designs,
cost estimates, or experimental proofs of principle are available.
Since much of the benefit of a polarized positron source would be
negated if luminosity were compromised, it is very important that
eventual designs have some margin on projected yields.  Also, the source
needs to be available for all collision energies.  The helicity of a
polarized positron source may be difficult to switch quickly and
provision needs to be made to allow this, with a strong motivation to
have helicity-switching capability at the train frequency.  Present
designs must be further developed and additional R$\&$D is needed to
pursue new schemes, some of which have been mentioned here.

Though a polarized positron source is not yet advanced enough to be included as
part of the baseline linear collider design,  it is an
attractive feature that should be pursued as an upgrade.  Site layout
and engineering for a linear collider baseline design should accommodate
such an upgrade at a later date.  This has been done for the TESLA
design and needs to be done for the NLC design as well.

 \emptyheads
\blankpage \thispagestyle{empty}
\fancyheads

\setcounter{chapter}{12}

\chapter{Photon Collider}
\fancyhead[RO]{Photon Collider}
\label{photon:section}

\section{Introduction}

The concept of producing $\gamma\gamma$ collisions through
Compton backscattering of laser photons in a linear
collider [1,2] was proposed in 1981.  The available laser
technology was barely adequate for the accelerators operating
at that time.  The linear colliders proposed since then
are orders of magnitude more ambitious and require equivalent
improvements in laser technology to produce a $\gamma\gamma$ collider.
Fortunately, breakthroughs in laser technology have made feasible lasers
capable of delivering the 10 kW of average power in short pulses
of 1 TW peak power that are required for the NLC.  The problem of
obtaining such high peak power was resolved in 1985
with the invention of Chirped-Pulse Amplification (CPA).   The high
average power requirement could not be met without a long technology campaign
that involved the development of diode-pumped lasers, adaptive optics and
high-power multilayer optics, plus all of the associated engineering for
thermal management.   Nevertheless, today  the laser and optics technology
is finally in hand to proceed with an engineering design of a photon collider.

In the past few years there has been a crescendo of interest and
theoretical activity in $\gamma\gamma$.  This work has 
focused particularly on the
precision measurement of the radiative width of the Higgs, the 
study of heavy neutral Higgs bosons, and on 
detailed studies of supersymmetric particles and the top quark.  
The $\gamma\gamma$
channel is also highly sensitive to new physics such as large extra
dimensions and the appearance of strong gravity at the 10 TeV scale.

With the publication of the TESLA Technical Design Report (TDR) and
the development of the NLC/JLC toward full conceptual design, it was
appropriate therefore to bring the photon collider from its highly
schematic state into parity with the mature design of the rest of the
accelerator.  A year ago, a team of scientists and engineers from
LLNL, SLAC, and UC Davis along with a FNAL-Northwestern theory consortium
began to develop a complete design that would be required for full
incorporation in the future NLC Conceptual Design Report.  This effort
involved a tightly integrated effort of particle theory and modeling,
accelerator physics, optics, laser technology and engineering.  The
guiding principle was to develop a design that was robust,
 relied on existing technology, involved a minimum of R\&D,
and posed the least risk.  Considerations of elegance, power
efficiency and cost, while not unimportant, were relegated to second
place.  A satisfactory design was also required to stay well away from
compounding detector backgrounds, and to involve minimal modification to
the existing Final Focus and detector geometries.  While this is still a
work-in-progress, the conclusion of the study so far is that a photon
collider can be built with confidence on existing technology, satisfying
these guidelines and criteria.  This chapter describes the principal
physics drivers for the $\gamma\gamma$ IR, and the basic design and 
technologies to implement it.

\section{Physics Studies at a $\gamma\gamma$ Collider}

\subsection{Production of Higgs bosons}

 Perhaps the most 
important physics that can be done 
at a $\gamma \gamma$ collider is in  probing the properties 
of the Higgs boson(s). 
At such colliders 
the Higgs bosons of the SM and the MSSM can be singly produced as 
$s$-channel resonances through one-loop triangle diagrams.  They will 
be observed in their 
subsequent decay to $b\bar b, \tau^+ \tau^-, WW^*, ZZ$, \etc\  Contributions 
to  this type of loop graph arise from all charged particles that receive
mass from  
the produced Higgs.
In the SM, the loop contributions are dominated 
by the $W$ and top. SUSY contributions may be as large as $10\%$ of the
SM amplitude.  In 
addition, other currently unknown particles may also contribute to the loop 
and their existence may be probed indirectly by observing a deviation 
from the 
SM value. (Since other particles, such as gravitons, can also appear in the 
$s$-channel, it will also be necessary to determine the spin of any 
resonances 
that are produced.) 
By combining measurements at both $e^+e^-$ and 
$\gamma \gamma$ colliders it will be possible to determine both the quantity 
$\Gamma_{\gamma\gamma}$ and the Higgs total 
width~\cite{higgs,higgs2}. 

\begin{figure}[htb]
\centerline{
\psfig{figure=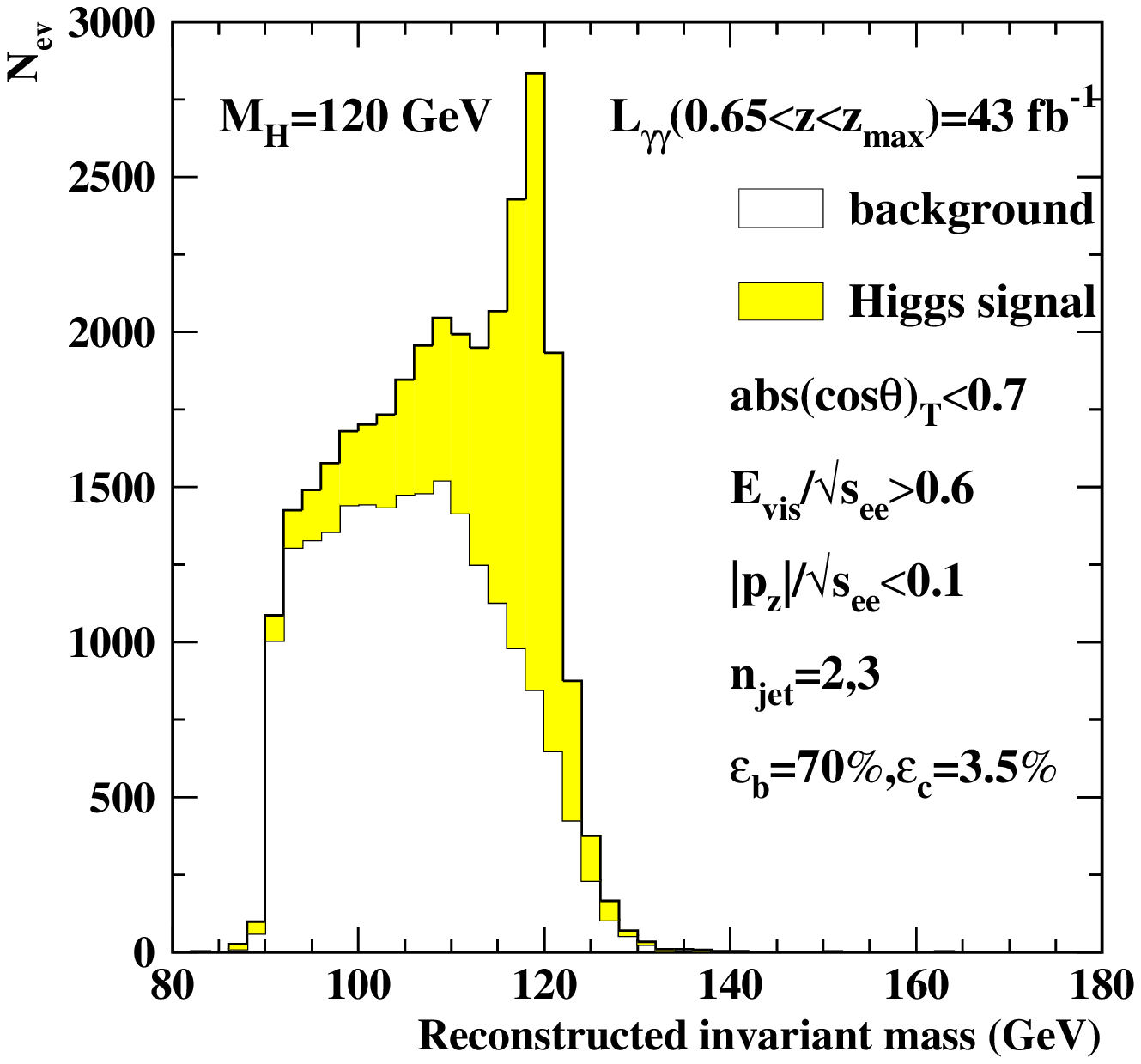,height=5.5cm,width=7cm,angle=0}
\hspace{-0.10cm}
\psfig{figure=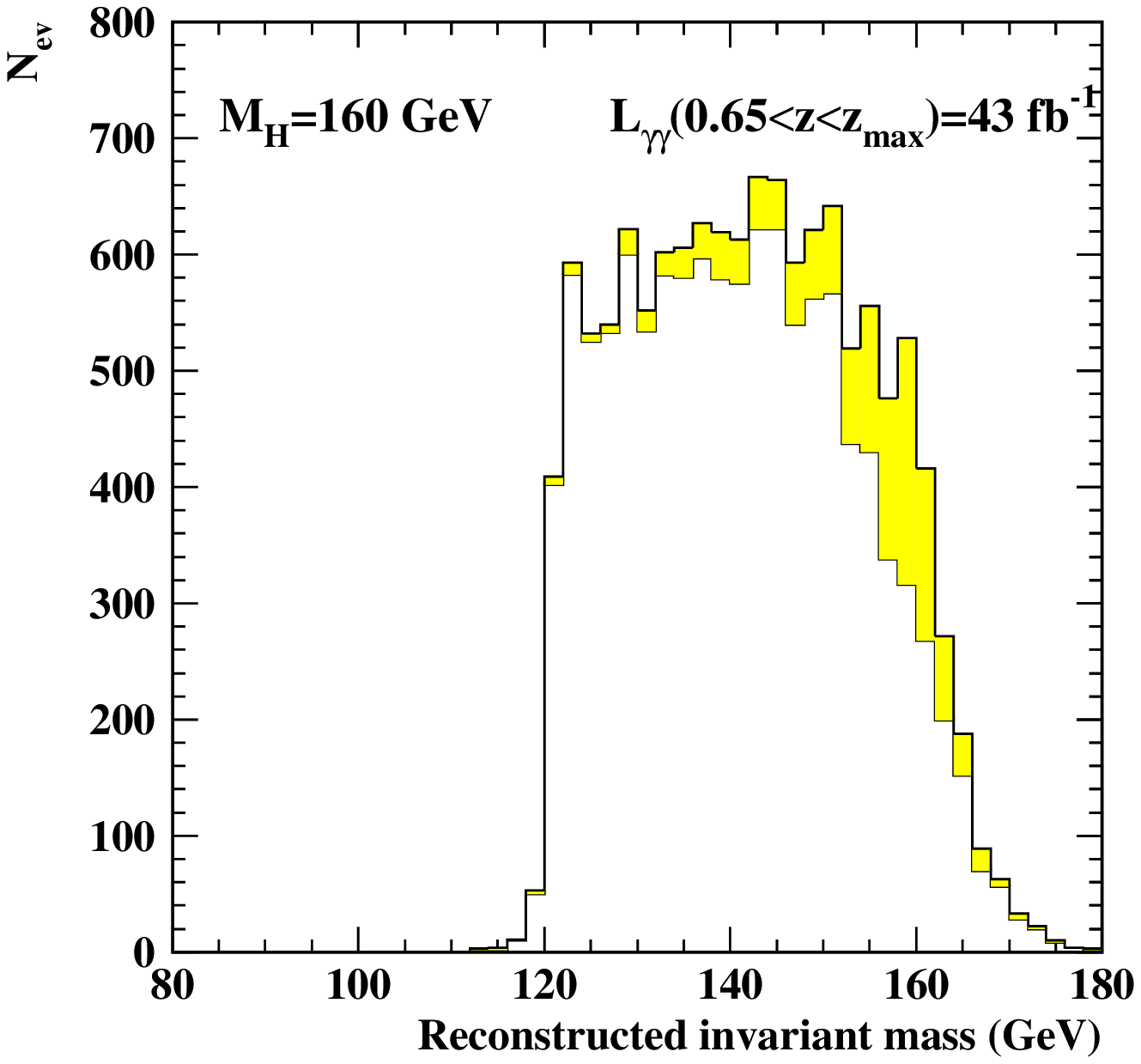,height=5.5cm,width=7cm,angle=0}}
\vspace*{0.1cm}
\caption{\label{fig:jikia}
Mass distributions for the Higgs signal and heavy quark background 
for a Higgs mass of 120(left) and 160(right) GeV from S\"oldner-Rembold and 
Jikia {\cite {jikia}}.  The reduced signal-to-background at 160 GeV reflects 
the diminished branching ratio to $b\overline{b}$ near the $WW$ threshold.}
\end{figure}

A light Higgs ($m_H \leq 135$ GeV) can be detected in the $b\bar b$ mode,
 with 
the main background due to the conventional 
QED $\gamma \gamma \to b\bar b, c\bar c$ continuum~\cite {higgs,higgs2}. 
Because of the relatively large $c\bar c$ cross section, excellent $b$ 
tagging is necessary.
The two initial photon polarizations can be chosen to produce spin-zero 
resonant states and to simultaneously 
reduce the cross section for the 
background which, at tree level, is suppressed by $m_q^2/m_h^2$. 
Unfortunately,  both 
QCD and QED radiative 
corrections remove this strong helicity suppression 
and must be well accounted 
for in both the Higgs and QED channels when comparing anticipated signals and 
backgrounds. Several detailed Monte Carlo simulations have been 
performed for this channel, with some typical results shown in 
Fig. \ref {fig:jikia} \cite {jikia}; these have demonstrated that the 
quantity 
$\Gamma_{\gamma \gamma}B(h\to b\bar b)$ can be determined with a relative 
error of $2\%$. Assuming that the $b\bar b$ branching fraction can be 
measured to the level of $1\%$ by combining $e^+e^-$ and
 $\gamma \gamma$ data, 
$\Gamma_{\gamma \gamma}$ will be determined at the level of $2\%$. This level 
of accuracy is sufficient to distinguish the SM and MSSM Higgs and to see 
contributions of additional heavy states to the triangle loop graph. 
If $e^+e^-$ colliders can also provide the branching fraction for $h\to 
\gamma \gamma$ at the $\sim 10\%$ level, the total Higgs width can be 
determined with a comparable level of uncertainty. A similar analysis
can be performed using either the $WW^*$ or $ZZ$ final state for 
Higgs masses up to 350-400 GeV, with comparable results \cite {higgs2}.

Very heavy Higgs bosons, such as those present in the MSSM, 
can also be produced as 
$s$-channel resonances in $\gamma \gamma$ collisions. In the MSSM, these
heavy states have suppressed
couplings to gauge bosons and may be most easily observed in $b\bar b$ or 
$t\bar t$ final states. These states may escape discovery at the 
LHC for intermediate values of $\tan \beta$.  At $e^+e^-$ colliders they 
can only be produced via associated production, $e^+e^- \to HA$, and thus lie 
outside the kinematic reach of the machine if their mass exceeds 240 
GeV. The single production mode of the $\gamma \gamma$ collider allows the 
discovery reach to be extended to over 400 GeV. The $\gamma \gamma$ collider 
also allows one to separate degenerate $H$ and $A$ states and to study 
possible CP-violating mixing between $H$ and $A$ using linear polarization.

\subsection{Supersymmetric particle production}

For production significantly above threshold,  
sfermion and charged Higgs boson pairs have production cross sections in
$\gamma\gamma$ collisions that are larger than those in $\ee$ annihilation.
Thus,  $\gamma \gamma$ collisions can 
provide an excellent laboratory for their detailed study. In addition, 
$\gamma \gamma$ production isolates the electromagnetic
 couplings of these particles, 
whereas in $e^+e^-$ the $Z$ and possible $t$-channel exchanges are also 
present. Thus complementary information can be obtained by combining data 
extracted from the two production processes. It should be noted that
 the search reach  for new particles is typically 
somewhat greater in $e^+e^-$ 
because of  the 
kinematic cut-off of the photon spectra. However, the SUSY process
$\gamma e \to \tilde{e}_{L,R} \chi_1^0$ shows that there are exceptions
to this rule; the threshold for this 
process can be significantly below that for $\tilde e$ pair production in 
$e^+e^-$ collisions when the $\chi_1^0$ is light. 
  In   the study  of this reaction,
both the $\tilde e$ and $\chi_1^0$ masses can 
be determined.

\boldmath
\subsection{$\gamma \gamma \to W^+W^-$ and $\gamma e \to W \nu$}
\unboldmath

\begin{figure}[htb]
\begin{center}
\epsfig{figure=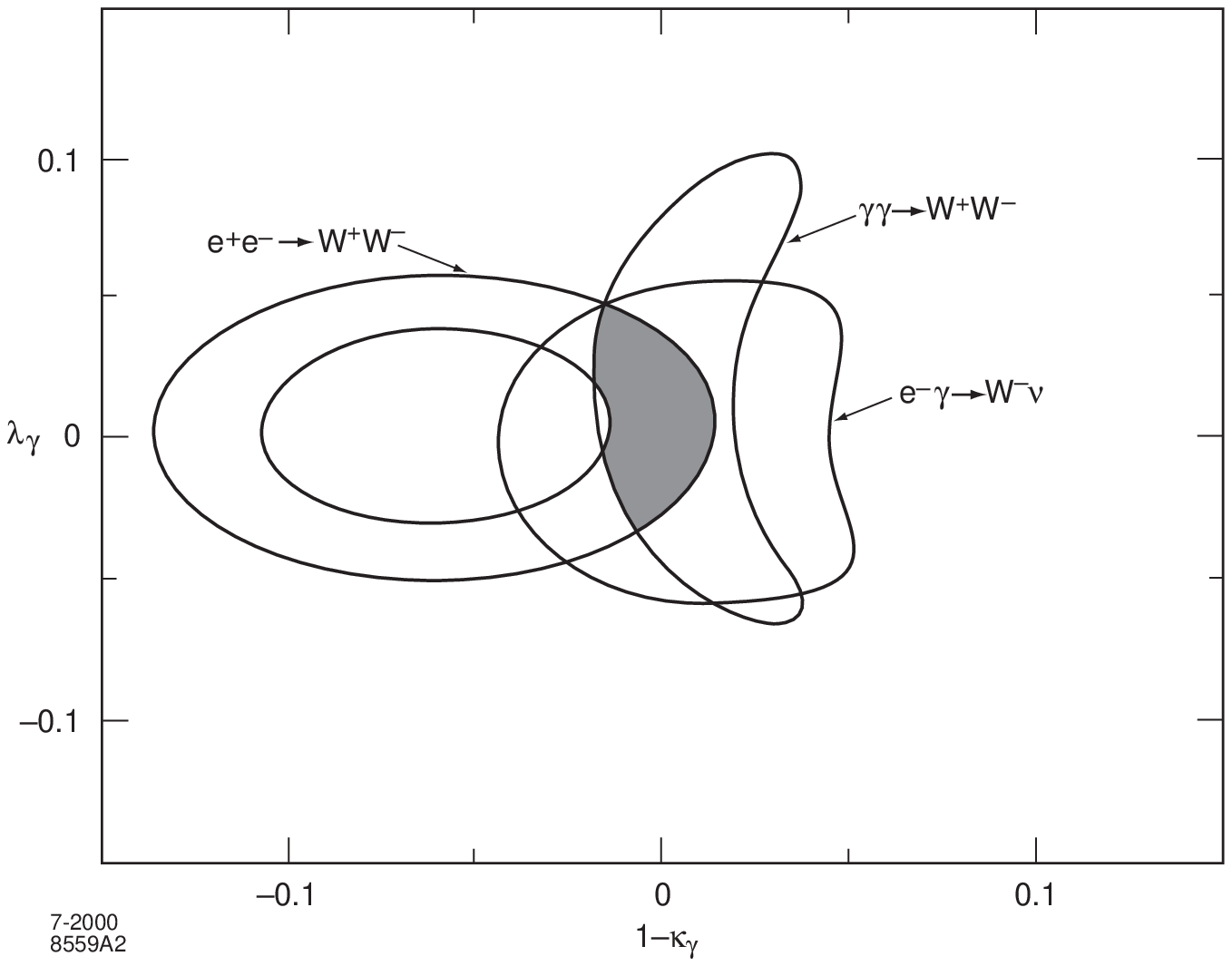,height=6cm,width=9cm,angle=0}
\caption{\label{fig:cands}
Allowed overlapping regions in the 
$\Delta \kappa_\gamma-\lambda_\gamma$
 anomalous coupling plane, from the analysis of Choi 
and Schrempp~\cite{choi}.}
\end{center}
\end{figure}

New physics beyond the SM can affect the expected values of the trilinear and 
quartic couplings of gauge bosons. These couplings can be studied in 
the reactions  $\gamma e \to W\nu$ and 
$\gamma \gamma \to WW$, as well as in $\ee\to WW$ \cite {choi}. 
 It is noteworthy that
the photon collider reactions
 isolate the anomalous photon couplings to the $W$, while   
$e^+e^- \to WW$ also involves anomalous $Z$ couplings. 
In 
addition, the process $\gamma \gamma \to W^+W^-$ allows access to the
quartic $\gamma \gamma W^+W^-$ coupling. 
The complementarity of the three reactions in determining the anomalous 
couplings is illustrated in Fig.~\ref {fig:cands}, taken from~\cite{choi}.
   Since the time of this study, it has
been understood how to achieve bounds on the anomalous couplings from 
 $\ee\to WW$ that are a factor of 30 better than those shown in the figure,
by taking advantage of more systematic event analysis and higher
luminosities.  Methods for that analysis are described in Chapter 5, Section 2.
A similar improvement should be possible for the constraints from 
 $\gamma e \to W\nu$ and 
$\gamma \gamma \to WW$, though the detailed study remains to be done.

\begin{figure}[htbp]
\begin{center}
\psfig{figure=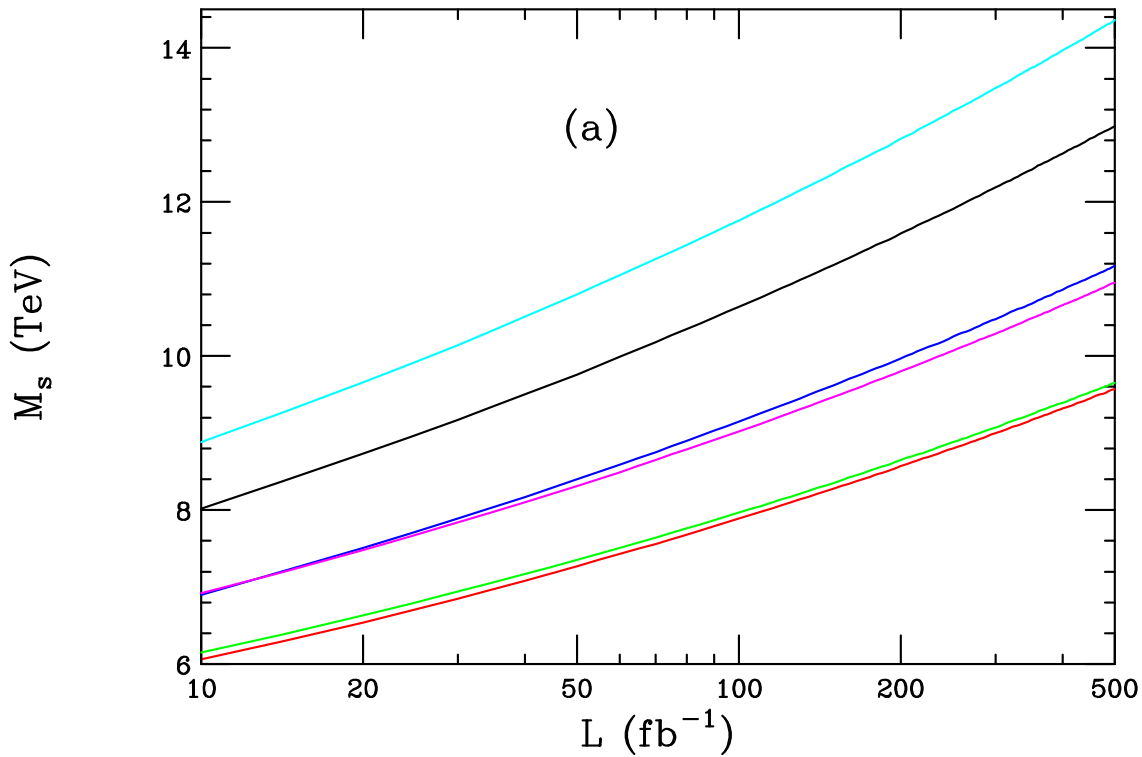,height=15cm,angle=90}
\caption{\label{fig:lims}
$M_s$ reach for the 
process $\gamma \gamma \to W^+W^-$ at a 1 TeV $e^+e^-$ collider 
as a function of the integrated luminosity 
for the different initial and final state polarizations. 
From top to bottom on the right hand side of the figure the 
polarizations are $(-++-)$, $(+---)$, $(++--)$, $(+-+-)$, $(+---)$,  
and $(++++)$.}
\end{center}
\end{figure}

The reaction 
$\gamma \gamma \to W^+W^-$ is also highly sensitive to other forms of new 
physics such as the exchange of virtual towers of 
gravitons that occurs in models  of millimeter-scale 
extra dimensions~\cite{ADD,tgr}.  
It has been shown that 
this is the most sensitive process to graviton exchange of all those so far 
examined. Such exchanges can 
lead to substantial alterations in cross sections, 
angular distributions, asymmetries and  $W$ polarizations.  These
effects make it possible to  probe the associated gravitational mass scale, $M_s$, 
to values as high as $13\sqrt s$ for 
the correctly chosen set of initial laser and electron polarizations. 
(For comparison, the 
reach in $e^+e^-$ is about $7\sqrt s$.) The 
search reach as a function of the $\gamma\gamma$ luminosity is shown in  
Fig. \ref {fig:lims} for the various polarization choices. This same process
can be used to search for graviton resonances such as those predicted
in the Randall-Sundrum model~\cite{rs}. 

\boldmath
\subsection{$\gamma \gamma \to t\bar t$}
\unboldmath

Since the top quark is the heaviest SM fermion, with a Yukawa coupling 
that is 
quite close to unity,  one might expect that its properties may be the most 
sensitive to new physics beyond the SM. For example, the top may have 
anomalous couplings to the SM gauge bosons, including the photon. 
The cross section for top pairs in 
$\gamma \gamma$ collisions is somewhat larger than in $e^+e^-$, thus this 
process may provide the best laboratory to probe new physics couplings 
to the top. In addition, 
while both $e^+e^-$ and $\gamma\gamma$ colliders can probe the anomalous 
$\gamma t\bar t$ 
couplings, these are more easily isolated in $\gamma \gamma$ 
collisions. As shown in {\cite {djouadi}}, there are 4 form factors that
describe this vertex, 
one of which is CP-violating and corresponds to the top quark electric dipole 
moment. By measurements of the $t\bar t$ angular distribution significant 
constraints on these form factors are possible with sensitivities to both 
electric and magnetic dipole moment couplings that are about an order of magnitude 
better in $\gamma \gamma$ colliders than in $e^+e^-$ machines. In addition, 
CP-violating couplings can be directly probed through the use of polarization 
asymmetries and limits superior to those obtainable from $e^+e^-$ colliders 
are possible.

\subsection{Other processes}

There are many other interesting processes that one can study in 
$\gamma \gamma$ collisions. As far as new physics is concerned, the 
$Z\gamma$ and $ZZ$ final states can be used to 
probe anomalous $ZZ\gamma$ and $Z\gamma \gamma$ couplings~\cite{gounaris} 
while the $\gamma \gamma$ final state can be used to search for 
non-commutativity and violations of Lorentz invariance in QED~\cite {hpr}. 
The couplings of leptoquarks discovered in $e^+e^-$ collisions can be more 
easily disentangled by using data from both $\gamma \gamma$ and $\gamma e $ 
collisions {\cite {lepto}}. 
It may also be possible to form resonances of stoponium, the 
supersymmetric version of toponium, with production rates that are 
significantly higher than in $e^+e^-$~\cite{stop}.

Within the SM there are a number of 
interesting QCD processes that can also be examined to obtain information on
topics such as
the gluon and quark content of the photon, the spin-dependent part of 
the photon structure function, and the QCD pomeron.  These topics are 
reviewed in Chapter 7, Section 3.

\section{Compton Backscattering for $\gamma\gamma$  Collisions}

\subsection{Introduction}

High-energy photons can be produced through two-body scattering
of laser photons from a high-energy electron beam.
For example, the scattering of 1 eV laser photons from an electron 
beam of 250 GeV can produce gammas of up to 200 GeV.
An electron linear collider can be converted to a $\gamma\gamma$ collider if a
high-power laser pulse intersects the electron beam just before the 
interaction point (IP).  The point where the laser beam intersects the 
electron beam---the conversion point (CP)---can be within 1 cm of the 
IP.
A high $\gamma\gamma$ luminosity comparable to that of
$e^+e^-$ can be achieved, since the
photons will focus to about the same spot size as the electron beam. 
The principles are reviewed in detail elsewhere~\cite{telnovintro}.

\subsection{Photon spectra}

\begin{figure}[bt]
\begin{center}
\psfig{file=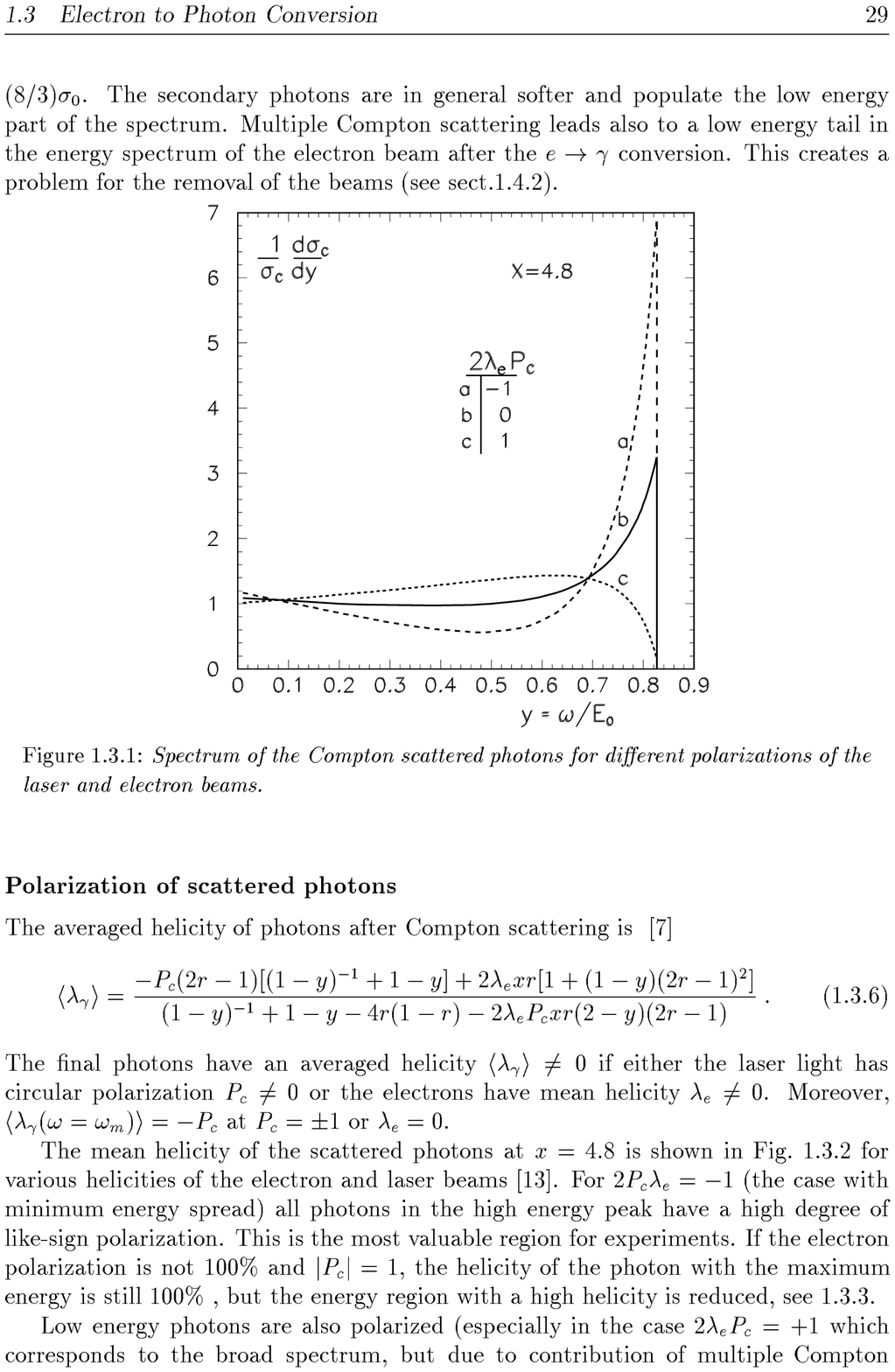,width=.48\textwidth,
clip=, bbllx=173, bblly=406, bburx=438, bbury=672}
\hfill
\psfig{file=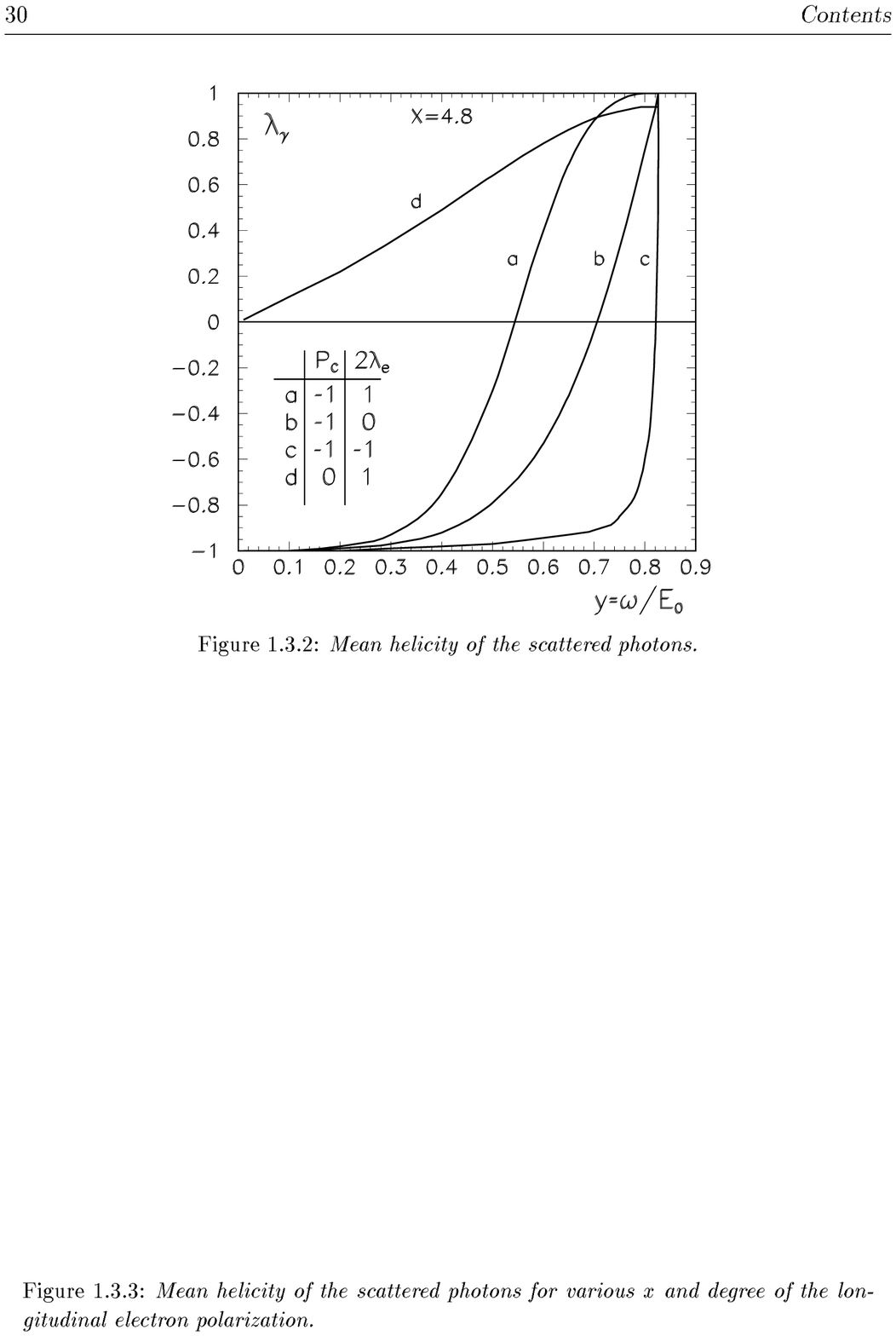,width=.48\textwidth,
clip=, bbllx=161, bblly=461, bburx=435, bbury=733}
\caption{\label{fig:bkscat}
The energy spectrum and helicity spectrum of the 
Compton-backscattered photons for various helicities of the incoming
electron beam with circularly polarized incoming photons~\cite{telnovintro}.
The variable $y$ is the photon energy as a fraction of the 
electron beam energy.  The laser photon and electron helicities are 
designated by
$P_c$ and $\lambda_e$.  The parameter $x = 4 E_e \omega_0/m^2c^4$. }
\end{center}
\end{figure}

For the case mentioned above---1 eV laser photons and 250 GeV electrons---the
energy spectrum of the backscattered photons ranges from 0 up
to 0.8 of the incoming beam energy.  Two-body kinematics creates a
correlation between the photon energy and the angle between 
the outgoing photon and the incoming electron.  The maximum photon 
energy occurs when the produced
photon is collinear with the incoming electron.
 
The exact energy spectrum is a function of the polarization of the 
incoming electron and laser beams.
Figure~\ref{fig:bkscat} shows the energy spectrum of the backscattered
photons for circularly polarized laser photons.  The population of the
high-energy peak is maximized when the electron beam is fully polarized
and of opposite helicity to the laser beam.  For that situation, the
high-energy photons are also fully circularly polarized.  While the lasers
naturally
produce linearly polarized photons,  any combination of circular and linear
polarization can be produced through the use of quarter-wave plates.

From Fig.~\ref{fig:bkscat} it can be seen that the ability to polarize
the incoming electron beam is crucial for producing high-energy
$\gamma\gamma$ collisions with polarized gammas.  Currently it is
foreseen that the electron beams will achieve 80\% polarization while
positrons will be unpolarized.  This makes it attractive to run in
an $e^-e^-$ mode rather than $e^+e^-$.  Many Standard Model backgrounds
are also suppressed by choosing $e^-e^-$ running.

\begin{figure}[thb]
\begin{center}
\epsfig{file=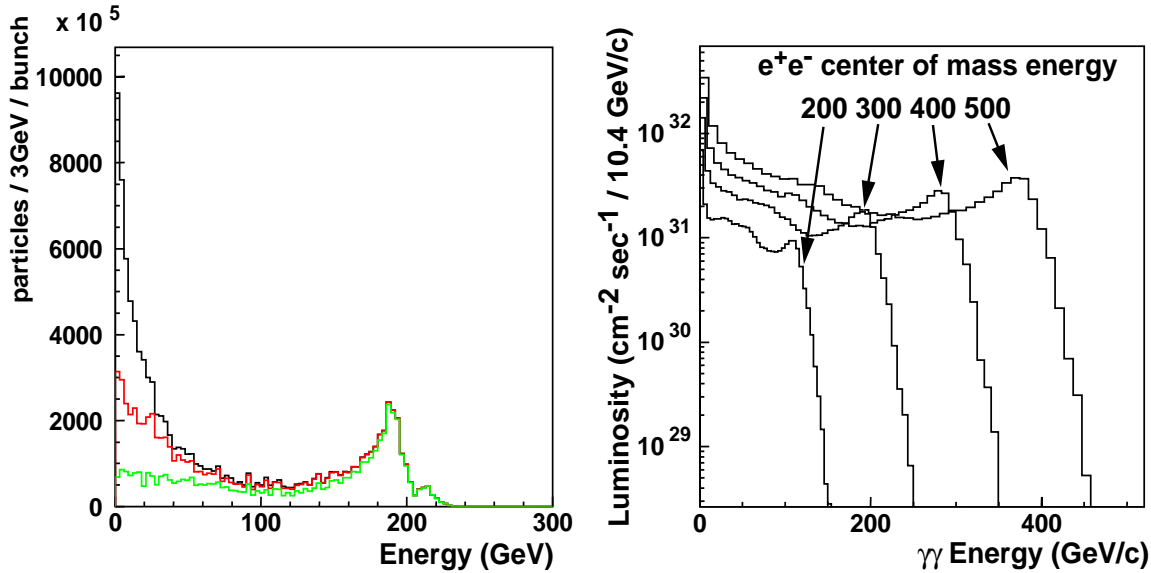,width=\textwidth,height=3in}
\caption{\label{fig:spectra}
The first plot shows the energy spectrum from Compton
backscattering when (respectively, from the bottom curve to the top)
primary, secondary, and all higher scatters are taken into account.
The second plot shows the $\gamma\gamma$ 
luminosity for $e^-e^-$ center-of-mass energies of 500, 400, 300, and 200 GeV
for the NLC-B machine parameters.}
\end{center}
\end{figure}

Calculating the $\gamma\gamma$ luminosity spectrum at the IP
is not as simple
as convoluting the single-scattering energy spectrum with itself.
There are additional sources of $\gamma$'s that must be included.  An
electron can Compton backscatter multiple times as it passes through
the laser beam.  This leads to a tail of low-energy photons, as
can be seen in Fig.~\ref{fig:spectra}.  Also,
the leftover electron beam arrives at the IP coincident with
the photons.  When the two electron beams interact they produce a 
large number of  beamstrahlung  photons.  All of these contribute
to the $\gamma\gamma$ luminosity.

As a result of the energy-angle correlation, the spot size for collision of 
soft photons will be larger than that for the collision of harder 
photons.  Thus the luminosity spectrum may be hardened by increasing 
the distance between the CP and IP.  In the following, we chose the CP
to be 5 mm from the IP.

To compute the $\gamma\gamma$, $e\gamma$ and $ee$ luminosities, we use
the program CAIN~\cite{CAIN}, which models all of the processes just
described. Results for various incident
electron beam energies are  shown in Fig.~\ref{fig:spectra}.  
The luminosity spectrum peaks at $\gamma\gamma$ CM energies close to  
0.8 times of the $e^-e^-$ CM energy. The 
decrease of luminosity with decreasing CM energy, apparent from the plot,
is primarily caused by  the increased
spot size of the electron beams and, secondarily, by a softer 
Compton-backscattering spectrum.  For a 120 GeV Higgs this leads to a 
situation where higher luminosities can be achieved by running
at 500 GeV $e^-e^-$ CM energy at the cost of having unpolarized photons.  For
measurements requiring definite states of $\gamma\gamma$ polarization,
on-peak running with 150 GeV $e^-e^-$ CM energy is required.  

\subsubsection{Accelerator modifications}

While no changes to the accelerator are required to produce $\gamma\gamma$
collisions, some changes can optimize performance.  
Beam-beam interactions are a major concern for $e^+e^-$ but are not
present in $\gamma\gamma$ collisions.  Therefore the $\beta$ functions of the
Final Focus should be as small as possible to achieve a minimum
spot size and maximum luminosity.  The luminosity improvements
from small $\beta$ functions are limited by chromatic aberrations
in the Final Focus and the hourglass effect, in which  the $\beta$ function
becomes comparable to the longitudinal spot size.  In addition, 
a small transverse
spot size tends to select unboosted events because of the
correlation between the energy and production angle of the
high-energy $\gamma$'s.  A Final Focus design with rounder beams
simplifies the final doublet stabilization and has been shown
to recover nearly a factor of two in luminosity by increasing
the contribution of boosted events.  However, these boosted events
suffer from reduced reconstruction efficiency and we have not
yet optimized the design for this effect.

\begin{table}[b!]
\begin{center}
\begin{tabular}{|l|c|} \hline\hline
$e^-e^-$ CM Energy (GeV) & 490 \\
Luminosity & $1.23 \times 10^{33}$ @  $>$65\% $e^-e^-$ energy \\
Bunch Charge & $1.5 \times 10^{10}$ \\
Bunches / pulse & 95 \\
Bunch separation & 2.8 ns \\
 $\gamma\epsilon_x$ at IP & $360 \times 10^{-8}$ m-rad\\
 $\gamma\epsilon_y$ at IP &  $7.1 \times 10^{-8}$ m-rad  \\
 $\beta_x$ / $\beta_y$ at IP &  0.76/1.81 mm  \\
 $\sigma_x$ / $\sigma_y$ at IP &  76/16 nm\\
$\sigma_Z$ at IP & 0.150 mm \\
\hline\hline
\end{tabular}
\caption{NLC-G parameter set.  Unless otherwise noted parameters are
identical to NLC-H. \label{tab:NLC-G}}
\end{center}
\end{table}

Achieving rounder beams requires only a change in the strength
of the Final Focus magnets.  It is useful also  to cut the number
of bunches in half and double the bunch charge, to better match the
laser technology.  This nominally increases the luminosity by a factor
of two, although this is not fully achieved due to the increased emittance
growth and the increased longitudinal spot size.  The parameters we use
are shown in Table~\ref{tab:NLC-G}.  These  have been reviewed and approved
by the NLC machine group. When we reduce the $e^-e^-$ CM energy such that
the $\gamma\gamma$ peak is at 120 GeV for Higgs running, 
the $\gamma\gamma$ luminosity becomes
$2.9 \times 10^{31}$ cm$^2$/s/GeV at 
$\sqrt{s_{\gamma\gamma}} = 120$~GeV,  with 80\% of events being spin 0.

\begin{figure}[p]
\begin{center}
\epsfig{file=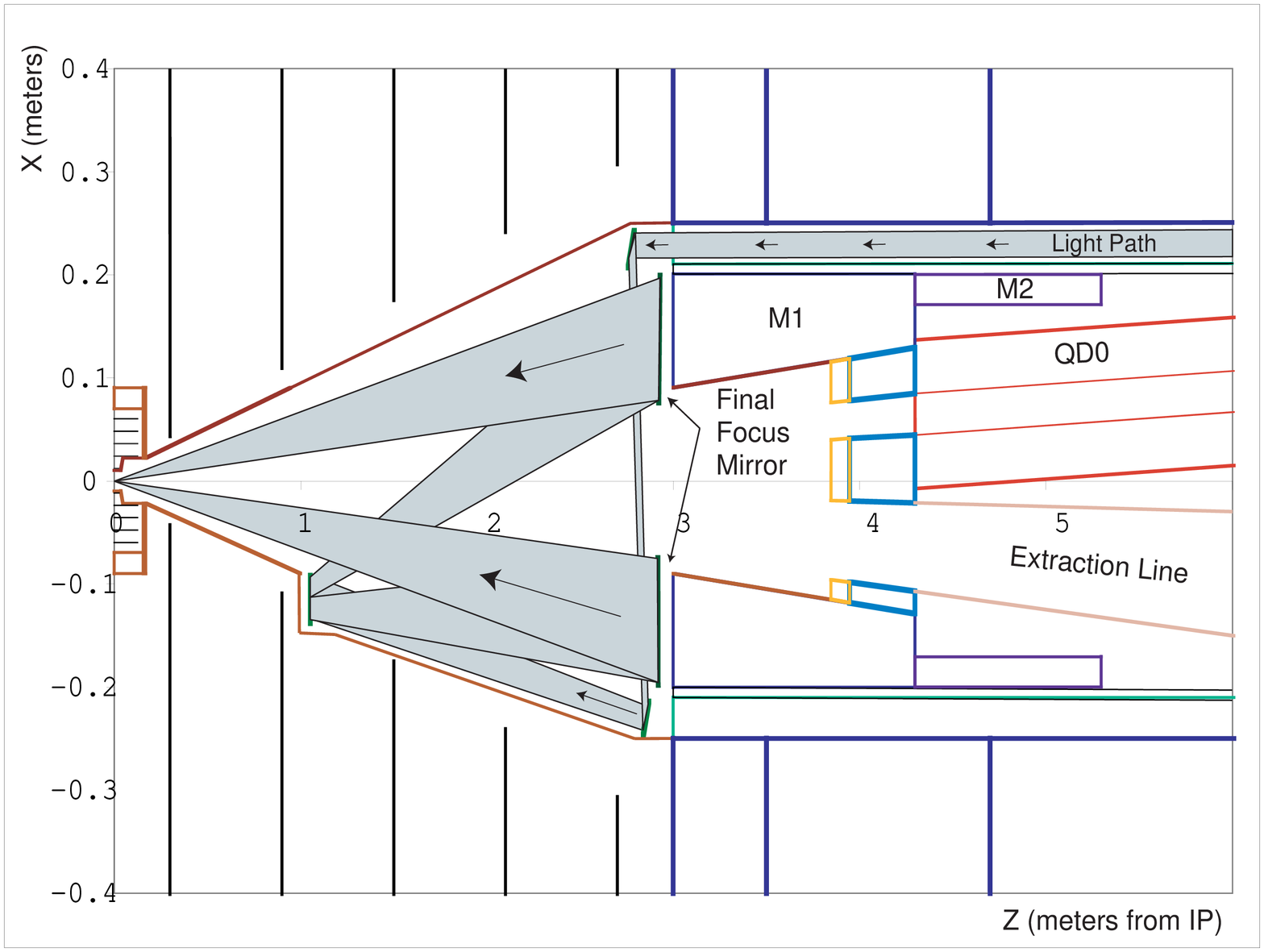,height=9cm}
\caption{\label{fig:ggIR}
Optical configuration to inject the laser light 
into the Interaction Region.
The high subpulse intensity requires all these optics to be 
reflective and mounted inside the vacuum enclosure.}
\end{center}
%\end{figure}
\vskip.1in
%\begin{figure}[bth]
\begin{center}
\psfig{file=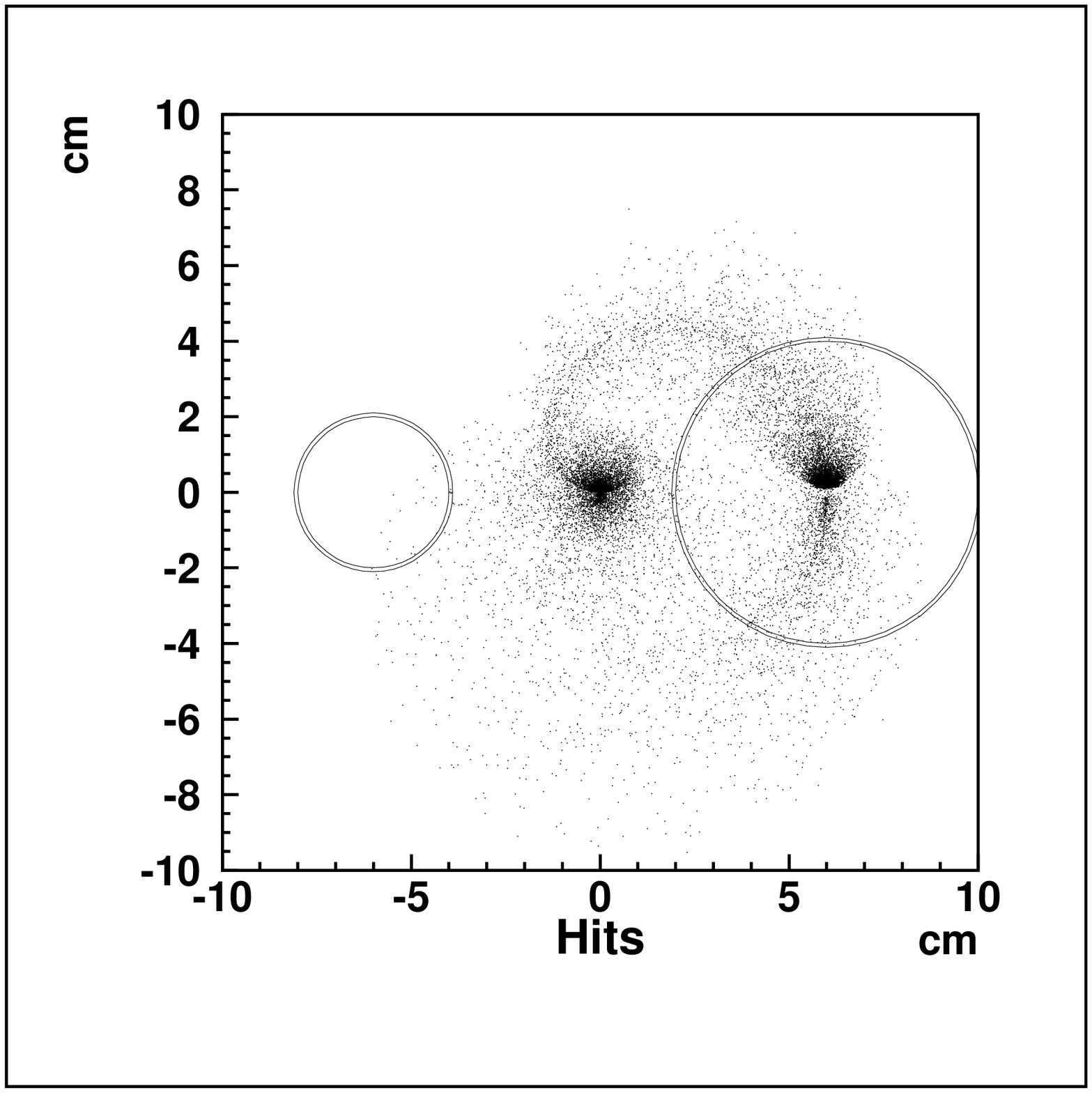,width=.48\textwidth,
clip=, bbllx=30, bblly=202, bburx=566, bbury=691}
\hfill
\psfig{file=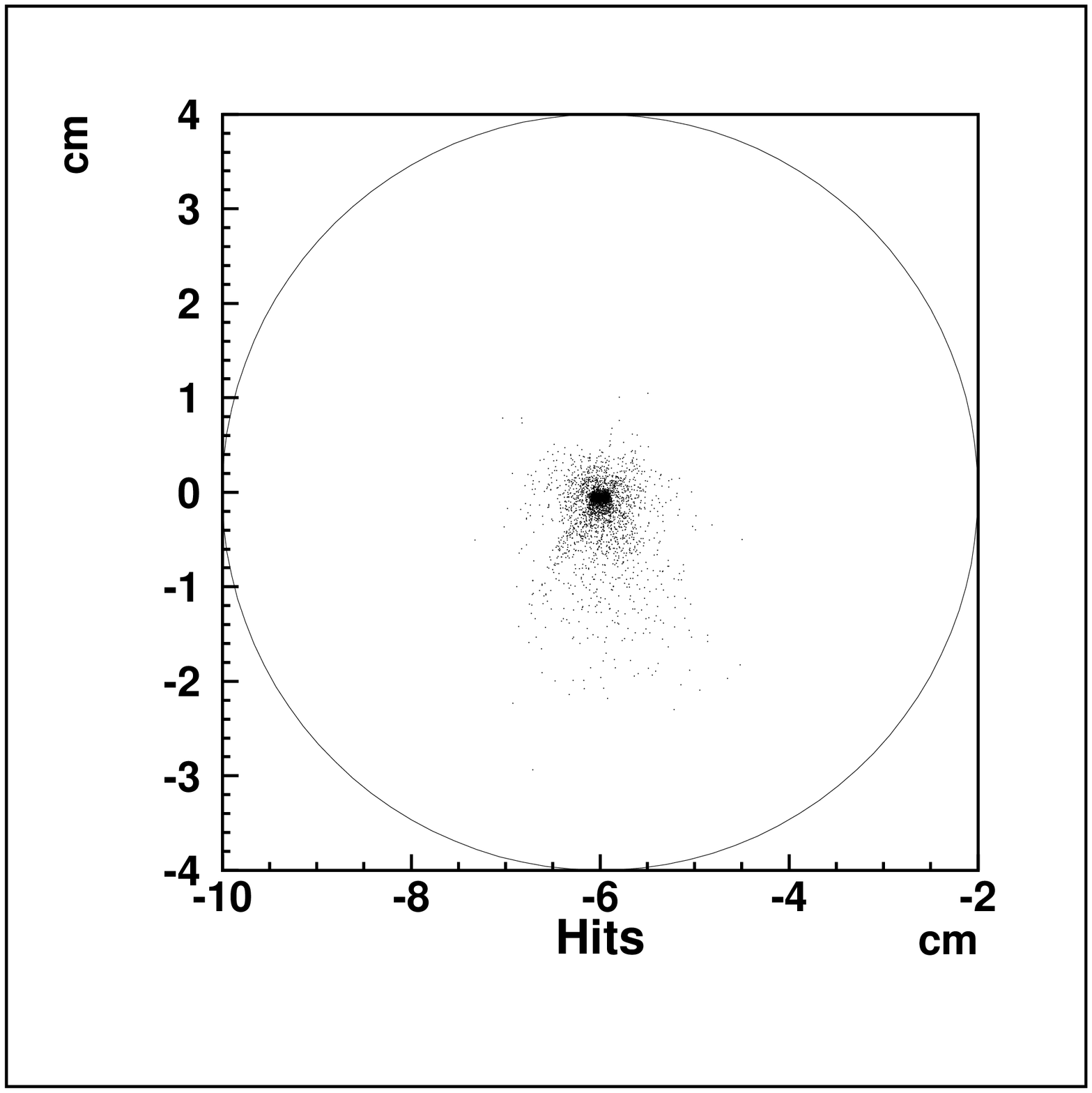,width=.48\textwidth,
clip=, bbllx=30, bblly=202, bburx=566, bbury=691}
\caption{\label{fig:backs}
The front face of the magnet at $z=4$ meters.  The first plot
shows the impact point of the pair background.  High-energy particles
travel out the extraction line.  Low-energy particles spiral in the
solenoidal magnetic field of the detector.  The second plot is an
expanded view of the extraction line aperture showing the location
of the outgoing beam.}
\end{center}
\end{figure}

\subsection{Interaction region design and backgrounds}

Figure~\ref{fig:ggIR} shows the interaction region for a 
$\gamma\gamma$ collider.  The design begins from the IR for 
$\ee$ collisions, but many modifications are needed to 
accommodate the laser beams.  The first of these 
 is the system of mirrors required to bring laser light into
the IR, described in detail in Section~\ref{sec:optics}.  The mirrors 
have been carefully placed to be outside the path of the beams and
the pair background.  The pair background consists of low-energy electrons
and positrons which spiral in the magnetic field of the detector.  Their 
transverse location at $z$ = 4~meters can be seen in Fig.~\ref{fig:backs}.
No additional backgrounds are generated by the presence of the mirrors.

The extraction lines for the spent beams must be modified for the
$\gamma\gamma$ interaction region.  The energy spectrum of electrons 
leaving the conversion point is composed of unscattered electrons at the
beam energy and scattered electrons which peak around 1/5 the beam energy.
The low-energy electrons receive a larger angular deflection from
the beam-beam interaction at the IP, necessitating an increased aperture
for the extraction line.  Additionally, just as in the case of the 
pair background, the low-energy electrons spiral in the magnetic field.
Figure~\ref{fig:backs} shows the position of the outgoing electrons
at the entrance to the extraction line.  An aperture of $\pm 10$ milliradians
accommodates these particles.  In order to prevent mechanical interference
between the extraction line and the last focusing quadrupole, the crossing
angle has been increased from 20 to 30 milliradians.

Increasing the extraction line aperture has a detrimental effect on
the neutron background levels at the IP.  The Silicon Vertex Detector,
1.2 cm away from the IP, now has a direct line of
sight back to the beam dump 150 meters away.  It experiences a fluence
of $10^{11}$ neutrons/cm$^2$/year.  The standard CCD technology chosen for the
$e^+e^-$ IR cannot withstand it.  The $\gamma\gamma$ IR would need
a rad-hard CCD or pixel design. 

We foresee no impact to the detector aside from the need for a rad-hard
vertex detector.  The machine backgrounds in the 
$\gamma\gamma$ IR are comparable to the $e^+e^-$ IR.  Still to be evaluated
is the effect of resolved photon events from the higher $\gamma\gamma$ luminosity.   

\section{IR optical system}

\subsection{Optics design}
\label{sec:optics}

The function of the optical system is to bring the laser beam to the 
IR while also minimizing the required laser 
pulse energy.  The requirement for efficient conversion of the 
electrons sets the laser photon density required at the interaction 
point.  The optical system will focus the laser beam at a point near the 
interaction point to maximize the conversion probability.  The size of 
the laser focal spot will be much larger than the electron beam; 
therefore the size of the focal point should be minimized in order to 
minimize the required laser pulse energy.

When bringing a beam to a focus, the size of the focal point 
is determined by the f-number of the focusing optic, defined as
the ratio of the 
focal length to the optic diameter.  The size of the useful laser spot is 
approximately the wavelength times the f-number. 
There is a limit, however, to how small one can usefully make the 
f-number.  The focal spot has a limited depth of focus.  When the 
electron-photon interaction region becomes longer than the depth of 
focus, the required laser energy becomes independent of the f-number. 
Lowering the f-number beyond this point results in no decrease in the 
required laser pulse energy.  Optimally, the laser pulse length should
be the same as the electron bunch length to minimize the required pulse
energy.  However, at such high intensities, non-linear effects degrade the
purity of the photon polarization.  We choose a pulse length of 2~ps FWHM,
which is well matched to the available laser technology.
For such a 2 ps laser pulse, decreasing the 
f-number below 7 gains little further energy reduction.  For the 
reference design the f-number is 8.

Figure~\ref{fig:ggIR} shows the optical design near 
the interaction region.  The 
final focusing optic is located at the 3 m station and is mounted 
adjacent to (or on) the 40 cm diameter tungsten plug (M1).  The optic 
has a 300 cm focal length and a 38 cm diameter, giving it an f-number 
of 8.  The central 15 cm hole provides a space for the electron beams 
and high-energy scattered electrons to pass through the Final Focus 
optic.  The secondary optic is mounted off-axis to minimize the 
obscuration of the laser beam.  Additional turning optics provide 
centering and pointing capabilities as well as beam injection to the
secondary optic.  The high subpulse intensity requires all these
optics to be reflective and mounted inside the vacuum chamber.

The laser beam enters the IR from one side.  A symmetric set of 
optics (not shown in Fig.~\ref{fig:ggIR}) takes the 
beam to a mirror that sends 
the beam back to a focus intersecting the second electron beam.  The 
difference in the image plane of the focal spots as well as the 
difference in arrival times can then be used to separate the incoming 
and exiting laser beams in the beam transport system.

\subsection{Beam pipe modifications}

The short pulse format of the laser results in beam intensities that 
cannot be propagated through air or transmissive optics. The pulse 
compression, beam transport and IR injection optics will all be 
reflective optics inside vacuum enclosures.  The small vacuum pipe 
that transports the electron beam must be expanded in the IR to 
contain the laser injection optics (as shown in 
Fig. \ref{fig:ggIR}).  The level of vacuum 
required will be determined by the electron beam since it will be 
higher than needed by the laser.  It should be noted that the vacuum 
requirement of the electron beam may place restrictions on the 
materials that can be used in the optics mounts and controls.

The laser beam transport pipe will contain isolation gate valves that 
will be open when the laser is operating.  These valves can be closed 
during maintenance and other operations when the laser in not 
operating.  They can also be used to prevent contamination or 
accidental pressurization of the linac and IR during 
shutdowns.

The optics and vacuum enclosures will be mounted on the same 
structures as the electron beam transport system.  The electron beam 
transport system in the IR region has not been designed in sufficient 
detail to begin the design of the laser system interfaces.  The 
seismic requirements for the laser optics are not as stringent as for 
the Final Focus magnets.  If both systems use the same 
supports, it will be important that the laser system does not feed 
excessive acoustic energy into the final quadrupole support structure.

\section{Laser system}
\label{photon:section:laser}

\subsection{Requirements and overview}

The laser system must match the pulse format of the electron beam and 
supply an adequate photon density at the IR to 
backscatter the laser photons efficiently to gamma rays.
For efficient conversion of 250 GeV
electrons, the optimal laser wavelength is one micron. 
The laser requirements for the NLC are summarized in Table~\ref{tab:laser-req}.

A picosecond-duration laser pulse cannot be amplified to the 
joule level directly.
The combination in the laser subpulse of a high pulse energy (1~J) and 
a short pulse duration (2~ps) generates field intensities that will 
damage laser materials.  This problem is solved by first stretching a 
very low-energy laser subpulse to 3 ns and then amplifying this long 
pulse.  The pulse is then compressed back to 2 ps for use in the IR. 
The procedure for stretching and compressing the laser pulse with 
diffraction gratings, known as Chirped-Pulse-Amplification (CPA)
\cite{Strickland:compression}, is 
discussed below.  The procedure requires the laser medium to have 
significant gain bandwidth.

 Efficiently energizing a laser with the very low required duty 
factor (300~ns/8~ms) requires the use of a `storage laser' material. 
Generally storage lasers are solid-state and, when used in a 
high-pulse-rate application, they are strongly limited by heat-removal 
capabilities.  LLNL has been developing a solid-state Yb:S-FAP laser 
with diode pump lasers and rapid helium gas cooling to address this 
issue as part of its Inertial Fusion Energy program.  The Mercury 
Laser Project is currently assembling a prototype.  The default Mercury laser 
pulse format differs from that required for $\gamma\gamma$ operation.  The necessary 
modifications of the laser are described below.

\begin{table}[b]
\begin{center}
\begin{tabular}{|l|c||l|c|} \hline  \hline
Wavelength & 1 $\mu$ & Format & $\sim 100$ subpulses/macro-pulse \\
Subpulse energy & 1 J & Repetition rate &  120 Hz  \\
Subpulse separation & 2.8 ns & Gain bandwidth &  10 nm  \\
Subpulse duration & 2 ps & Beam quality &  $< 1.5$ diffraction limit  \\
\hline\hline
\end{tabular}
\caption{\label{tab:laser-req}
$\gamma\gamma$ collider laser requirements.}
\end{center}
\end{table}

\subsection{Laser system front end}

The laser system front end must generate a low-power laser signal 
with a temporal format matched to that of 
the electron linac.  This signal will then be delivered to the 
Mercury amplifiers to generate the high pulse energies needed to 
interact with the electron pulses.

A laser oscillator will be required with an approximately 350 MHz 
pulse rate and 2 ps pulse duration.  With pulse energies of 
1.0 nJ, the average power will only be only 0.35 W.  The laser must be 
tuned to the 1.047 micron wavelength which overlaps the gain bandwidth of 
the Yb:S-FAP laser amplifiers.  Commercial Ti-sapphire lasers will be 
appropriate for this task.  The laser oscillator must have high 
frequency stability and must be locked to the master clock of the 
linac so that the laser pulse timing matches that of the 
electron pulses.

The beam from the oscillator will pass through a Pockels cell slicer 
that will cut out 300~ns pulse trains at 120 Hz.  These
batches will match the electron bunch trains, which contain 
approximately 100 subpulses.  The pulse trains will then be passed 
through an electro-optic modulator that will impose a moderately 
increasing amplitude ramp on the macro-pulse.  This amplitude ramp is 
designed to offset the decreasing gain ramp that will be experienced 
in the amplifier as the stored energy is extracted during the laser 
macro-pulse. The low power (about  1 $\mu$W) is easily handled by current EO 
modulators.

The gain in the amplifier will have frequency variations as well as 
amplitude diminution during the macro-pulse.  To avoid strong 
amplitude variations at different frequencies in the amplified laser 
signal, the amplitude of the input laser beam will be sculpted in 
frequency space~\cite{Perry:shaping} to offset the effects of the gain variation.  The 
short pulse length of the subpulses gives them a frequency bandwidth 
such that a diffraction grating will spread the beam over a range of 
angles.  The different frequencies are then passed through a 
programmable liquid crystal display that provides different 
attenuation for different positions (frequencies) in the beam.

The laser beam is next passed through a diffraction grating pulse 
stretcher, described in a later section, that stretches the 2 ps 
subpulses to 3 ns.  The spectral sculpting and pulse stretching might 
be combined into a single device if appropriate.

The stretched laser pulses can now be passed through a high-gain, 
low-power preamplifier.  A laser optical parametric amplifier (OPA) will 
provide the high bandwidth needed to preserve the frequency profile 
of the laser pulse.  A high-pulse energy green laser will pump a BBO 
crystal to provide the gain needed.  The laser beam will be amplified 
to 500 $\mu$J/subpulse.  The beam will be split into twelve 10 Hz beams and 
then injected into the Mercury amplifiers.

\begin{figure}[t]
\begin{center}
\psfig{file=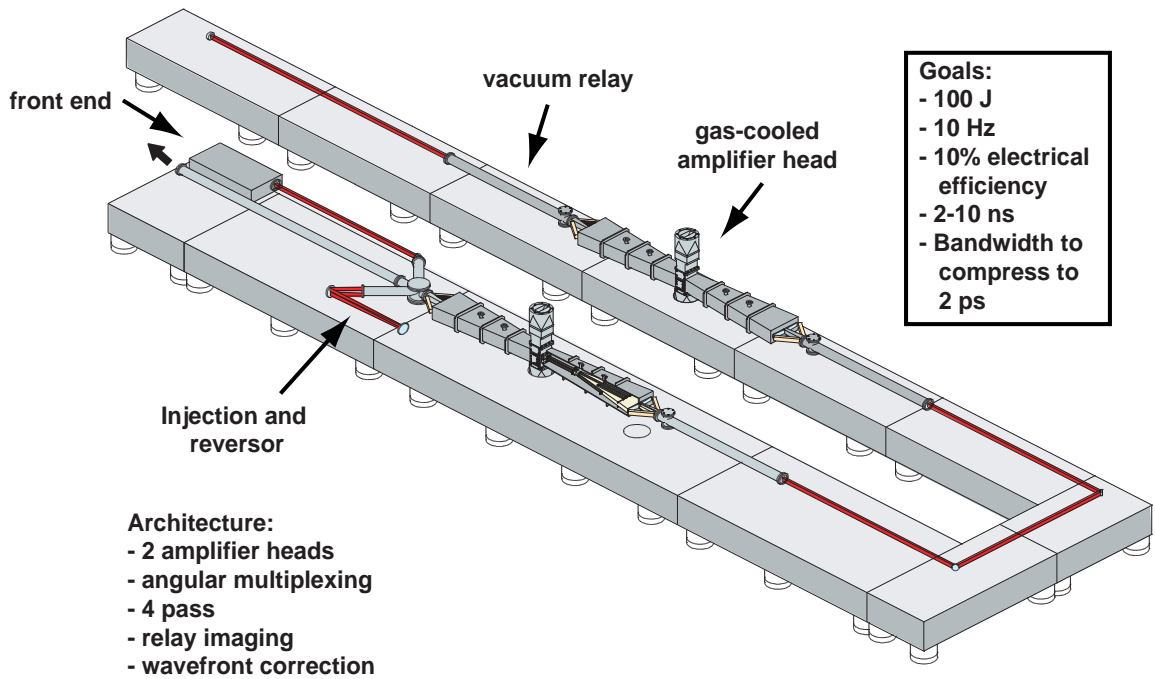,width=\textwidth}
\caption{\label{fig:mercury}
The diode-pumped solid state Mercury laser is a high-pulse 
rate, next-generation laser fusion driver.}
\end{center}
\end{figure}

\subsection{Mercury amplifier}

The Mercury laser (Fig.~\ref{fig:mercury}) will operate at 10 Hz with 100 J pulses. 
Twelve such lasers would have to be time-multiplexed to achieve the 
$\gamma\gamma$ laser requirements.  The major challenge will be the modification of 
the Mercury laser pulse format,  which is currently a single 
several-nanosecond-long pulse.  Achieving the desired diffraction-limited 
beam quality will also be an important challenge.

\begin{figure}[bth]
\begin{center}
\psfig{file=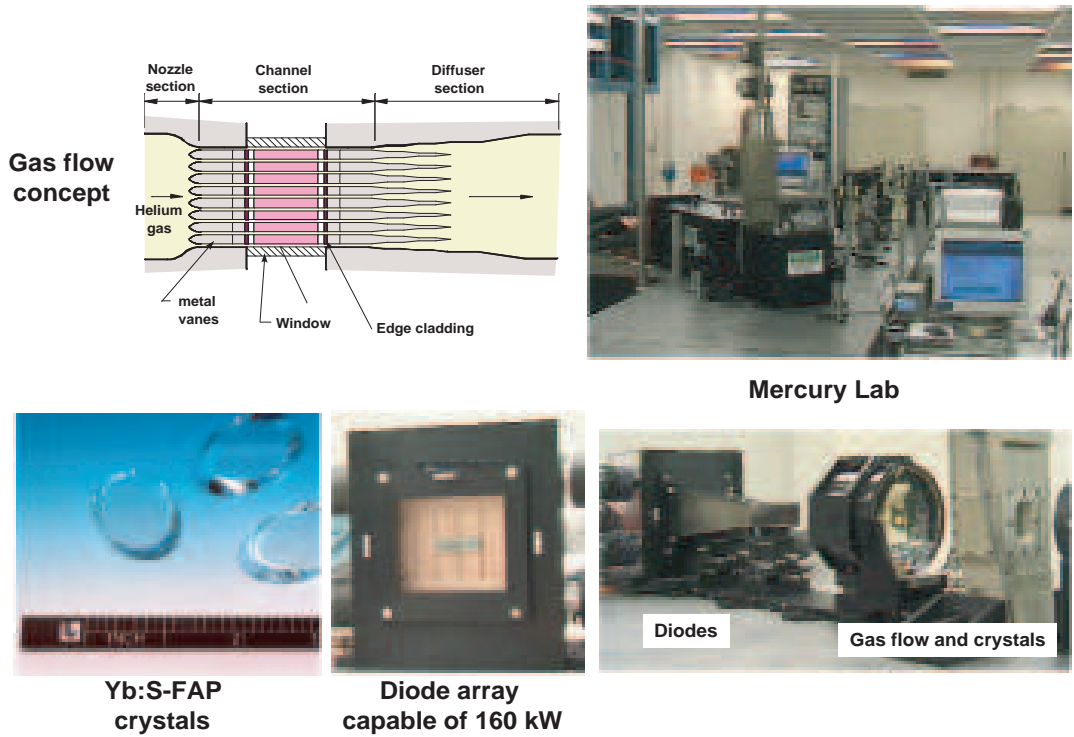,width=\textwidth}
\caption{\label{fig:early3}
The Mercury laser will utilize three key technologies: gas 
cooling, diodes, and Yb:S-FAP crystals to deliver 100 J at 10 Hz with 
10\% efficiency.}
\end{center}
\end{figure}

The Mercury laser utilizes three primary innovations to achieve the 
goal of a high-efficiency, high-repetition-rate laser driver for 
laser fusion experiments.  The first is that the 
removal of heat from the laser media 
is accomplished by flowing helium at high speed over the surface of 
thin laser slabs.  The thermal gradients in the laser media are 
oriented both in
the short dimension, for effective conductive cooling, and in the 
direction of the laser propagation, to minimize the optical 
distortion.  The low index of refraction of helium minimizes the 
helium thermal-optical
distortions that must later be removed with adaptive optics.  
Figure~\ref{fig:early3} shows the arrangement of thin laser 
slabs embedded in flow vanes 
within the helium flow duct.  Full-scale demonstrations have 
validated the flow and
thermal models have confirmed that the design meets the optical 
system requirements.

The second innovation is the use of diode lasers rather than flash 
lamps to energize the laser media.  The narrow frequency output of 
the diode laser is matched to the absorption band of the laser media. 
The efficient coupling and the efficiency of diode lasers result in 
significantly higher pumping efficiency of the laser media and also 
significantly lower waste heat that must be removed by the 
helium cooling system.  The primary challenge for the diode laser 
design is minimizing the high capital cost of the diode laser and 
its packaging design.  LLNL has developed a low-cost packaging 
design that also efficiently couples the diode light into the laser 
slabs.  This design has been produced under commercial contract and 
will be tested this year in the Mercury laser laboratory.

The third innovation is the use of Yb:S-FAP as the laser media 
instead of the usual Nd-glass.  This crystalline media has better 
thermal conductivity for cooling,  longer storage lifetime for 
efficient pumping, and a high quantum efficiency to minimize waste 
heat.  The growth of these new crystals (Fig.~\ref{fig:early3}) with adequate size 
and optical quality has been the primary technical challenge in the 
Mercury project.  Crystals grown recently may satisfy these 
requirements, but some testing remains to be done.

The Mercury laser has two amplifier heads and a four-pass optical 
system.  This year one amplifier head and the full optical 
configuration will be tested in the Mercury laboratory.  A second 
amplifier head must be constructed before full-power extraction can 
be demonstrated.

\subsection{Multiplexer and beam transport}

The beams from twelve Mercury lasers, each operating at 10 Hz, must 
be combined into a single co-aligned beam to produce the required 
120 Hz beam.  The beam combination should occur before the pulse 
compressor to minimize  the stress on the combiner optics.  At these 
low pulse rates the simplest beam combination scheme is a simple 
rotating faceted optic.  %~\cite{JNAI:genuflector}.

The beam combination optic is a 4 cm-diameter optic with twelve flat 
facets each covering a thirty degree sector. Each facet is ground at a 
slightly different angle.  The optic is rotated on its axis at 10 Hz 
(600 rpm).  The twelve incoming laser beams arrive at slightly 
different angle, such that they are all aligned after reflection off 
the optic.  The angle differences are sufficiently large to allow the 
incoming laser beams to be projected from spatially separated optics. 
The incident laser beam diameter of 0.5 cm will give a power density 
of 5 kW/cm$^2$ on the optic.  This will be below the damage threshold of 
10 kW/cm$^2$.  The optic can be made larger if  a larger damage margin 
is desired.

The combined beam is then transported to the pulse compressor.  The 
pulse compressor can be located in the laser facility or
close to the detector, just prior to the final transport optics into the IR.  
For the reference design it is assumed that the 
compressor is located in the laser facility and that the laser 
facility is located a nominal 100 meters from the detector hall.  The 
transport of the laser beams will be in vacuum pipes from the exit of 
the Mercury laser modules.  To minimize the evolution of amplitude 
variations due to diffraction or phase aberrations, the laser beam 
will be expanded to a nominal 10~cm and image-relayed.  The 
vacuum tubes should be 15 cm to allow for errors in initial alignment 
procedures.

\subsection{Compressor / stretcher}

The basic concept of compressing long pulses into short pulses after 
amplification is well known and 
widely used~\cite{Treacy:compression,Perry:compression}.  
The challenge is 
in designing and fabricating high-efficiency gratings that can handle 
high-power laser beams. The specifications for the stretcher and 
compressor systems are given in Table~\ref{tab:comp-req}.

\begin{table}[bth]
\begin{center}
\begin{tabular}{|l|l|l|}  \hline\hline
                                  & Stretcher    & Compressor \\ \hline
Substrate material                & silica       & silica  \\
Coating material                  & gold         & Multi-layer  \\
First grating size (cm)           & 4 x 15       & 30 x 84  \\
Second grating size (cm)          & 4 x 15       & 30 x 84  \\
Roof mirror size (cm)             & 4 x 8 (flat) & 30 x 40  \\
Grating separation (m)            & 5            & 15       \\
Lines per mm                      & 1740         & 1740     \\
Laser beam diameter (cm)          & 1            & 10       \\
Cut bandwidth (nm)                & 2.0          & 2.0      \\
Exit subpulse duration (ps)      & 3000         & 2.2      \\
Efficiency-single bounce (\%)     & 90.          & 96.0     \\
System efficiency (\%)            & 60           & 80       \\
Laser macro-pulse fluence (J/cm$^2$) & $10^{-7}$    & 1.3      \\
Damage fluence (J/cm$^2$)            & 0.4          & 2.0      \\
\hline\hline
\end{tabular}
\caption{\label{tab:comp-req}
Specifications for stretcher and compressor optical systems.}
\end{center}
\end{table}

The subpulses from the oscillator are 2 ps and 1.0 nJ. Their 
transform-limited full-width-half-maximum is 0.9 nm. The gratings in the 
stretcher give the beam an angular spread.  
Light of different wavelengths within the bandwidth of the laser follows 
optical paths of different length, thus introducing a frequency-time 
correlation to the subpulse (``chirping'').
The laser subpulse has a 3 ns halfwidth duration upon 
exiting.  The finite size of the grating results in the truncation of 
some frequencies and gives the exiting pulse a truncated spectral 
distribution and a temporal pulse with side lobes.  The 100 
subpulses that are separated by 2.8 ns will overlap to form a 300 ns 
macro-pulse that has some ($\sim 10$\%) time/amplitude modulation.  Since 
the beam in the stretcher is of such low power, there are no 
technical issues with this system.  The system efficiency will be 
limited by the reflectivity of the gratings in the first order and 
the frequency clipping due to finite grating size.

\begin{figure}[bth]
\begin{center}
\psfig{file=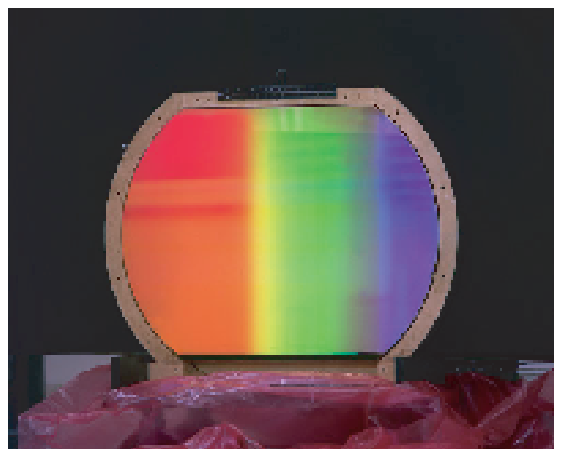,width=.48\textwidth}
\hfill
\psfig{file=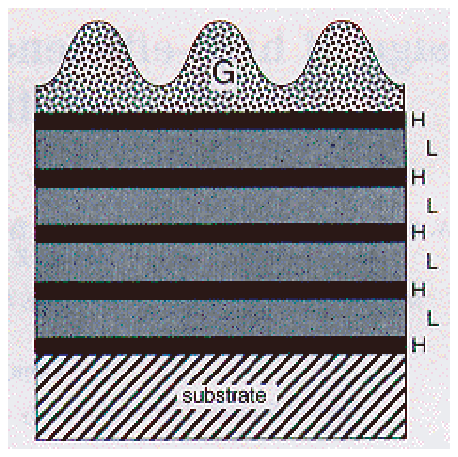,width=.48\textwidth}
\caption{\label{fig:early45}
The 94 cm aperture gold-coated diffraction grating 
used for pulse compression on the Petawatt laser is shown on the
left.  A multilayer dielectric grating design of 
high-index (H) and low-index (L) layers 
and groove corrugations (G) is shown on the right. 
Layers form a high-reflectivity stack under the corrugations.}
\end{center}
\end{figure}

The compressor gratings must be designed to handle the full 100 J 
macro-pulses without damage.  The 100 Hz pulse rate will also generate 
an average-power thermal concern.  The large gold coatings used in 
laser fusion experiments 
(Fig.~\ref{fig:early45}) have too large an absorption and 
would have thermal distortion problems.  LLNL has also developed 
multi-layer dielectric diffraction gratings with high 
efficiency~\cite{Shore:gratings}. 
Their low absorptivity removes the thermal concerns while also 
increasing the system efficiency. Figure~\ref{fig:early45} 
shows the design of these 
gratings. Alternating layers of hafnia and silica are placed on the 
substrate to give a high-reflectivity, high-damage fluence coating. 
The grating is etched  in the silica overcoating.

\begin{figure}[bth]
\begin{center}
\psfig{file=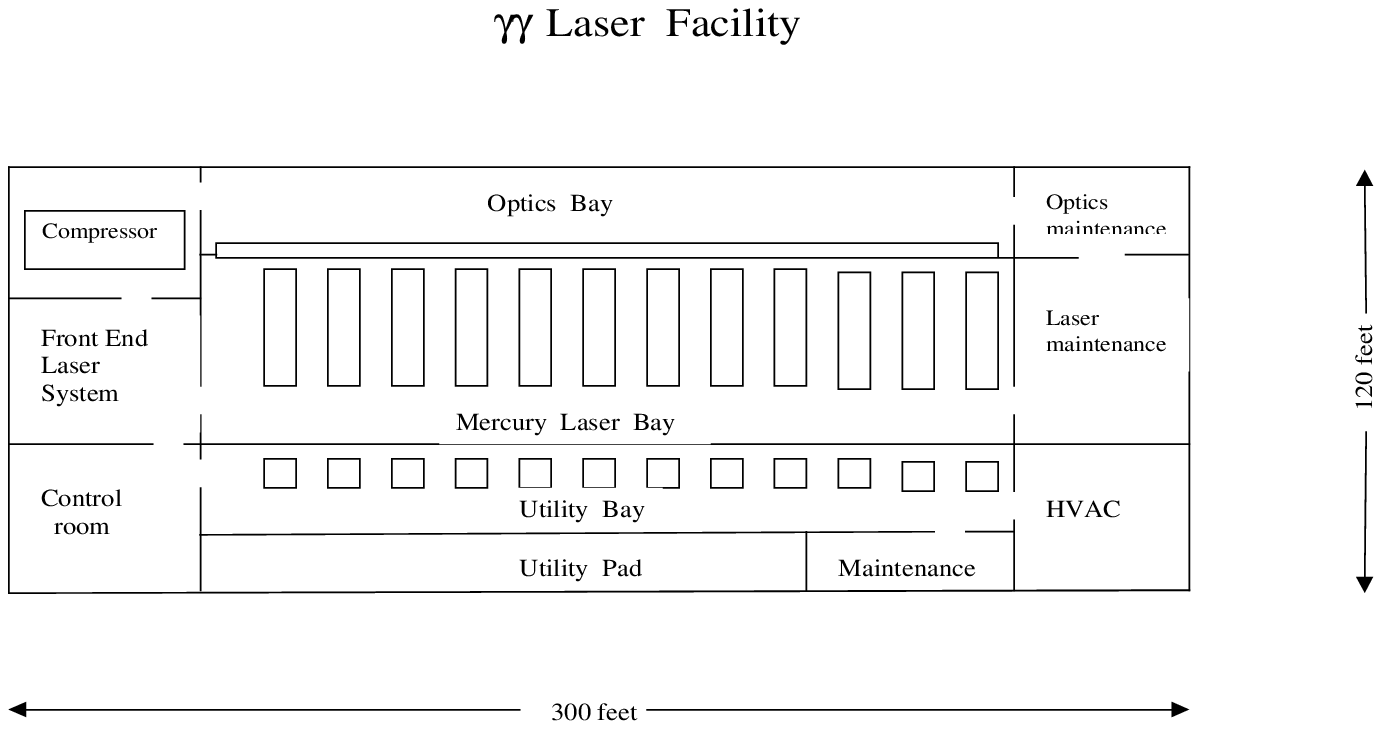,width=\textwidth}
\caption{\label{fig:laser_fac}
Floor plan of the laser physical plant.}
\end{center}
\end{figure}

\subsection{Laser facility, systems design and risk reduction}

The general layout of the laser facility is shown in 
Fig.~\ref{fig:laser_fac}.  The 
facility is dominated by the operating bays for the laser amplifiers 
and their utilities.  The operating strategy will be to do no laser 
repairs in these operation bays.  The laser systems will be designed 
with quickly removable Line Replaceable Units (LRUs) for all the 
major subsystems, as in the NIF project.  
The equipment will be monitored by computer 
during operation. When a system needs special or preventive 
maintenance, the LRU is quickly removed and moved to a separate 
repair facility.  A new LRU is inserted, and the laser is immediately 
returned to service.  This repair strategy allows for high system 
availability without requiring excessive component lifetimes or 
redundancy.  Some long-lifetime components such as the optics vacuum 
chamber may have to be occasionally repaired in place.

A systematic cost estimate has not yet been done.  The expected 
capital cost is of order of \$200M, and the operating budget of 
approximately \$20M/yr.  The largest 
uncertainties  in the capital costs are  the diode costs and the 
laser size needed to meet the performance requirements.  The 
operating cost uncertainties are dominated by diode laser lifetime 
and cost uncertainties.

The cost risk reduction strategy is to identify the main cost 
drivers.  Since diode lifetime is expected to be the primary cost 
risk driver, efforts will be made to acquire diode lifetime data.

The technical risks are dominated by the laser beam quality 
uncertainties and the lack of prototype demonstrations of some of the 
subsystems.  The Mercury laser being built for the fusion program 
will serve as the main laser amplifier prototype.  Other critical 
systems such as the laser system front end will be prototyped as part 
of a risk reduction program.

\setcounter{chapter}{13}

\chapter{\boldmath $e^-e^-$ Collisions}
\fancyhead[RO]{$e^-e^-$ Collider}

\section{General characteristics of $e^-e^-$ collisions}

The primary goal of the linear collider program will be to elucidate
new physics at the weak scale.  The $e^-e^-$ collider brings a number
of strengths to this program.  Electron-electron collisions are
characterized by several unique features:

\noindent
$\bullet$ {\em Exactly Specified Initial States and Flexibility.}  For
precision measurements, complete knowledge of the initial state is a
great virtue.  This information is provided optimally in $e^-e^-$
collisions.  The initial state energy is well-known for both $e^+$ and
$e^-$ beams, despite small radiative tails due to initial state
radiation and beamstrahlung. For $e^-$ beams, however, 85\%
polarization is routinely obtainable now,  and 90\% appears to be within
reach for linear colliders.  The three possible polarization
combinations allow one to completely specify the spin $S_z$, weak
isospin $I^3_w$, and hypercharge $Y$ of the initial state.  One may
also switch between these combinations with ease and incomparable flexibility.

\noindent
$\bullet$ {\em Extreme Cleanliness.}  Backgrounds are typically highly
suppressed in $e^-e^-$ collisions.  The typical annihilation processes
of $e^+e^-$ collisions are absent. In addition, processes involving
$W$ bosons, often an important background in $e^+e^-$ collisions, may
be greatly suppressed by right-polarizing {\em both} beams.

\noindent
$\bullet$ {\em Dictatorship of Leptons.}  In $e^+e^-$ collisions, particles
are produced `democratically'.  In contrast, the initial state of
$e^-e^-$ collisions has lepton number $L=2$, electron number $L_e = 2$,
 and electric charge
$Q=-2$.

With respect to the first two properties, the $e^-e^-$ collider takes
the linear collider concept to its logical end.  The third property
precludes many processes available in $e^+e^-$ interactions, but also
provides unique opportunities for the study of certain types of new
physics, such as supersymmetry.  The physics motivations for the
$e^-e^-$ collider have been elaborated in a series of workshops over
the past six years~\cite{ee-95,ee-97,ee-99}.  In the following, we
briefly describe a number of possibilities for new physics in which
$e^-e^-$ collisions provide information beyond what is possible in
other experimental settings.  We then review the accelerator and
experimental issues relevant for $e^-e^-$ collisions.

\section{Physics at $e^-e^-$ colliders}

\subsection{M{\o}ller scattering}

The process $e^-e^- \to e^-e^-$ is, of course, present in the standard
model.  At $e^-e^-$ colliders, the ability to polarize both beams makes
it possible to exploit this process fully.

One may, for example, define two left-right asymmetries
\begin{eqnarray}
A_{LR}^{(1)} &\equiv&
\frac{d\sigma_{LL} + d\sigma_{LR} - d\sigma_{RL} - d\sigma_{RR}}
     {d\sigma_{LL} + d\sigma_{LR} + d\sigma_{RL} + d\sigma_{RR}}
\nonumber \\[2ex]
A_{LR}^{(2)} &\equiv&
\frac{d\sigma_{LL} - d\sigma_{RR}}
     {d\sigma_{LL} + d\sigma_{RR}} \ ,
\end{eqnarray}
where $d\sigma_{ij}$ is the differential cross section for $e^-_i
e^-_j \to e^- e^-$ scattering.  There are four possible beam
polarization configurations.  The number of events in each of the four
configurations, $N_{ij}$, depends on the two beam polarizations $P_1$
and $P_2$.  Given the standard model value for $A_{LR}^{(1)}$, the
values of $N_{ij}$ allow one to simultaneously determine $P_1$, $P_2$,
and $A_{LR}^{(2)}$.  For polarizations $P_1 \simeq P_2 \simeq 90\%$,
integrated luminosity $10~\ifb$, and $\sqrt{s}=500~\gev$, the beam
polarizations may be determined to $\Delta P / P \approx
1\%$~\cite{ee-Cuypers:1996it,ee-Czarnecki:1998xc}.  Such a measurement
is comparable to precisions achieved with Compton polarimetry, and has
the advantage that it is a direct measurement of beam polarization at
the interaction point.

This analysis also yields a determination of $A_{LR}^{(2)}$, as noted
above.  Any inconsistency with the standard model prediction is then a
signal of new physics.  For example, one might consider the
possibility of electron compositeness, parameterized by the dimension-six
 operator ${\cal L}_{\rm eff} = \frac{2\pi}{\Lambda^2} \bar{e}_L
\gamma^{\mu} e_L \bar{e}_L \gamma_{\mu} e_L$.  With $\sqrt{s} =
1~\tev$ and an $82~\ifb$ event sample, an $e^-e^-$ collider is
sensitive to scales as high as $\Lambda = 150~\tev$~\cite{ee-Barklow}.
The analogous reach for Bhabha scattering at $e^+e^-$ colliders
with equivalent luminosity is roughly $\Lambda = 100~\tev$.

\subsection{Higgs bosons}

The Higgs boson production mechanism $e^+e^- \to Zh$ in the $e^+e^-$
mode is complemented by production through $WW$ and $ZZ$ fusion in both
$e^+e^-$ and $e^-e^-$ colliders.  The study of $e^-e^- \to e^-e^- h^0$
through $ZZ$ fusion has a number of
advantages~\cite{ee-Minkowski:1998cv,ee-Gunion:1998jc}. The cross section is
large at high energy, since it does not fall off as $1/s$.  The
usual backgrounds from $e^+e^-$ annihilation are absent.
  The final electrons typically have transverse
momenta of order $m_Z$.  Thus, one can reconstruct the recoil mass
and observe the Higgs boson
in this distribution, as shown in Fig.~\ref{fig:mink}.  Invisible decays
of the Higgs boson, and branching ratios more generally, can be studied
by this technique.

\begin{figure}[htbp]
\begin{center}
\epsfig{file=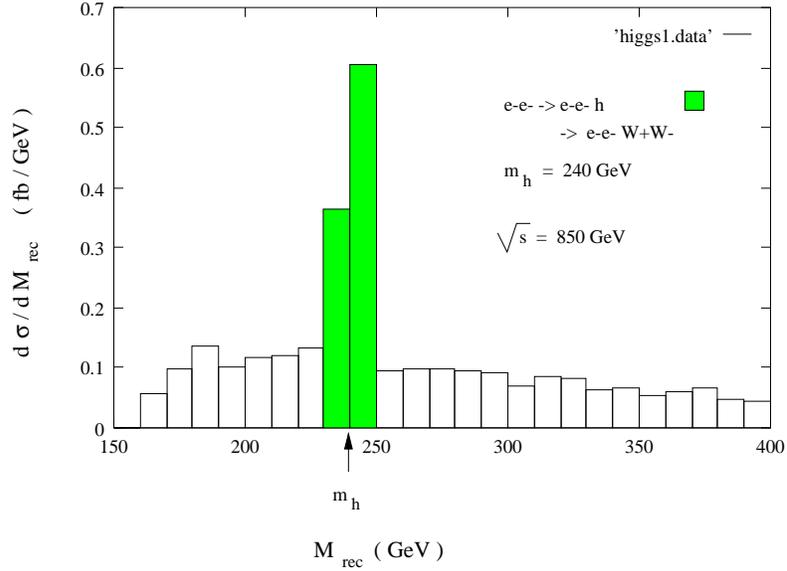,height=3.in}
\end{center}
\caption{\label{fig:mink}
Differential cross sections as functions of recoil mass for
$e^-e^- \to e^- e^- h$ and its principal standard model background
$e^-e^- \rightarrow e^-e^- W^+ W^-$.  The Higgs boson mass is $m_h =
240~\gev$, $\sqrt{s} = 850~\gev$, and each electron satisfies an
angular cut $\theta_{e^-} > 5^{\circ}$.  {}From~\protect\cite{ee-Minkowski:1998cv}.}
\end{figure}

\subsection{Supersymmetry}

The $e^-e^-$ mode is an ideal setting for studies of sleptons.  All
supersymmetric models contain Majorana fermions that couple to
electrons---the electroweak gauginos $\widetilde{B}$ and
$\widetilde{W}$. Slepton pair production is therefore always
possible~\cite{ee-Keung:1983nq}, while all potential backgrounds are
absent or highly suppressed.  Precision measurements of slepton masses,
slepton flavor mixings, and slepton couplings in the $e^-e^-$ mode are
typically far superior to those possible in the $e^+e^-$ mode.  Studies of
all of these possibilities are reviewed in Chapter 4, Section 6.1.

The $e^-e^-$ collider may also be used to determine the properties of
other superpartners.  For example, the production of right-handed
selectron pairs is highly sensitive to the Majorana Bino mass $M_1$
that enters in the $t$-channel (see Fig.~\ref{fig:M1dependence}).  As
a consequence, extremely high Bino masses $M_1$ may be measured
through the cross section of $\tilde{e}_R^-$ pair
production~\cite{ee-Feng:1998ud}.  This region of parameter space is
difficult to access in other ways.

\begin{figure}[htbp]
\begin{center}
\epsfig{file=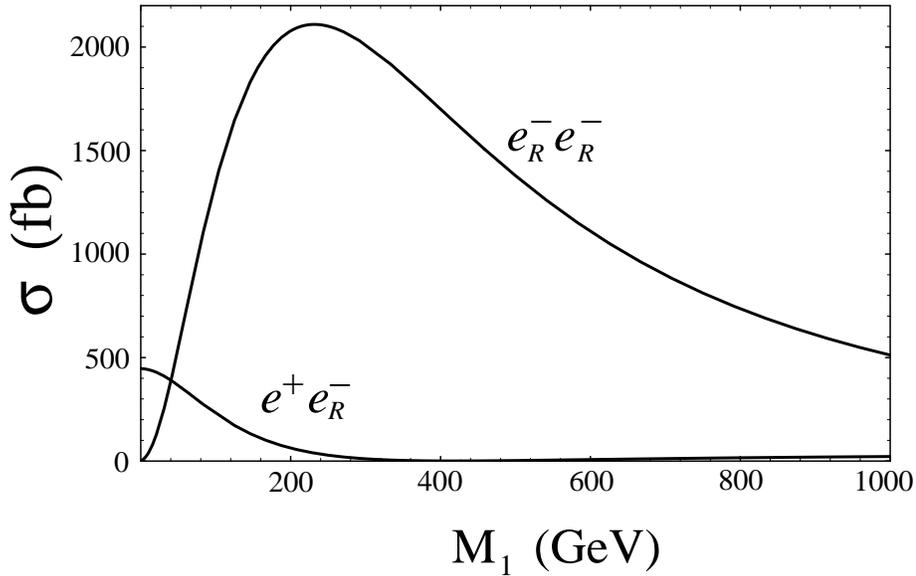,width=5in,height=3in}
\end{center}
\caption{\label{fig:M1dependence}
The total selectron pair production cross sections for the
$e_R^- e_R^-$ and $e^+ e_R^-$ modes with $m_{\tilde{e}_R}=150~\gev$
and $\protect\sqrt{s}=500~\gev$, as functions of the Bino mass
$M_1$. {}From~\protect\cite{ee-Feng:1998ud}.}
\end{figure}

\subsection{Bileptons}

The peculiar initial state quantum numbers of $e^-e^-$ colliders make
them uniquely suited for the exploration of a variety of exotic phenomena.
Among these are bileptons, particles with lepton number $L=\pm 2$.
Such particles appear, for example, in models where the SU(2)$_L$
gauge group is extended to SU(3)~\cite{ee-Frampton:1998gc}, and the
Lagrangian contains the terms
\begin{equation}
{\cal L} \supset \left(\begin{array}{ccc} \ell^-& \nu & \ell^+
\end{array}\right)_L^*
\left( \begin{array}{ccc}
 &  & Y^{--} \\
 & & Y^- \\
Y^{++} & Y^+ & \end{array} \right)
\left( \begin{array}{c} \ell^- \\ \nu \\ \ell^+ \end{array} \right)_L
 \ ,
\end{equation}
where $Y$ are new gauge bosons.  $Y^{--}$ may then be produced as an
$s$-channel resonance at $e^-e^-$ colliders, mediating background-free
events like $e^-e^- \to Y^{--} \to \mu^- \mu^-$.  Clearly the $e^-e^-$
collider is ideal for such studies.

Bileptons may also appear in models with extended Higgs sectors that
contain doubly charged Higgs bosons $H^{--}$.  In these models, both
types of particles are produced as resonances in $e^-e^-$ scattering.
However, the types of states are clearly distinguished by initial state polarization:
bileptons are
produced from initial polarization states with $|J_z| = 1$, while doubly
charged Higgs particles
are produced in channels with $J_z = 0$. The potential of
$e^-e^-$ colliders to probe the full spectrum of these
models is reviewed in~\cite{ee-Gunion:1998ii}.

\subsection{Other physics}

In addition to these topics, the potential of $e^-e^-$ colliders has
also been studied as a probe of strong $W^-W^-$ scattering, anomalous
trilinear and quartic gauge boson couplings, heavy Majorana neutrinos,
leptoquarks, heavy $Z'$ bosons, TeV-scale gravity and Kaluza-Klein
states, and non-commuting spacetime observables.  These topics and
other possibilities are discussed in~\cite{ee-95,ee-97,ee-99}.

\section{Accelerator and experimental issues}

\subsection{Machine design}

There are at present two well-developed approaches to linear collider
architecture in the 0.35 to 1 TeV energy range: the NLC/JLC and
TESLA designs. Both approaches are easily adaptable to make both
$e^+e^-$ and $e^-e^-$ collisions available with relatively little
overhead.

The general layout of the NLC design is given in Fig.~\ref{fig:NLC}.
The careful inclusion of the $e^-e^-$ design is described in
~\cite{ee-Tenenbaum:2000kv}. The installation of a second
polarized electron source presents no difficulty, but magnet polarity
reversals and potential spin rotators need to be carefully optimized.

\begin{figure}[htbp]
\begin{center}
\epsfig{file=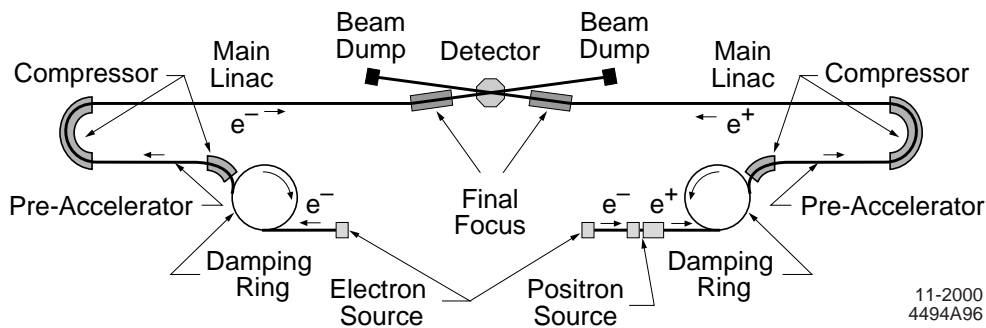} % ,height=3.5cm}
\end{center}
\caption{\label{fig:NLC}
Schematic of the NLC. {}From
~\cite{ee-Tenenbaum:2000kv}.}
\end{figure}

Three different modifications for the injection area on the
``positron" side have been investigated~\cite{ee-Larsen:2000pr}.  We
show one of these in Fig.~\ref{fig:linac}.  In this scheme, the damping
ring and bunch compressor for the $e^+$ beam are used for an $e^-$ beam
which circulates in the opposite direction.  A new electron gun and some
additional components for injection and extraction are needed, but the cost
of these is modest, and the switchover from $e^+$ to $e^-$ operation can be
accomplished without significant manual intervention.

\begin{figure}[htbp]
\begin{center}
\epsfig{file=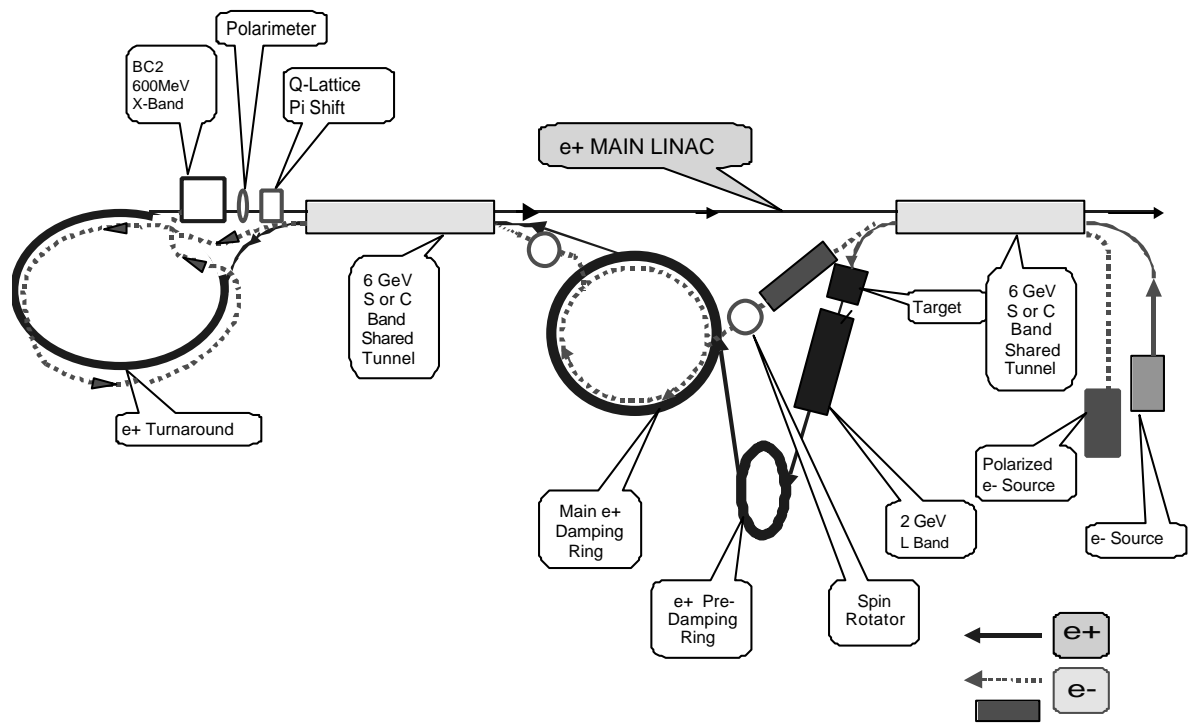,width=\hsize}
\end{center}
\caption{\label{fig:linac}
The direction reversal model. {}From
~\cite{ee-Larsen:2000pr}.}
\vskip-1pc
\end{figure}

For the TESLA project, it is even simpler to introduce polarized $e^-$
through the $e^+$ injection system.  A new polarized electron source is needed, and
new components are needed for injection and extraction from the
existing positron `dogbone' damping ring \cite{ee-TESLATDRdogbone}.
The positions of these
new devices mirror the positions of the electron injection and extraction
points on the other side of the machine.

Similar considerations apply to the higher-energy CLIC
 proposal~\cite{ee-delahaye}.  As with NLC/JLC and TESLA, the main
difficulties involve the injection scheme; once appropriate components are
provided, the acceleration of $e^-$ beams and the switchover from
$\ee$ to $e^-e^-$ should be straightforward.

\subsection{Interaction region}

Although $e^-e^-$ operation is straightforwardly incorporated in
linear collider designs, experimentation at $e^-e^-$ colliders is not
entirely equivalent to that at $e^+e^-$ colliders.  This is because
the luminosity of the collider is decreased significantly by
beam disruption due to the
electromagnetic repulsion of the two $e^-$ beams.

Clever manipulation of the beam parameters can minimize the relative
luminosity loss; see, for example, \cite{ee-Thompson:2000ij}.  The resulting
parameters give about a factor 3 loss for NLC/JLC  and a factor 5 loss for
TESLA, and
do not much reduce the merits of the proposed $e^-e^-$ studies.
A plasma lens~\cite{ee-Chen:1996jg,ee-Ng:2000xh} has been proposed
to reduce the disruption effects, but this would introduce
a serious level of beam-gas backgrounds.

The beamstrahlung effect in $e^-e^-$ is somewhat larger than that in
$\ee$ due to the larger disruption, leading to a stronger effective
field from the opposite beam.  The effect is still modest in size for
500 GeV CM energy.  Figure~\ref{fig:R-S} shows a comparison of the
 $e^-e^-$ and $\ee$ cases for the TESLA machine design~\cite{ee-Reyzl:2000ps}.

\begin{figure}[htbp]
\begin{center}
\epsfig{file=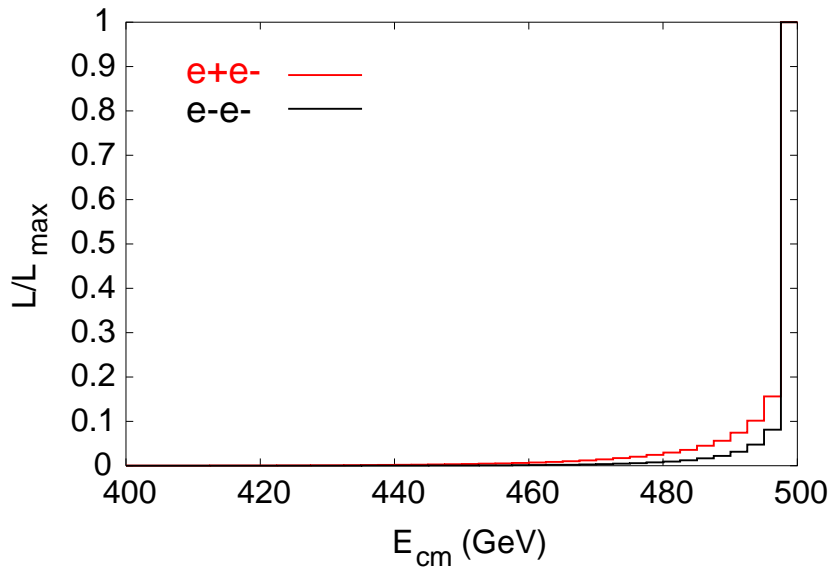,height=3in}
\end{center}
\caption{\label{fig:R-S}
Normalized luminosity spectrum for $e^-e^-$ collisions
compared to $e^+e^-$. {}From~\cite{ee-Reyzl:2000ps}.  }
\end{figure}

\subsection{Detectors}

It is important to realize that the detector configuration is easily
shared for $e^+e^-$ and $e^-e^-$ experimentation. A caveat exists for
beam disposal downstream of the interaction point: if there is any
bend upstream of this point, like-sign incoming beams will {\em not}
follow the incoming trajectories of the opposite side, and special
beam dumps may have to be configured.

If the linear collider program plans to incorporate
$e\gamma$ and $\gamma\gamma$ collisions, with backscattered photon beams,
the photon beams must be created from $e^-$ rather than $e^+$ beams, so
that the electron beam polarization can be used to optimize the energy
spectrum and polarization of the photon beams.  Photon colliders of course
have
their own, very different, requirements for interaction regions and
detectors.  These are described in Chapter 13, Section 3.

\section{Conclusions}

For a number of interesting physics scenarios,
the unique properties of $e^-e^-$
colliders will provide additional information through new channels and
observables.  While the specific scenario realized in nature is yet to
be determined, these additional tools may prove extremely valuable in
elucidating the physics of the weak scale and beyond.  Given the
similarities of the $e^+e^-$ and $e^-e^-$ colliders,  it should be
possible with some thought in advance to guarantee the compatibility of these
two modes of operation and the ease of switching between them.  For
many possibilities for new physics in the energy region of the linear collider,
the small effort to ensure the availability of $e^-e^-$ collisions should
reap great benefits.

\emptyheads

\begin{center}\begin{large}{\Huge \sffamily Detectors for the Linear Collider}
\end{large}\end{center}

\addtocontents{toc}
{\bigskip\noindent{\huge Detectors for the Linear Collider}}

\blankpage \thispagestyle{empty}

\fancyheads

\setcounter{chapter}{14}

\chapter{Detectors for the Linear Collider}
\fancyhead[RO]{Detectors for the Linear Collider}

\section{Introduction}

The linear collider detector must be optimized for physics performance,
taking consideration of its special environment.  To plan for this 
detector, we consider the physics
requirements of the linear collider  and build on the
experience of operating SLD at the SLC.

  The detector must be hermetic,
with good charged-track momentum and impact parameter resolution.  The
calorimeter must provide good resolution, with good granularity,
particularly in the electromagnetic section.  Electron and muon
identification must be done efficiently.

The beamline conditions of the linear collider motivate a strong
solenoidal magnetic field to contain the vast number of low-energy
electron-positron pairs.  There must be provision for an
 accurate measurement of the differential
luminosity, and for timing information that will be useful to
separate interactions from separate bunches within a bunch train.

This chapter begins with a discussion of the major issues for the linear
collider detector, starting from the beamline conditions and working
through the subsystems.  Following this discussion, three potential
detectors developed for the NLC are described, two designed for the
higher-energy IR, and the third for the second IR, where the 
lower-energy operation is foreseen.  Other detectors have been considered in
Europe \cite{tesla-tdr} and Asia \cite{jlc-det}.

These detector studies have been undertaken to understand how well the diverse
physics measurements at a linear collider can be accomplished, to provide
preliminary guidance on costs, and to highlight areas where R\&D is needed.
The specific choices of technology and full detector optimization will await
the formation of LC experimental collaborations.

\section{Interaction region issues for the detector}

\subsection{Time structure.} The NLC is expected to  operate with
trains of 190 bunches with 1.4 ns bunch spacing.  This time structure
requires that the beams cross at an angle.  It also affects the number
of bunches seen within the integration time of any detector subcomponent
and has a strong influence on the types of feedback schemes that can be
used to keep the beams in collision.

{\it Crossing angle and parasitic collisions.} In order to avoid
parasitic collisions, a crossing angle between the colliding beams is
required.  The minimum angle acceptable for this beam-beam limit is
approximately 4 mrad for the NLC parameters.  A larger angle is
desirable because it permits a more straightforward
 extraction of the spent beams
(see Fig.~\ref{fig:layout}), but an excessively large crossing angle
will result in a luminosity loss.  The angle between
the beams chosen in the NLC design  is 20 mrad.

 The bunches must interact head-to-head or there will
be a substantial loss of luminosity.  RF cavities that rotate each
bunch transversely will be located 10--20 m on either side of the IP.  At
20 mrad crossing angle, the relative phasing of the two RF pulses must
be accurate to within 10 $\mu$m to limit the luminosity loss to less
than 2\%.  This corresponds to 0.04 degrees of phase at S-band (2.8
GHz).  The achievable resolution is about 0.02 degrees, which sets an
upper limit on the crossing angle of 40 mrad.

{\it Solenoid field effects.} The crossing angle in the $x$--$z$ plane
causes the beam to see a transverse component of the detector's solenoid
field.  If uncorrected, this field will deflect the beams so they do not
collide.  Likewise, the deflection would cause dispersion that would
blow up the beam spot size.  Both of these effects can be cancelled by
judiciously offsetting the position of the last quadrupole, QD0, and
steering the beam appropriately. Synchrotron radiation emission in the 
transverse field leads to  an irreducible increase in spot size.  This
effect is proportional to $(L^*B_S \theta_C)^{5/2}$, where $L^*$ is the
distance between the IP face of the last magnet and the interaction
point.  While it is small at the values of $L^*$, $B_S$, and $\theta_C$
considered to date, this effect might someday limit the design of the
detector and IR.

After the beams collide at the IP, they are further bent by the solenoid
field.  Since compensating for this energy-dependent position and angle
change with independent dipoles is difficult, the extraction line must
be adjusted appropriately for the chosen beam energy.  Realignment will
be required if the extraction line does not have adequate dynamic
aperture to accommodate the full range of beam energies used in experiments.

Finally, if the permeability of QD0 is not exactly unity, the field
gradient of the solenoid in the detector endcap region will result in
forces on QD0 that will need to be compensated.  This may influence the
schemes considered to compensate for nanometer-level vibration
compensation of the magnet.

\subsection{IR layout}

\begin{figure}[htb]
\begin{center}
\epsfig{file=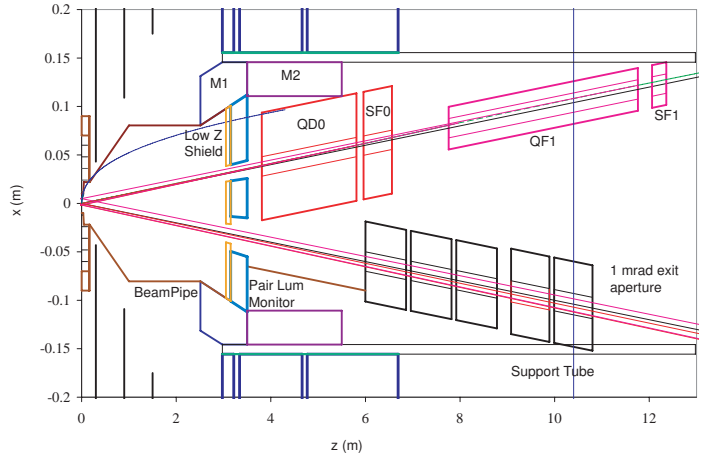,height=2.5in}
\caption{IR Layout for the NLC Large Detector.}
\label{fig:layout}
\end{center}
\end{figure}

{\it Magnet technology.} The NLC/JLC and TESLA designs have chosen to use
different technologies for the final quadrupole doublet.  The choices are
dictated by the choice of crossing angle, and by the  scheme to extract the
spent beam after the collision.  The NLC approach is to extract outside
the outer radius of a compact Rare Earth Cobalt (REC) magnet into an
extraction line that begins 6 m from the IP.  The REC of choice is Sm$_2$
Co$_{17}$, because of its radiation-resistant properties.  Since the final
quadrupoles can be made light and stiff and have no external power
connections, they are well suited for vibration stabilization.  The
downside to this choice is a lack of flexibility.  Other issues that
need to be explored further are the compatibility of the REC material
with the solenoid field of the detector and the variation of the magnetic
field with temperature.

In the current design $L^*$ = 3.8 m. An additional 30 cm of free space
has been left in front of the pair luminosity monitor to allow for
different magnet configurations as the beam energy is increased.
Increasing $L^*$ provides more transverse space for the final quads and
moves their mounting points further outside the detector, where
 they can presumably be better stabilized.  By keeping $L^*$ larger than
the minimum $z$ of the endcap calorimeter, the heavy W/Si-instrumented
mask described below can be better incorporated into the detector's
acceptance and mechanical structure.  By increasing the distance between
the IP and the first piece of high-$Z$ material seen by the beam,
one can minimize  the
effect of backscattered debris from the interaction of off-energy
$e^+e^-$ pairs created when  the beams interact.

On the other hand, increasing $L^*$ tightens the tolerances of the Final
Focus optics and reduces its bandwidth.  Synchrotron radiation produced
by beam halo particles in the final lenses determines the minimum radius of
the beam pipe inside the vertex detector.  The larger $L^*$ is, the
larger will be the fan of photons shining on the vertex detector.

{\it Masks.} The electrons and positrons produced in pairs in the
beam-beam interaction have a mean energy of about  13 GeV at $\sqrt{s} =$ 1
TeV.  These off-energy particles spiral in the detector's solenoid
field and strike the pair-luminosity monitor or the inner bore of the
extraction line magnet.  The main purpose of the masking is to shield
the detector from the secondary particle debris produced when an $e^\pm$
interacts.  There are three masks foreseen.  M1 begins at the back of
the pair luminosity monitor and extends 0.64 m in $z$ beyond its front
face; its inner angle is set by the requirement that it stay just
outside the so-called ``dead cone" through which the pairs coming from
the IP travel.  With the mask tip at 2.5 m, a 3 T field requires an
inner angle of 32 mrad.  This mask would ideally be made of W/Si and be
fully integrated with the detector's calorimetry.  M2 is a simple
tungsten cylinder.  The last mask near the IP is a 10--50 cm layer of 
low-$Z$ material (\eg, Be or C)  that absorbs low energy charged
particles and neutrons produced when the pairs hit the front face of the
W/Si pair luminosity monitor.  The very low-energy charged secondaries
would otherwise flow back along the solenoid's field lines toward the
vertex detector (VXD) and produce unacceptable backgrounds.

\subsection{Small spot size issues}

The beams must be held stable with respect to one another in the
vertical plane at the level of one nanometer.  Measurements in existing
detectors imply that the mounting of the final quadrupoles 
 may have to correct as
much as 50 nm of vibration, caused mostly by local vibration sources and
to a much lesser extent by naturally occurring seismic ground motion.
Concerns about vibrations caused by moving fluids lead to the choice of
permanent magnets for QD0 and QF1.  These magnets will be mounted in
cam-driven mover assemblies and the beam-beam interaction used to
control their position to compensate for disturbances at frequencies
below about $f$/20, where $f$ is the beam repetition rate of 120 Hz.

For frequencies above 5--6 Hz, the NLC strategy for stabilizing
luminosity relies on a combination of passive compliance (minimizing and
passively suppressing  vibration sources while engineering to avoid resonant
behavior) and active suppression techniques.  Quad motion will be
measured either optically relative to the surrounding bedrock or
inertially, and a correction will be applied to either the final doublet
position (via an independent set of magnet movers) or its field center
(via a corrector coil).  Finally, there will be feedback based on the
measured  beam-beam deflection.  Such a system
 can respond sufficiently rapidly (within 15 ns) to
correct the trailing bunches in a train, once the first few are used to
measure any collision offset.

\subsection{The beam-beam interaction}

The two main experimental consequences of the beam-beam interaction are
a broadening of the energy distribution, due to the emission of photons
by one beam in the field of the oncoming beam, and the subsequent
background generated by  interactions of those photons.  The beamstrahlung
contribution to the energy spread must be considered together with the 
intrinsic energy spread of the accelerator and the effect of initial
state radiation.  These effects have been taken into account in the 
discussion of the 
 various physics process.  Below we discuss the beam-beam
interaction as a potential source of backgrounds.

{\it $e^+e^-$ pairs and the minimum solenoid field.} The incoherent
production of $e^+e^-$ pairs arising from Bethe-Heitler
($e^\pm\gamma\rightarrow e^\pm e^+e^-$), Breit-Wheeler
($\gamma\gamma\rightarrow e^+e^-$), and Landau-Lifshitz
($e^+e^-\rightarrow e^+e^-e^+e^-$) processes is the main source of
background at the present generation of planned linear colliders. 
At CM energies of 1~TeV, roughly
$10^5$ particles are produced each bunch crossing, with a
mean energy of  13 GeV.  Very few particles
are produced at a large angle and the dominant deflection is due to the
collective field of the oncoming beam.  The so-called `dead cone' that
is filled by these particles is clear in the $R_{\rm max}$ vs. $z$ plot in
Fig.~\ref{fig:rmax}.  The beam pipe inside the VXD innermost layer
must be large enough and short enough that it does not intersect this
region.

\begin{figure}
\begin{minipage}[htbp]{3.0in}
\begin{center}
\vspace{-1in}
\mbox{\epsfig{file=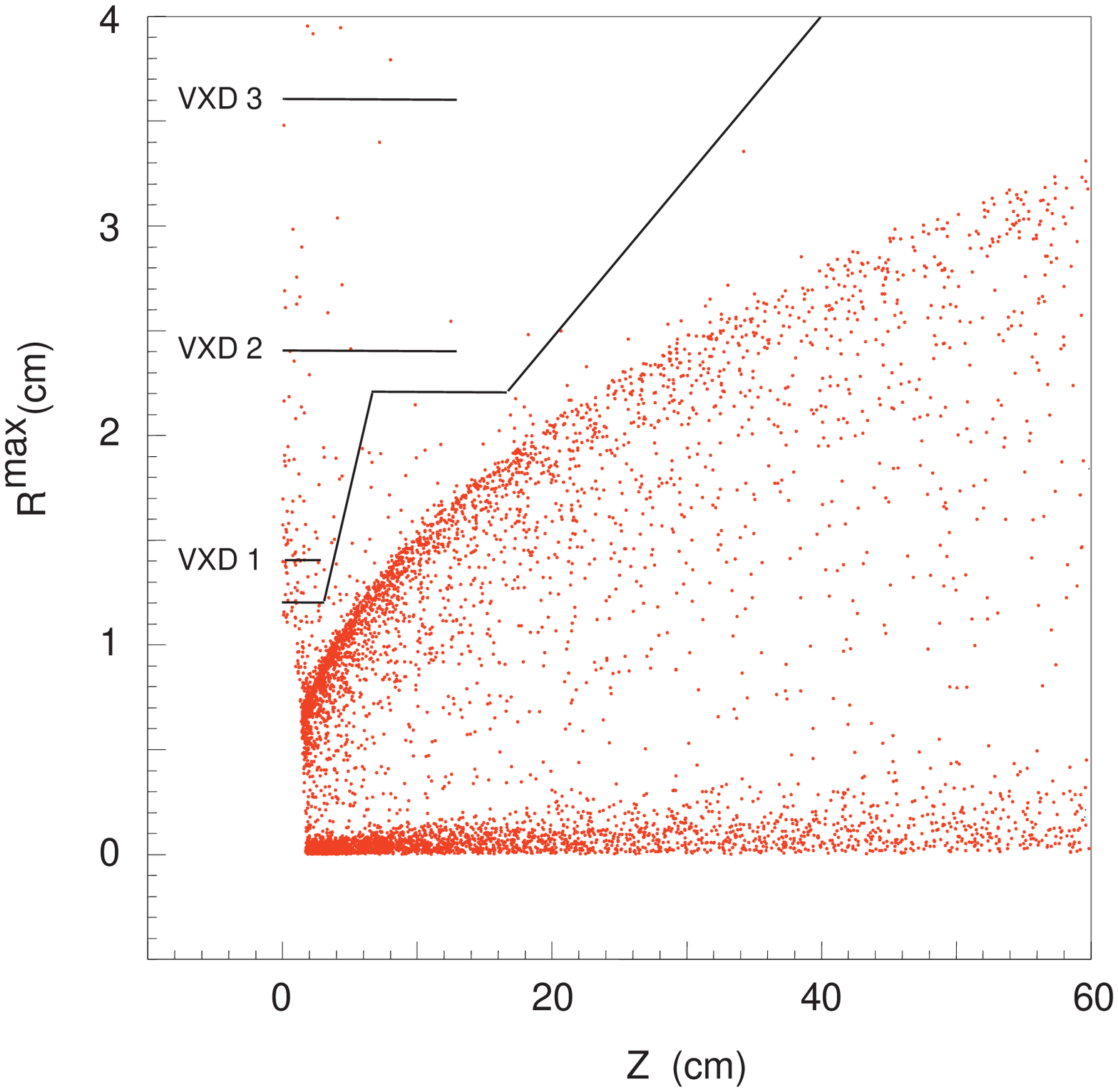,width=2.5in}}
\end{center}
\caption[rmax]{
\label{fig:rmax}
$R_{\rm max}$ vs. $z$ distribution of pairs in a 3 Tesla solenoid field.
$R_{\rm max}$ is the maximum radius the particle travels from the IP,
plotted at the $z$ corresponding to the first apex of its helical trajectory.
}
\end{minipage}
\begin{minipage}[t]{0.2in}~~\end{minipage}
\begin{minipage}[t]{3.0in}
\begin{center}
\mbox{\epsfig{file=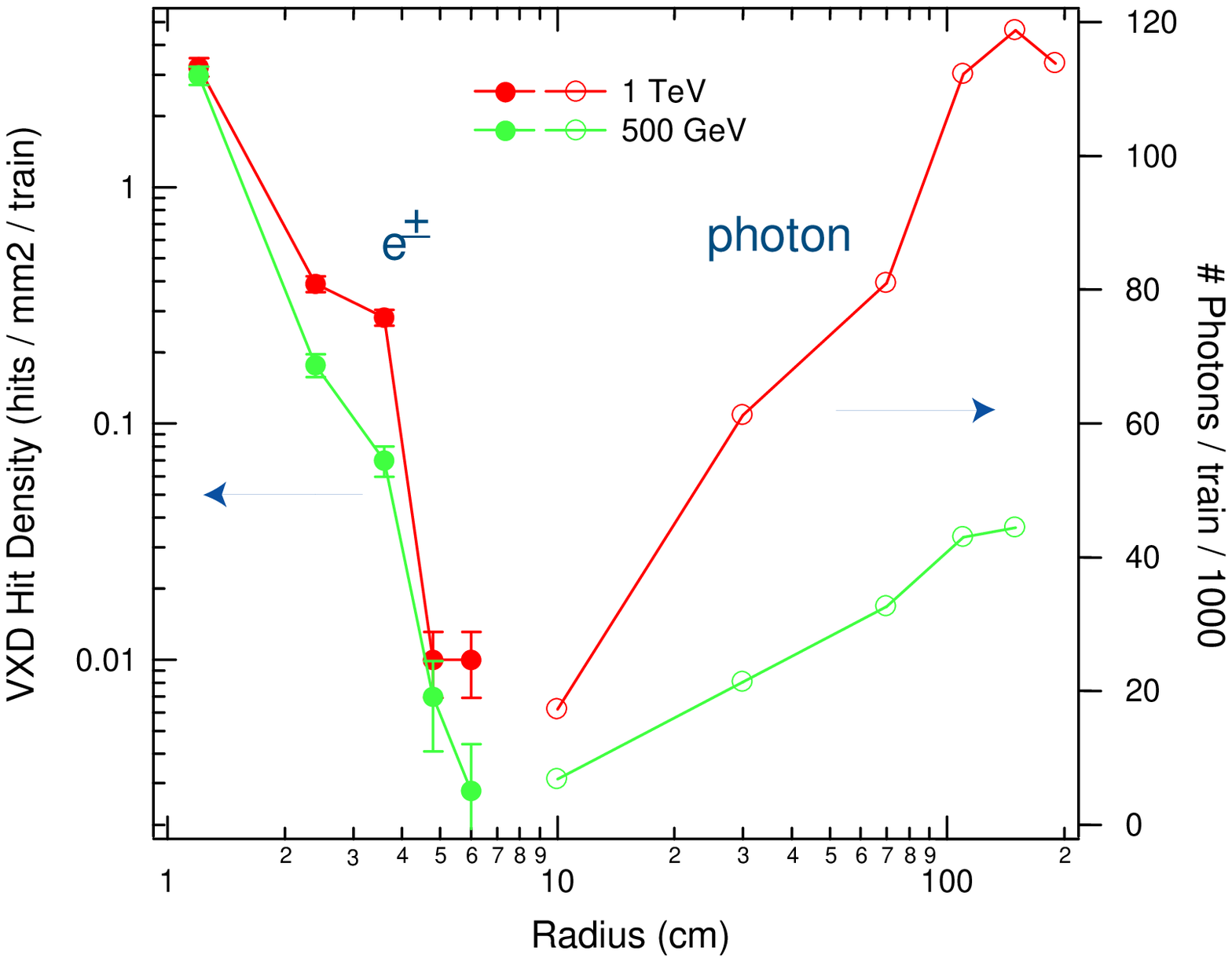,width=3.0in}}
\end{center}
\caption{Charged particle hit density per train in the VXD,
and the absolute number of photons per train entering the TPC
within $|\cos\theta|<0.92$, as function of radius.
\label{fig:hits}
}
\end{minipage}
\end{figure}

{\it Secondary particles and
their sources.} Secondary particle backgrounds---from 
neutrons, photons, and charged particles---can be a problem for the
detector whenever primary particles or particles from the
collision are lost close to the IP. 
  The main purpose of the masking described
earlier is to limit the backgrounds these secondaries produce.  Figure
\ref{fig:hits} shows the charged particle hit density per train in the
VXD as a function of radius, and the absolute number of photons per train
entering the TPC within $|\cos\theta|<0.92$.  The most important sources
of secondary particles are as follows:
\begin{itemize}
\item  $e^+e^-$ pairs striking the pair luminosity monitor are the
most important source of secondaries as the pairs are well off the
nominal beam energy, spiral in the detector's field and strike high-Z
materials close to the IP.  Backgrounds from this process are controlled
by the masks described above.
\item  Radiative Bhabhas are a source of
off-energy particles that are outside the energy acceptance of the
extraction line.  However, they are sufficiently few in number and leave
the beam line sufficiently far from the IP that they are not an
important background for the main detector elements.
\item The low-energy tail of the disrupted beam
cannot be transported all the way to the dump.  The current design of
the extraction line includes a chicane to move the charged beam
transversely relative to the neutral beam of beamstrahlung photons.  The
bends at the beginning and the end of the chicane are the primary
locations where particles are lost.  The number of particles lost,
$\sim$ 0.25\% of the beam, and the separation of the loss point from the
IP makes this an unimportant background source for the main detector,
but calls into question the viability of sophisticated instrumentation,
such as a polarimeter and an energy spectrometer, in the extraction
line.
\item  Neutrons shining back on the detector from the
dump are controlled by shielding immediately surrounding the dump,
placing concrete plugs at the tunnel mouths, maximizing the distance from the
dump to the IP, and minimizing window penetrations in the concrete.  The
detector of most concern is the VXD, which can look into the dump with
an aperture equal to that provided to accommodate the outgoing beamstrahlung
photons and synchrotron radiation.
\end{itemize}

{\it Beamstrahlung photons.} At 500 GeV, 5\% of the beam power is
transformed into beamstrahlung photons; this rises to 10\% at 1~TeV.
 The IR is designed so that these photons pass unimpeded to a dump.  This 
consideration, along with the angular spread of the synchrotron radiation
(SR) photons, determines
the exit aperture of the extraction line, currently set at 1 mrad.   The
maximum transverse size of the dump window that can be engineered and
the beamstrahlung angular spread set the maximum distance the dump can
be located from the IP.  That distance and the size of the aperture in
the concrete blockhouse surrounding the dump determine the level of
neutron backshine at the detector.

{\it Hadrons from $\gamma\gamma$ interactions.} Beamstrahlung photon
interactions will also produce hadrons.  For the TESLA 500 GeV IP
parameters it is estimated that there is a 2\% probability per bunch
crossing of producing a hadronic event with ${p_T}_{\rm min} >$ 
2.2~GeV \cite{tesla-ir}.  The average number of charged tracks is 17 per
hadronic $\gamma\gamma$ event, with 100 GeV deposited in the
calorimeter.  This study needs to be repeated for the NLC IP parameters
and detector acceptance.  Nonetheless, we can estimate the severity of
this background by scaling the rate from the TESLA study 
by the square of $n_\gamma$, the 
average number of photons produced by beamstrahlung, giving 
a factor ($(1.2/1.6)^2$),
and also  taking the bunch structure (190/1) into
account.  This leads to an event probability of 2.2 events/train with
220 GeV in the calorimeter at $\sqrt{s} = 500$ GeV.  It would clearly be
advantageous to be able to time-stamp the hit calorimeter cells and tracks
with the bunch number that produced them and thereby limit the
background affecting a physics event of interest.

{\bf Muons and synchrotron radiation.} SR photons arise from the beam halo
in the final doublet, as shown in Fig. \ref{fig:fans}.  The limiting
apertures of the IR layout determine the maximum angular divergence of
the charged particles that can be tolerated.  Particles above the
maximum divergence must be removed by the accelerator's collimation
system.  If the VXD radius is too small, the apertures in the collimation
system required to remove the beam halo will be unreasonably small and
will produce wakefields that will lead to beam spot size increases and
a loss of luminosity.  As particles are scraped off by the collimation
system, muons are produced.  Depending on the level of the halo and the
robustness of the detector against background muons, a magnetic muon
spoiler system may be required.

\begin{figure}[htb]
\begin{center}
\epsfig{file=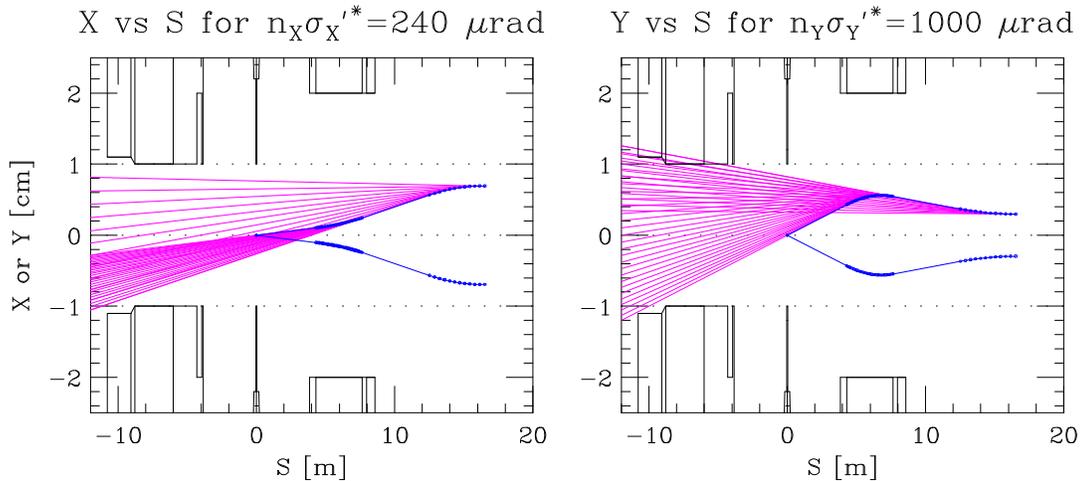,height=2.5in}
\caption{Synchrotron radiation fans from beam halo particles
.}
\label{fig:fans}
\end{center}
\end{figure}

\section{Subsystem considerations}

\subsection{Vertex detector}

Recent experiments have benefited enormously from investments in
excellent vertex detectors.  An important lesson has been the immense
value of a pixelated detector. This technology enabled
 SLD to match many of the
physics measurements at LEP with a much smaller data sample.  The
physics goals of the linear collider will also demand optimal
vertex detection.  The physics signals are rich in secondary vertices,
and event rates are limited, demanding highly pure and efficient
tagging.

Physics processes requiring vertex detection include the Higgs
branching ratios, SUSY Higgs searches such as A$\rightarrow
\tau^+\tau^-$, searches for staus, top studies, improved measurement of
$W$ pairs, $Z'$ studies such as $\tau$ polarization, and $Z$ pole
physics.  Some processes will involve several heavy quark decays,
complicating the reconstruction, and increasing the demand for pixelated
detectors.  The physics will require highly efficient and pure $b$ and
$c$ tagging, including tertiary vertex reconstruction, and charge
tagging (as needed for $b/\overline{b}$ discrimination, for example).
Optimal performance calls for point resolutions better than 4 $\mu$m,
ladder thickness under 0.2\%~$X_0$, inner layers within 2 or 3 cm of the
interaction point, coverage at least over  $|\cos \theta| < 0.9$, and good
central tracking linked to the vertex detector.  The accelerator time
structure and radiation environment will constrain the design, and must
be carefully considered.

A pixel CCD vertex detector was developed at the SLC.  The SLD vertex
detector, VXD3 \cite{sld}, comprised 307 million pixels on 96 detectors,
and achieved 3.8~$\mu$m point resolution throughout this large system.
With such exceptional precision, extremely pure and efficient flavor
tagging at the $Z$-pole was possible:  60\% $b$ tagging efficiency with
$>$98\% purity, and  better than 20\% $c$ tagging efficiency with 60\%
purity.  SLD also achieved exceptional charge separation between $b$ and
$\overline{b}$.  The value of the pixel detector has been clearly
established, even in the relatively clean environment of the SLC, where
the hit occupancy in VXD3 was about $10^{-4}$.  These successes motivate
the choice of CCDs for the next-generation linear collider, where even better
performance is foreseen.

The main weaknesses of the CCD approach to vertex detection are the slow
readout speed and the radiation sensitivity.  The speed issue can be
managed at the linear collider, as SLD demonstrated.  The hit density is
maximal at the inner radius, where one expects about 3 per mm$^2$ per
bunch train at 1.2 cm.  This rate of $\sim 10^{-3}$ per pixel is
challenging, but manageable, especially when the inner layer hits are
matched to tracks reconstructed outside this layer.

With regard to the radiation background, the neutrons create the major
challenge.  Fluences greater than $10^{9}$/cm$^2$/year are expected.  CCDs are
expected to withstand this level of radiation.  However, since the
neutron backgrounds could be larger, CCDs with engineered
rad-hard enhancements are being studied \cite{sinev}.

Despite the established performance of the CCD vertex detector, active
pixels do provide interesting alternatives.  They can be inherently less
sensitive to radiation damage (hence the interest in using them at the
LHC), but generally have been less precise, and they contain
 more material leading to
multiple scattering.  Efforts are underway to close the gap between the
demonstrated CCD performance and the state of the art in active pixels.
These efforts will be followed closely.

Central tracking is vital to the performance of the vertex detector.
With severely limited momentum resolution of its own, the vertex
detector relies on the momentum measurement of the tracker for inward
projection of tracks.

\subsection{Tracking}

Tracking of high-energy isolated charged particles will be important at
a linear $\ee$ collider.  Isolated leptons are prevalent in many new physical
processes, including production of sleptons, heavy leptons, and leptoquarks,
and in many 
interesting Standard Model processes, notably in associated $hZ$
production where the $Z$ decays into charged leptons.  While the
calorimeter may provide a good measure of electron energy (but not
electric charge), excellent tracking will be needed to measure high muon
energies and the charged decay products of $\tau$'s.

Reconstruction of hadron jets will also be important, both in searching
for new physical processes and in understanding Standard Model channels.
Compared to the high-energy leptons discussed above, charged hadrons in
jets have much lower average energies, relaxing the asymptotic
$\sigma(1/p_t)$ requirements.  But tracking these hadrons well requires good
two-track separation in both azimuth ($\phi$) and polar angle ($\theta$).
Aggressive jet energy flow measurement also requires unambiguous
extrapolation of tracks into the electromagnetic calorimeter, again demanding
good two-track separation and also  good absolute
precision.

Forward-angle tracking is expected to be more important at a linear
collider than has been traditionally the case for $\ee$ detectors.  Some
supersymmetry processes have strongly forward-peaked cross sections.
Furthermore, in order to monitor beamstrahlung adequately, it is likely
that precise differential luminosity measurement will be necessary,
including accurate (0.1~mrad) polar angle determination of
low-angle scattered electrons and positrons~\cite{fwdtrack}.

The central tracker cannot be considered in isolation.  Its outer radius
drives the overall detector size and cost.  Given a desired momentum
resolution  the tracker's spatial resolution  and sampling drive the
required magnetic field.  This affects the solenoid design, including the flux
return volume.

For a detector with a compact silicon vertex detector and a large gas
chamber for central tracking, an intermediate tracking layer can improve
momentum resolution, provide timing information for bunch tagging, and
serve as a trigger device for a linear collider with a long spill time.

The most important technical issue for the tracking system
is designing to meet a desired
resolution in $1/p_t$ of order $10^{-5}$ GeV$^{-1}$. This goal is driven
by mass resolution on dileptons in
Higgsstrahlung events and by end-point resolution in leptonic
supersymmetry decays.  There are tradeoffs among intrinsic spatial
resolution, the number of sampling layers, the tracking volume size,
 and the magnetic
field.  The choices affect many other issues.  For example, pattern
recognition is more prone to ambiguities for a small number of sampling
layers, with in-flight decays a particular problem.  Matching to the
vertex detector and achieving good two-track separation is more
difficult for large intrinsic spatial resolution.  A large magnetic
field distorts electron drift trajectories for several tracking
technologies.  High accelerator backgrounds may lead to space charge
buildup in a time projection chamber (TPC), degrading field uniformity
and hence resolution.  More generally, though, 
high backgrounds tend to favor choosing a TPC or another device which
makes
3-dimensional space point measurements (such as a 
silicon drift detector) over a device with 
2-dimensional projective measurements
(such as an axial drift chamber or silicon microstrips).  On the other hand, a
pixel-based vertex detector may provide adequate `seeds' for tracks,
even in the presence of  large backgrounds.

Material in the tracker degrades momentum resolution for soft tracks and
increases tracker occupancy from accelerator backgrounds due to Compton
scattering and conversions.  Because front-end electronics can be a
significant source of material, readout configuration can be quite
important, affecting detector segmentation and stereo-angle options.
Achieving polar angle resolution comparable to the azimuthal angle
resolution may be expensive and technically difficult.

As mentioned above, accelerator backgrounds can degrade track
reconstruction via excessive channel occupancy.  One possible way to
ameliorate the effects of this background is via bunch tagging (or
bunch-group tagging) of individual tracking hits, but such tagging may
place strong demands on the tracker readout technology.

\subsection{Calorimetry}

\subsubsection{Energy flow}

The first question for calorimetry at the linear collider is one that 
not only influences the overall philosophy of this system but also has
ramifications for
other detector subsystems and for the  overall cost:  Should the
calorimeter be optimized to use the `energy flow' technique for jet
reconstruction?  The promise of substantial improvement in resolution
using this technique is appealing.  However, quantitative measures of
this improvement are still being developed, and it is likely that an energy
flow calorimeter will be relatively complicated and expensive because of the
fine segmentation and high channel count.

Clearly, multi-jet final states will be important for LC physics.
Examples from the physics program include separation of $WW$, $ZZ$, and
$Zh$ in hadronic final states, identification of $Zhh$, and $t\bar{t}h$
in hadronic decays,  and full reconstruction of $t\bar{t}$ and $WW$
events in studies of  anomalous couplings and 
strongly-coupled EWSB.  A further example comes at
high energy from the processes $e^+e^-\rightarrow \nu\bar{\nu} WW$
and  $e^+e^-\rightarrow \nu\bar{\nu} t\bar t$, where
because of low statistics and backgrounds, one would need good jet-jet
mass resolution without the benefit of a beam energy constraint.
Indeed, one of the often-stated advantages of the $e^+e^-$ environment
is the possibility to reconstruct many types of final states accurately.
In some instances, this is the key to the physics performance. 

The energy flow (EF) technique makes use of the fact that the modest
momenta of charged hadrons within jets are more precisely determined in
the tracking detectors, than with a calorimeter.  On the other hand,
good energy resolution for photons (from $\pi^0$ decay) is achieved
using any standard technique for electromagnetic calorimetry.
Long-lived neutral hadrons (mostly $K^0_L$) are problematic using any
technique, but they cannot be ignored.  Therefore, a calorimeter
designed to take advantage of EF must efficiently separate neutral from
charged particle energy depositions.  Such designs are characterized by
a large tracking detector (radius $R$), a large central magnetic field
($B$), and an electromagnetic calorimeter highly segmented in 3-D.  A
figure of merit describing the ability to separate charged hadrons from
photons within a jet is $BR^2/R_m$, where $R_m$ is the Moliere radius of
the electromagnetic calorimeter (EMCal).  The EMCal's transverse
segmentation should then be less than $R_m$ in order to localize
the photon showers accurately and distinguish them from charged particles.
Similarly, the separation of the long-lived neutral hadrons from charged
hadrons improves with $BR^2$ and a finely segmented hadron calorimeter
(HCal).  The reconstruction process involves pattern recognition to
perform the neutral-charged separation in the calorimeter, followed by a
substitution of the charged energy with the corresponding measurement
from the tracker.

The advantage of EF is clear in principle.  Whether the advantage is
borne out with realistic simulation is not yet resolved, as the tools
required to do justice to the technique are still under development.
With their silicon/tungsten EMCal, the TESLA group currently
finds \cite{LCWS00-Brient} $40\%/\sqrt{E}$ for jet energy resolution (where
$E$ is the jet energy in GeV).  They expect this to improve to
$30\%/\sqrt{E}$ with progress in pattern recognition.  Assuming that
such good performance is indeed achievable with EF, it is useful to
identify how this would improve the physics outlook, and at what cost.

\subsubsection{Resolution, segmentation, and other requirements}

There is no compelling argument from LC physics that demands
outstanding photon energy resolution, resulting for example from an EMCal
using high-$Z$ crystals.  Furthermore, such an optimization would not be
consistent with the high degree of segmentation required for excellent
jet reconstruction.  Instead, the requirements for calorimetry from LC
physics are jet energy and spatial resolution, and multi-jet invariant
mass resolution.  The required jet energy resolution depends, of course,
on specific physics goals.  A recent study \cite{LCWS00-Gay} indicates
that  a resolution of $40\%/\sqrt{E}$ is necessary to measure the 
Higgs self-coupling using $Zhh$ final states.  One benchmark for jet-jet mass
resolution is the separation of $W$ and $Z$ hadronic decays in $WW$,
$ZZ$, and $Zh$ events.  Both of these requirements may be achievable
using energy flow reconstruction.

Segmentation is a critical parameter, since an EF design
requires efficient separation of charged hadrons and their showers from
energy depositions due to neutrals.  The typical charged-neutral
separation, $\Delta x$, is dervied from the particle density in jets after
they pass through
the tracking detectors.  This depends upon the physics process and
$\sqrt{s}$, as well as the tracker radius and the detector magnetic field.
Studies show that the minimum $\Delta x$ is typically 1--4 cm in the EMCal
and about 5--10 cm in the HCal.  The EMCal should be
very dense, with Moliere radius of a few cm or less, and should have transverse
segmentation that is smaller still, in order to  localize the
photon showers accurately.  Fine longitudinal segmentation, with each layer read
out, is also essential in order to track the charged particles through
the EMCal and to allow charged-neutral separation in 3-D.  This will also
benefit the energy resolution for photons and electrons.  There is no
reason to organize the layers in towers, and, in fact, this probably
should be avoided.  The fine transverse segmentation provides excellent
electron identification and photon direction reconstruction.  The latter
is also  useful for measuring photons which result from a
secondary vertex.  This is relevant, 
for example, in  gauge-mediated SUSY, which can lead to 
secondary vertices with a photon as the only visible decay particle.

For EF in the HCal, it is desirable to track MIPs throughout.  One would
need to identify shower positions with a resolution of a few cm.
Because of the relatively diffuse distribution of deposited energy for
hadron-initiated showers, the solution for charged/neutral identification
is not as obvious as for the EMCal case, and different ideas are under
consideration.  In any scheme, one requires a high degree of
segmentation.  This might be implemented, for example, using
scintillator tiles roughly 5--10 cm on a side.  Another idea is to push
to finer segmentation, using, for example 
resistive plate chambers (RPCs), but without providing pulse
height in the readout.  Such a `digital' hadron calorimeter is one of
the options being considered for TESLA.  This provides increased
resolution for pattern recognition, but perhaps with poorer neutral
hadron energy resolution.

As with this segmentation issue, many of the other properties of the HCal
in an EF calorimeter remain uncertain.  One example is the necessary
total calorimeter depth in interaction lengths.  Another is the
placement of the solenoid coil.  Since the fields are typically large,
and the coils are at large radius, their thickness is not negligible.
Qualitatively, for good performance one would prefer to have the coil
outside the HCal.  But the tradeoffs are not yet well understood
quantitatively.

The EF jet resolution is dominated by the tracker momentum resolution,
the calorimeter pattern recognition efficiency, and by the 
purity of charged/neutral identification.
Hence, single-particle resolutions are less important.  However, the
current EF designs yield energy resolution $A/\sqrt{E}$ in the range
$A=12$--20\% for photons, and in the range $A=40$--50\% for single
hadrons.

For a detector not designed to use energy flow, there are, of course,
many traditional choices available.  Assuming that jets are to be
reconstructed using the calorimeter only, one might choose a
compensating, sampling calorimeter with a tower geometry.  One or more
layers of detector with finer segmentation may be required at the front
of the EMCal, or at shower maximum,
 to aid with electron and photon identification.  Such a
calorimeter would certainly be cheaper than an EF device at a
similar radius.  At low $\sqrt{s}$, especially at the $Z$, this may
suffice.

One also needs to consider Bhabha scattering in the calorimeter design.
First, the final state $e^\pm$ at $\sqrt{s}/2$ determines the upper end
of the dynamic range of the EMCal readout.  For example, for a dense EMCal,
the ratio of deposited energy for Bhabha electrons to MIPs can be 
$10^3$ to $10^4$, depending on segmentation.  Secondly, the Bhabhas are
used for luminosity measurements of two types.  First, the Bhabha rate
can be used to measure the absolute luminosity.  Since this rate at
intermediate to large angles (endcap and barrel) will be large compared
to (known) physics processes, it would not be necessary to rely on a
small-angle luminosity monitor (LUM), although a LUM would still be
useful for crosschecks and operations.  Running at the $Z$ is an exceptional case
 where a precise LUM would be required.  The Bhabhas
also provide probably the best measurement of the luminosity spectrum,
$d{\cal L}/dE$, because the Bhabha acolinearity is closely related to
the beam energy loss.  This is ideally measured at intermediate angles,
and the EMCal endcap will need to be able to aid the tracker with this
measurement.

In addition to Bhabha scattering, two other types of measurement have
been discussed for the small-angle region.  One is a measurement of the
flux of pairs produced in the collision beam-beam interactions.  This
would provide immediate feedback to operators of a quantity closely 
related to the instantaneous luminosity.
The other is small-angle tagging of the
forward-scattered electron or positron resulting from a two-photon
interaction.  This is useful both in the study of the two-photon 
process itself and in reducing background in the study of processes
such as slepton pair production which resemble two-photon reactions.
Such a device
would need to tag a single high-energy electron within the angular
region flooded by low-energy pairs from the beam-beam interaction.

Finally, the small-angle elements of any calorimeter design must reflect
the requirement to limit the detector contribution to the missing
transverse momentum resolution.  This contribution is roughly
$E_b\theta_{\rm min}$, where $E_b$ is the beam energy.  Given the
limited angular coverage of the central tracking systems, one should
consider carefully what type of calorimetry should be used near
$\theta_{\rm min}$.

\subsubsection{Technology options}

For the dense, finely segmented electromagnetic calorimeter required for energy flow,
layers composed of a tungsten radiator with silicon detectors (Si/W) are
a natural choice.  The Moliere radius of tungsten is small (9 mm), and
the silicon is thin and easily segmented transversely.  Si/W EMCal's
are currently incorporated in two LC detector designs,  the TESLA
detector  and
the NLC SD detector described in Section~\ref{sec:sddetector}.  
This option has one
outstanding drawback, the cost of the silicon detectors.  Both TESLA and
SD assume that a cost of 
 roughly \$3/cm$^2$ can be achieved in the future with a very
large order.  This is about a factor two cheaper than current costs.
There are a number of cost and performance optimization possibilities.
For example, one would probably not need to sample the EMCal uniformly in
depth, reducing the sampling frequency after about 12~$X_0$.  One could
also improve the photon energy resolution by sampling with thicker
silicon, at some small loss of Moliere radius.

Perhaps it is possible to design a competitive energy flow electromagnetic
calorimeter at lower cost using an alternative to silicon, for example,
scintillator tiles.  The transverse segmentation
is limited using present techniques by the inability to couple
sufficient light to a readout fiber.  Perhaps this can be improved.
However, given the larger cells, sufficiently large $B$ and $R$ may
compensate for the segmentation disadvantage.  This is the rationale for
the NLC L design described in  Section~\ref{sec:ldetector}.  Another
alternative being considered for TESLA is a Shashlik EMCal.  Beam test
results \cite{LCWS00-Checchia}, using fibers of two lifetimes in order to
achieve some longitudinal segmentation, have been impressive, but it
is unclear whether the segmentation is sufficient for EF.

The hadron calorimeter for an EF detector is not as highly constrained
as the EMCal.  Here, scintillator tiles can be of size similar to present
applications, say 8--10 cm on a side, with coupling to an optical fiber.
Such a scheme is under consideration for the TESLA and NLC L and P
 designs.  (The last of these is described in Section~\ref{sec:pdetector}.)
Other possibilities include the
`digital' option mentioned above, which might use, for example,
double-gap RPC readout layers or extruded scintillator.  The spatial
resolution per layer might be about 1~cm.

If it were possible to relax the need for precise jet reconstruction,
then one might forego EF, and save some money with a more
traditional calorimeter.  For example, the NLC P design uses modestly
segmented towers built up from Pb/scintillator layers.  This might also
be implemented using liquid argon.

\subsection{Muon detection}

The main purpose of the LC muon system is to identify muons and provide
a software muon trigger.  A secondary purpose is to use the muon
detector as backup calorimetry for those particles that penetrate beyond
the normal hadron calorimeter.  The signature for muons is their
penetration through the calorimetry and the 
instrumented iron flux-return for the solenoid field.

The momentum of muons is determined from the central and forward
tracking systems.  This requires the association of tracks found in the
instrumented flux-return with hits/tracks in the central and forward
tracking detectors.  Two conditions permit this:  a reasonable density
of hits in the inner layers of the tracking detectors and limited
confusion from multiple scattering due to the electromagnetic and
hadronic calorimeters between the inner tracking detectors and 
the front face of the muon detectors.  These conditions are satisfied,
since the maximum density of tracks, at a radius of 3 m, is about
1/cm$^{2}$~\cite{piccolo} and the $r$--$\phi$ rms multiple scattering of a
10~GeV/c muon is approximately 2 cm.  The number of radiation lengths
$X_0$ of material in front of the muon system for the three candidate
detectors L, SD and P are $200, 88$ and $125$, respectively.

Muons are identified by their ionization in tracking chamber
panels~\cite{piccolo} or scintillator strips~\cite{para} in 2 cm gaps
between $5$ or $10$ cm thick Fe plates that make up the barrel and end
sections of the Fe return yoke for the central solenoidal magnetic
field.  RPCs are taken as the example
technology.  These planar devices can be built with appropriate
perimeter shapes, and they do not contain wires that could break.
Tracking hits from the avalanche produced in the RPC gaps are read out
with strip electrodes that run in the $\phi$ and $z$ directions.  The
spatial resolution of these strips is 1 cm per detector plane.

For the case of the L detector, it can be seen in Fig.~\ref{fig:muhits} 
that the number of hits as a function of momentum for $W$
pair production, plateaus at about 5 GeV with 25 instrumented gaps.  The
plot shows that in the 3 T field
there will be very good efficiency if 15 or more hits are required in
the muon tracking algorithm.

\begin{figure}[ht]
  \begin{center}
  \epsfig{file=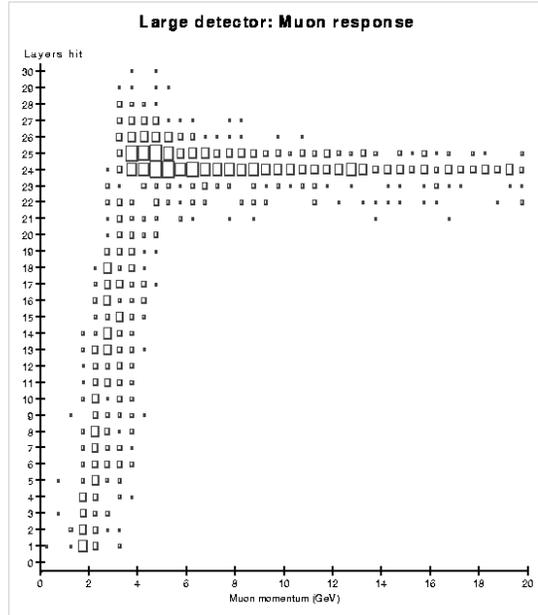, height=4.0 in}
  \caption{Hits in the  muon system as a function of  momentum for the 
L detector.  The pllot shows 10000 $\ee\to W^+W^-$ events in which
    one $W$ decays to a muon. \label{fig:muhits}}
   \end{center}
\end{figure}

The Fe plate and strip readout for the muon system can be used as
additional coarse hadron calorimetry, since the number of interaction
lengths $\lambda$ for the L, SD and P options are, respectively, 6.6, 6.1
and 3.9.  The muon Fe adds 7, 6, and 6~$\lambda$ that can be
used in the determination of residual hadronic energy with a resolution
that will be about $1/\sqrt{E}$.

\subsection{Solenoid}

The detector is assumed to be a classical solenoidal design.  The field
in the tracking region ranges from 3 to 5 T for the various designs.
The solenoid is assumed to be of the CMS type, based on a relatively
thick, multi-layer superconducting coil.  The radial thickness of the
complete assembly is about 85 cm.  The CMS vacuum shell has a total
thickness of 12 cm, and a cold mass thickness of 31 cm (aluminum).  It
is likely that the cold mass thickness will scale roughly as
$B^2R$. Then, the coil of the SD detector would be about 35\% thicker.

The iron serves as the flux return, the absorber for the muon tracker,
and the support structure for the detector.  The (perhaps debatable)
requirement of returning most of the flux drives the scale of the
detector.  At this stage of preliminary design, it is assumed that the
steel is in laminations of 5 cm with 1.5 cm gaps.

The door structure  very likely runs along the beamline past L$^*$, the
position of the downstream face of the last machine quadrupole.  Thus it
is essentially certain that the Final Doublet (FD) is inside the
detector, and quite possibly within the Hadronic Calorimeter.  For this
reason,  the
FD cannot be mounted on a massive column going directly to bedrock.

\subsection{Particle ID}

The physics topics of the linear collider do not demand hadron ID in a
direct way, though the information may prove valuable for some analyses.
Pions, kaons and protons are produced in the ratio  of about
8:1:0.6 in high-energy $\ee$ colliders.  The momentum spectrum of kaons
in $q\bar q$ events at $\sqrt{s} = $500 GeV extends up to
150-200 GeV/c, posing a possibly unsurmountable ID measurement challenge.
However, the average kaon momentum is only 10--17 GeV/c, and more than
half of all kaons have momenta below 7 GeV/c.  In $t$-quark and
multi-$b$ jet Higgs events, the multiplicity is higher, and so kaons have a
slightly lower mean momentum.

The measurement of particle species distributions provides information
on QCD processes and permits model tests, but the most important use of
hadron ID may be to assist the application of other techniques, such as
$B$ tagging.  As an example, two studies \cite{wilsonSitges,lcws-fnal00}
have discussed the use of net kaon charge to tag the flavor of neutral $B$
mesons produced in \( q\overline{q} \) events.  They find that with
perfect knowledge of decay product identities in vertex-tagged neutral $B$
mesons, roughly a quarter are correctly tagged by the net charge of
kaons. The efficiency is much lower if all undiscriminated hadrons are
used.  It is a detailed, and so far unanswered, question whether the
use of hadron ID with realistic detector efficiencies can be an
important tool to unscramble complex events that contain multiple $b$- or
$c$-quark jets.

The geometric and, ultimately, the cost constraints 
limit the choice of technology for a hadron ID system
of a linear collider detector.
Ideally, it should take up no space and introduce no additional mass in
front of the calorimeter.  Traditional ionization measurement ($dE/dx$) in
gas-based tracking chambers comes close to meeting these criteria.

The Time Projection Chamber (TPC) technology that appears in the
TESLA  and L tracker designs may be an optimal choice
for combined tracking and ionization measurements for particle ID.  The
energy resolution that has been achieved with existing non-pressurized
TPCs ({\it e.g.},  ALEPH at LEP) is 4.5\%, which would yield $\pi$/K separation
of better than 2$\sigma$ for $p<0.8$ GeV/c and 2-3 $\sigma$ for
$1.7<p<65$ GeV/c.  One can improve the capability of a TPC by using
pressurized gas to achieve 2.5\% resolution, as demonstrated by the TPC
at PEP.  According to a recent model \cite{yamamotoSitges}, this could
provide 4$\sigma$ $\pi$/$K$ separation in the range $1.75<p<30$ GeV/c.

In practice, experiments that desire a high degree of species separation
have supplemented ionization measurements with specialized devices such
as time-of-flight, threshold Cerenkov or ring-imaging Cerenkov devices.
The major drawback of a specialized
hadron ID subsystem is its collateral impact on
the tracking and calorimetry.  All supplementary techniques take up
radial space between the tracker and calorimeter, which means either
shorter tracking volume or increased calorimeter radius with consequent
cost and performance implications.  Without a clearly defined need for
the capability, it is difficult to justify a significant impact on the
rest of the detector.  For example, in the $B^0$ tagging study, even
though the best performance was provided by an SLD-style CRID or a 
high-pressure TPC, relatively inexpensive improvements to an ALEPH-type TPC
could achieve a sensitivity within a factor of two of these more
complicated options but with little impact on the calorimetry.

In summary, at this stage there is no compelling argument to include a
specialized hadron ID system in the high energy detector design, though
in the process of optimizing the design this assumption may be
reexamined.

\subsection{Electronics and data acquisition}

The NLC beam consists of 190 bunches spaced 1.4 ns apart, in trains that
repeat at 120 Hz.  There are variations with a doubled
bunch spacing and an increased train frequency of 180 Hz, but these
variations do not affect the basic theme.  For most of the detector
subsystems it will neither be possible, nor particularly desirable, to
resolve bunches in a train. The train repetition rate of 
 120 Hz is a low frequency compared with
Level 1 or Level 2 trigger rates at many other machines. There
is no need for a hardware trigger, and (zero-suppressed, 
calibration-corrected) data can flow from the detector at this rate. A
traditional Level 3 Trigger (software on a small set of processors) can
select events for storage.

The time horizon for a detector is roughly 8 to 10 years away,
which is at least 5 Moore's Law generations.  To be sure, Moore's Law
refers to computing power per dollar, but there are clearly related trends
in most areas of silicon technology.  At this time it seems most
appropriate to sketch plausible architectures to help generate cost
estimates, and to avoid detailed designs.

Perhaps the clearest distinction that should be made is the role of
interconnections that are not on silicon.  Rather
inexpensive systems have been developed for large CCD detectors.
The costs strongly reflect the number of output nodes that must be
serviced, and correspond only weakly to the number of pixels being
transmitted through that node.  In addition, 
because of the train spacing, there is no penalty to serial multiplexing
of the data from very large numbers of pixels.  This is in contrast to
the LHC, where there are many interactions associated with each beam
crossing, which occurs every 25 ns.  This is not to say there are no
limits to the serial multiplexing. The readout of the SLD Vertex
Detector crossed about 8 beam crossings at SLC, and it would be
desirable to avoid this at the next-generation linear collider.

Consequently, we have developed the concept of clusters rather than
channels.  A cluster is a set of detector elements that can conveniently
be processed and serialized into a single data stream, presumably an
optical fiber.  In the CCD example, each node might correspond to a
cluster, although it might even be possible to handle multiple nodes in
a single cluster.  For the CCD case, we think of an ASIC located
millimeters from the CCD and bonded to the CCD.
This ASIC might handle the clock generation and the gate drives as well
as the amplification and digitization of the CCD data.  For silicon
strip detectors, we foresee a single chip servicing a cluster of strips,
presumably a complete detector a few cm wide.  For a calorimeter
utilizing scintillator and Hybrid Photo Diodes or Multi-Anode
Phototubes, a cluster would correspond to all the outputs from each such
device.  In all cases, we avoid, as much as possible, all low-level
cables and interconnects.  The cluster reflects the mechanical nature of
the detector.  Some cases are less obvious.  For a tungsten-silicon
calorimeter, a cluster might correspond to a large area board carrying
many close packed wafers of silicon diodes. It may cover 
perhaps a square meter or
so.  Variations on this concept would cover  readout sectors of the 
TPC and the muon tracking detectors.

Thus the detector proper carries all the front end processing, and a
relatively modest set of fibers carries data off the detector.  We
envision the fibers delivering the data to processors, perhaps based on
VME, although there are hints that crate systems based on optical serial
backplanes may arrive in time.
  These processor arrays would complete the signal
processing, build the events, and pass those events to the system responsible
for the Level 3 decision.

\section{Detectors}

Three detector models are now being studied as potential
detectors for the NLC.  These include two options for the high-energy
IR, called L and SD, and one for the lower-energy, second IR, called P.
Here we describe each of these detectors, and present some of their
performance curves.

\subsection{L detector for the high-energy IR}
\label{sec:ldetector}

The L detector design is driven by the desire to provide a large
tracking volume, to optimize tracking precision.  This leads to a 
large-radius calorimeter and limits the magnetic field strength to about 3~Tesla.

The L detector is illustrated in Fig.~\ref{fig:ldet}.  Table~\ref{tab:dimen} 
presents the dimensions of the L detector, along with
those for the SD and P detectors, described below.

\begin{figure}[htbp]
\begin{center}
\epsfig{file=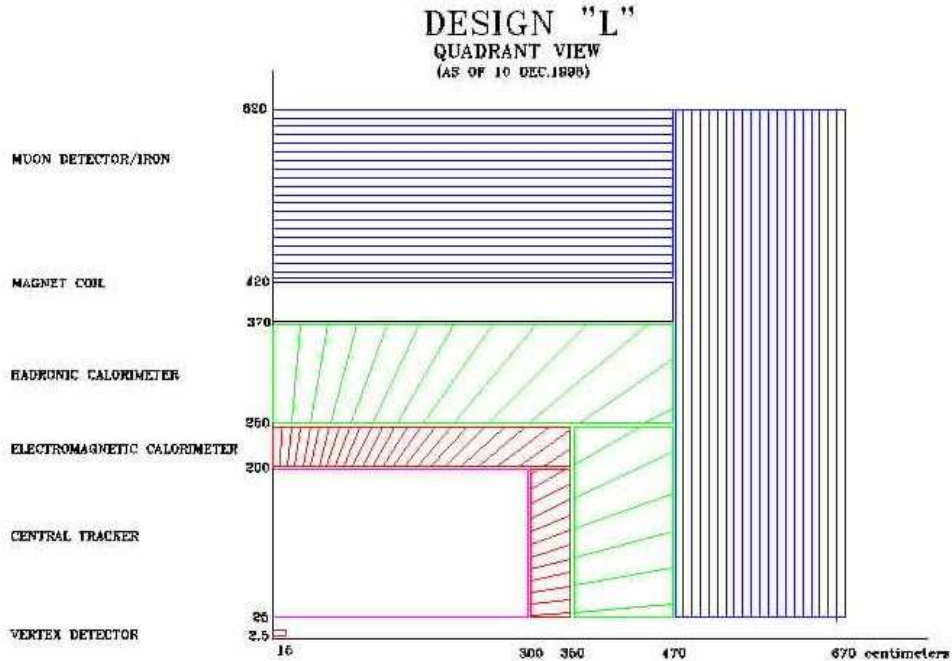,height=3.5in}
\caption{Quadrant view of the L detector.}
\label{fig:ldet}
\end{center}
\end{figure}

\begin{table}[htbp]
\footnotesize
\begin{center}
\begin{tabular}{|l|cc|cc||cc|cc||cc|cc|}
\hline
\hline &\multicolumn{4}{c||}{L Detector}& \multicolumn{4}{c||}{SD Detector}
& \multicolumn{4}{c|}{P Detector}\\
        Component & \multicolumn{2}{c|}{R(cm)} & \multicolumn{2}{c||}{Z(cm)}
& \multicolumn{2}{c|}{R(cm)} & \multicolumn{2}{c||}{Z(cm)}
& \multicolumn{2}{c|}{R(cm)} & \multicolumn{2}{c|}{Z(cm)}\\                                                                                                                                                                                               &Min &  Max  &Min &   Max &Min &  Max  &Min &   Max
&Min &  Max  &Min &   Max\\ \hline

Vertex Det. & 1.0  & 10 &  0 &  15
& 1.0  & 10 &  0 &  15
& 1.0  & 10 &  0 &  15
\\

C.Track.  &25 &200 & 0 & 300
&20& 125 & 0& 125
&25 & 150 & 0 & 200
\\
           ECal &&&&&&&&&&&& \\

\ \ Barrel& 200 & 250   &  0 & 350
& 127 & 142 & 0 & 187
& 150 & 185 & 0 & 235
\\

\ \ EndCap &  25 & 200 & 300  & 350
& 20 & 125 & 172 & 187
 & 25 & 150 & 205 & 240
\\

 HCal &&&& &&&&&&&&\\

\ \ Barrel & 250 & 370 & 0  & 470
& 143 & 245 & 0 & 289
 & 215 & 295 & 0 & 320
\\

\ \ EndCap  &  25   &  250  & 350  &  470
 & 20 & 125 & 172 & 187
 & 25 & 175 & 240 & 320
\\
       Magnet  & 370 &   420  &  0   &   470
& 248 & 308 & 0 & 289
 & 185 & 215 & 0 & 235
\\

Iron/Muon &&&& &&&&&&&& \\

\ \ Barrel & 420 & 620 & 0 & 470
& 311 & 604 & 0 & 290
 & 295 & 425 & 0 & 320
\\

\ \ EndCap &  25  &   620 &   470   &  670
& 20 &  604 & 290 & 583
 & 25 & 425 & 320 &  450
\\ \hline
\hline
\end{tabular}
\caption{Dimensions of the L, SD, and P Linear Collider Detectors.}
\label{tab:dimen}
\end{center}
\end{table}

The vertex detector is a five-barrel CCD vertex detector, based on the
technology developed for SLD.  The beam pipe radius of 1 cm allows the
inner barrel of the detector to reside 1.2 cm from the IP.  The inner
barrel extends over 5 cm longitudinally.  The other barrels have radii of 2.4
cm, 3.6 cm, 4.8 cm, and 6.0 cm, and they each extend 25 cm longitudinally.  The
barrel thicknesses are 0.12\% $X_0$ and the precision is assumed to be 5
$\mu$m.  (This is taken as a conservative assumption, since SLD has
achieved 3.8 $\mu$m.)  The entire system comprises 670,000,000 pixels of
$20\times 20 \times 20\  \mu{\rm m}^3$.

Figure~\ref{fig:vxd} illustrates this system.  The detector operates in
an ambient temperature of 190$^\circ$K, created by boil-off nitrogen.  It is
enclosed within a low mass  foam cryostat.  The same five-barrel CCD design
has been assumed for the SD and P detectors below.

\begin{figure}[htbp]
\begin{center}
\epsfig{file=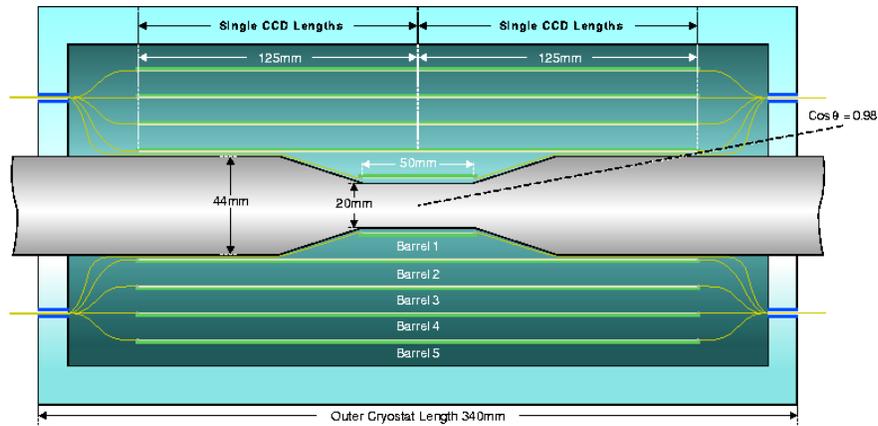,height=3.5in}
\caption{The five-barrel CCD vertex detector proposed for the linear collider.
\label{fig:vxd}}
\end{center}
\end{figure}

The performance of the vertex detector is illustrated in Figs.~\ref{fig:drvtx}
 and \ref{fig:vtagging}, where it is also compared to
the SD and P Detector performance.  The impact parameter resolution
shown in Fig.~\ref{fig:drvtx} is shown to surpass the
performance of SLD's VXD3.  The bottom and charm tagging performance,
shown in Fig.~\ref{fig:vtagging}, is also seen to be exceptional.

\begin{figure}[htbp]
\begin{center}
\epsfig{file=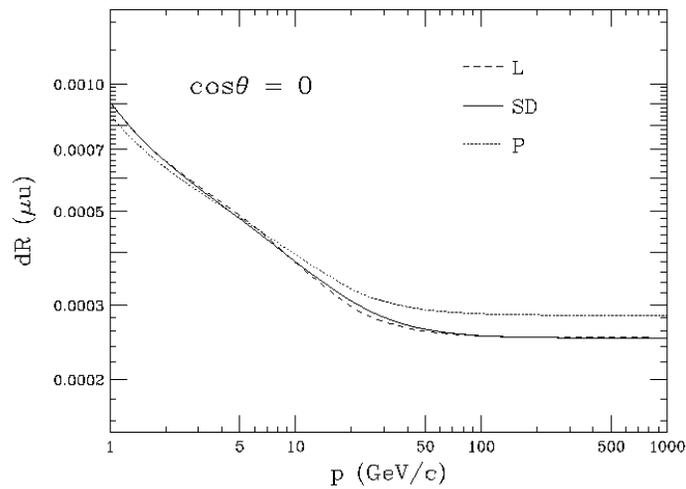,height=2.5in}
\caption{Impact parameter resolution versus momentum for the vertex 
      detector shown in Fig.~\ref{fig:vxd}.
\label{fig:drvtx}}
\end{center}
\end{figure}

\begin{figure}[tbhp]
%\begin{center}
\makebox[3.1in][l]{
\epsfxsize 2.5in
\epsfbox{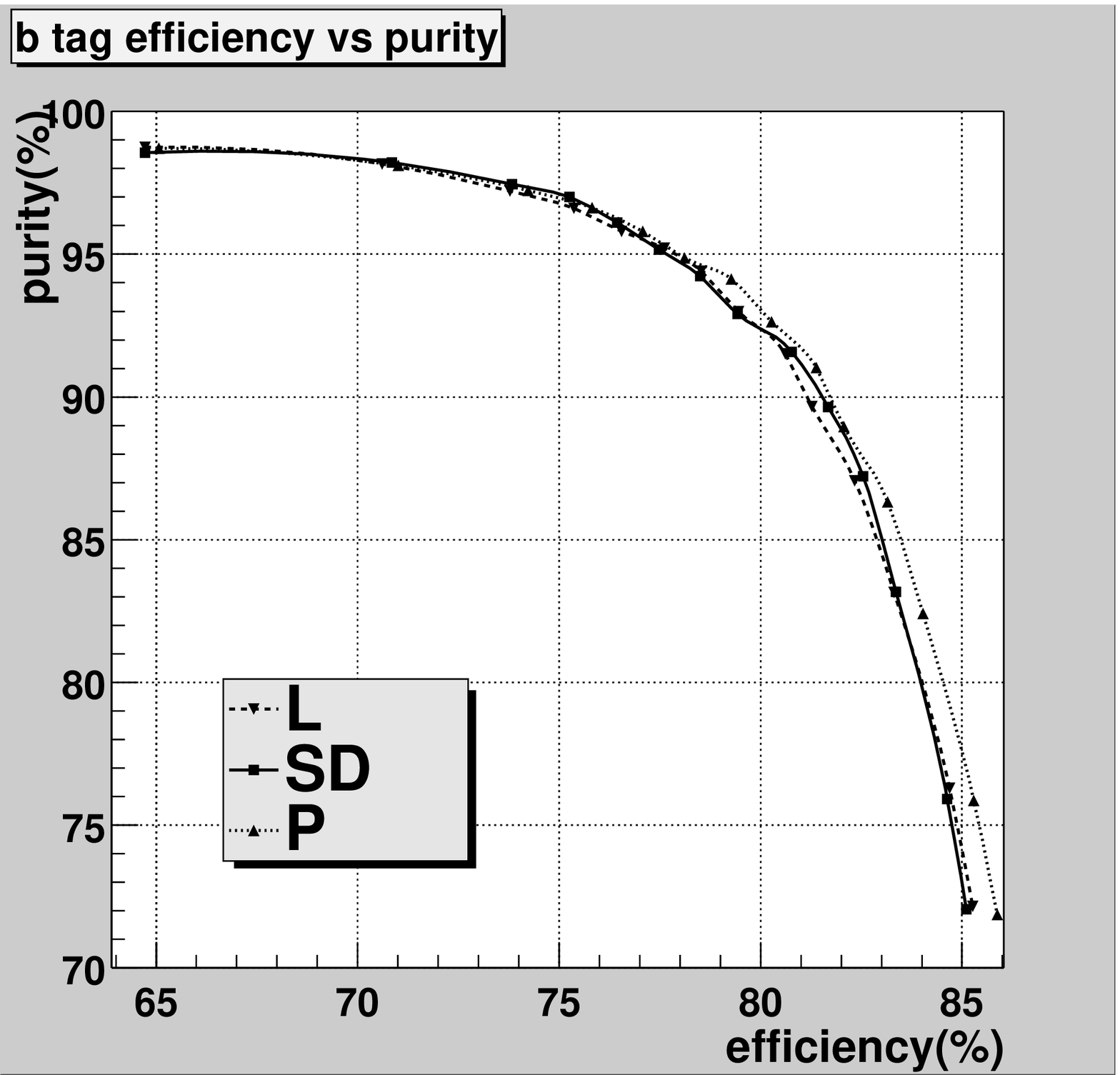} }
\makebox[3.1in][r]{
\epsfxsize 2.5in
\epsfbox{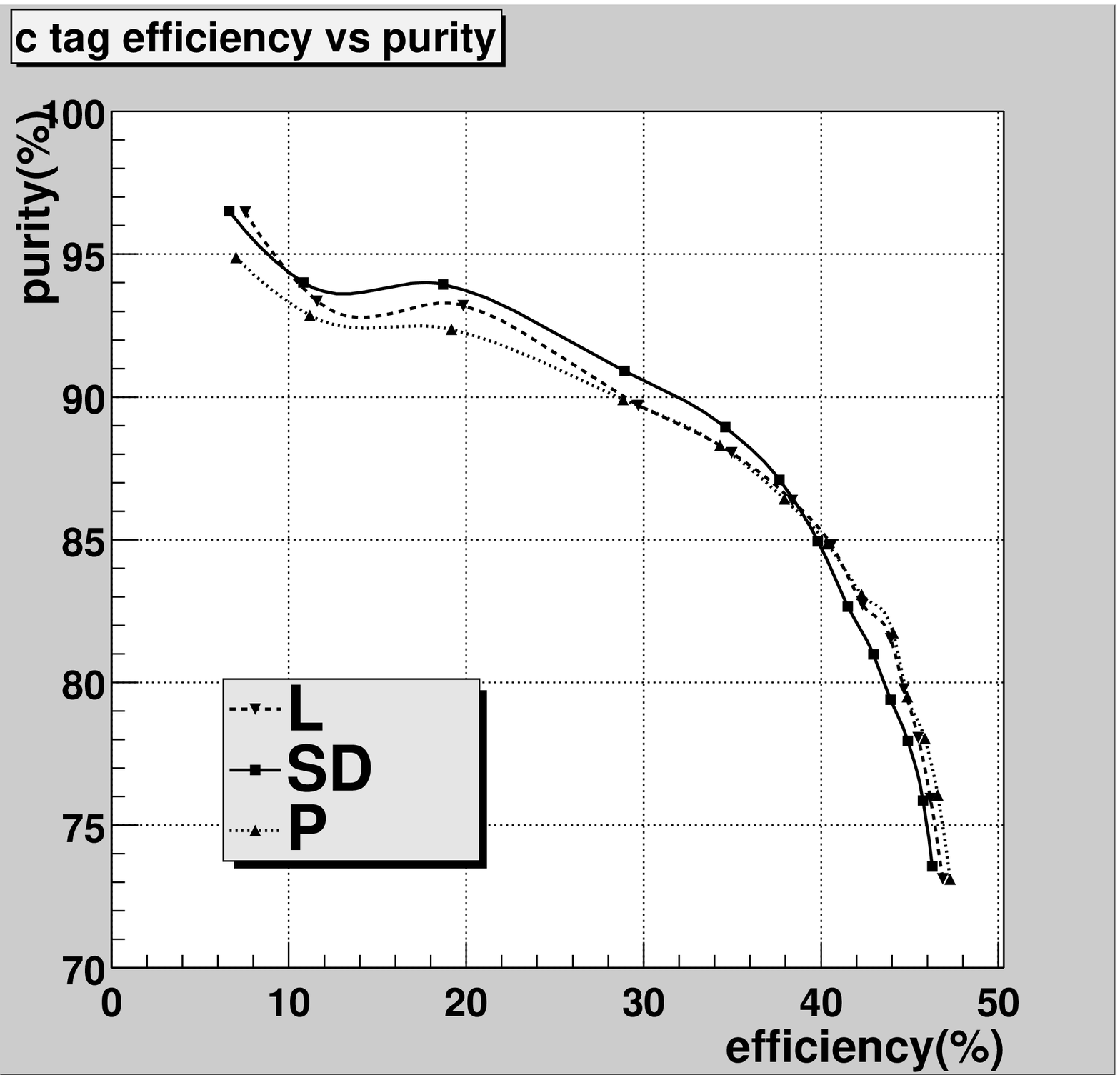} }
%\end{center}
\caption{
Vertex-tagging purity versus efficiency for $b$ (left) and $c$ (right),
evaluated for decays of the $Z^0$ at $E_{\rm CM} = 91.26$ GeV.
\label{fig:vtagging}}
\end{figure}

The L detector central and forward trackers consist of a large-volume
TPC, an intermediate silicon tracking layer (silicon drift detector or
double-sided silicon microstrips), and five layers of double-sided,
silicon microstrip disks in the forward regions.  An additional
scintillating-fiber intermediate tracker option has also been proposed
to provide precise bunch timing.  Figure~\ref{fig:ltracker} shows a
sketch of the L detector tracking system.

\begin{figure}[htb]
\begin{center}
\epsfig{file=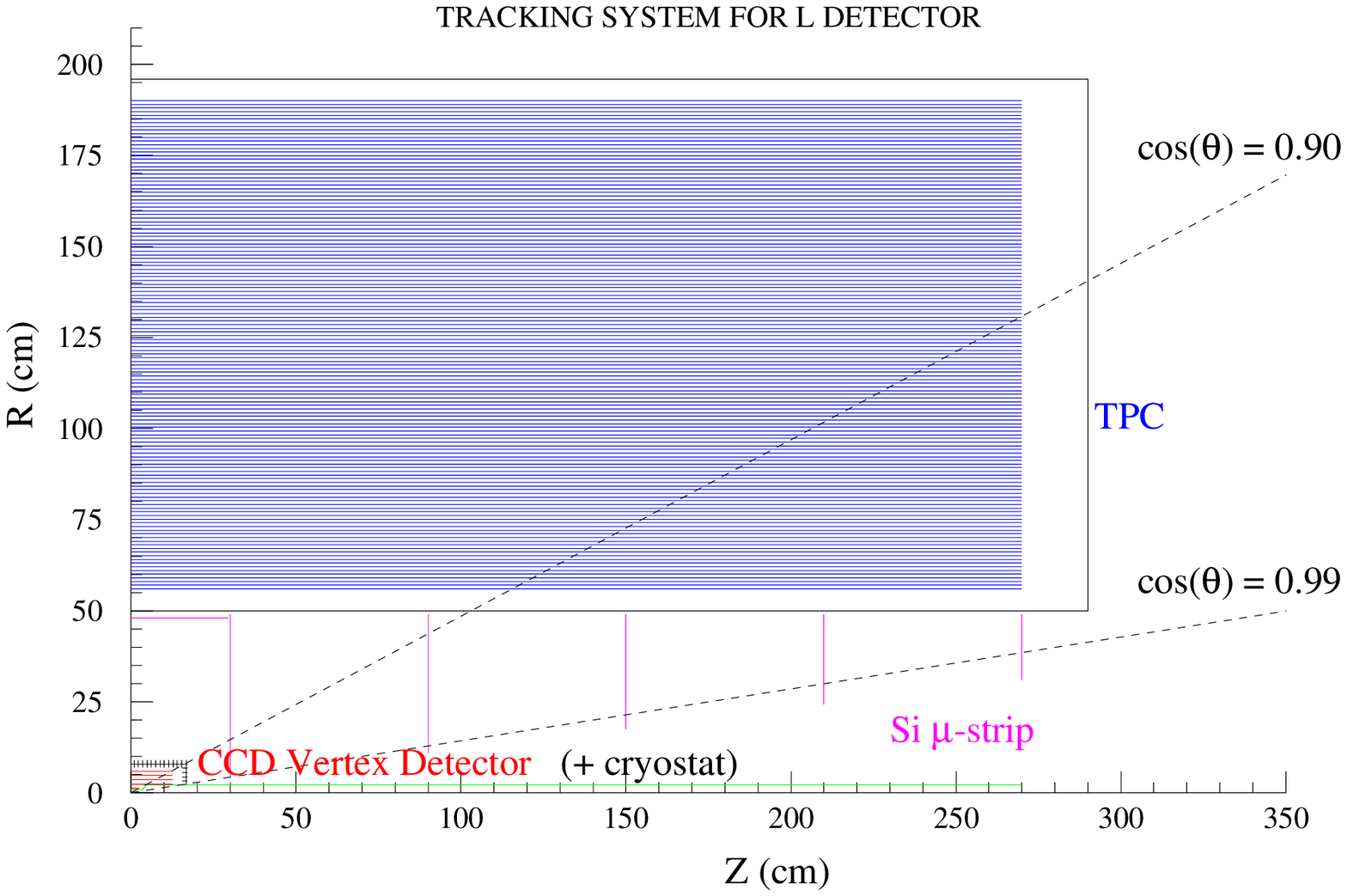,height=3.0in}
\caption{Sketch of L detector tracking system.}
\label{fig:ltracker}
\end{center}
\end{figure}

A large-volume TPC with three-dimensional space point measurements along
charged particle trajectories provides excellent pattern recognition
(including recognition of 
long-lived particles that decay in the tracking volume)
and good particle identification via $dE/dx$ measurements.  The baseline
L detector TPC~\cite{ronan} has 144 tracking layers enclosed in a
cylindrical volume of inner and outer radii = 50 and 200 cm, respectively,
 and of half-length
290 cm.  The assumed resolutions on each hit are 150 $\mu$m in
$r$--$\phi$ and 0.5 mm in $r$--$z$.  A GEM-based readout has the
potential to reduce the $r$--$\phi$ resolution to 100 $\mu$m.  The
small transverse diffusion for TPC operation in the 3~T magnetic field
 requires very narrow cathode pads and large total channel
counts.  Longer pads or the use of induced signal on adjacent pads may
be considered to reduce the channel count.  Good track timing
resolution is obtained by requiring individual charged tracks to point
back to a reconstructed vertex in the $r$--$z$ plane.  This timing
resolution helps in reducing accelerator backgrounds.

The TPC in the STAR detector at RHIC has over 138,000 electronics
channels and includes several design innovations.  To reduce the
required cable plant, low-noise low-power front end electronics
are mounted on the TPC end planes.  The analog signals are amplified,
sampled and digitized before being sent to the DAQ system over fiber
optics.  A similar scheme is assumed here, with 20-MHz sampling, a
200-ns peaking time and 9-bit digitization.

The TPC analog front end electronics would consist of a
high-bandwidth preamplifier and shaper amplifier (8-16 channels/ASIC
chip), providing a 200~nsec peaking time pulse to the analog sampling
and digitization section.  The analog signals from the preamplifier and
shaper amplifier would be sampled and stored with a high-frequency
20-MHz clock as they come in, and then digitized on a longer (10
$\mu$sec) time scale as new samples are being taken.  The recognition of
charge cluster signals on a central cathode-pad channel triggers a
switched capacitor array (SCA) to sample the channel and its nearest
neighbors.

Gas mixtures of argon with methane and carbon dioxide are being
considered, with Ar(90\%):CH$_4$(5\%):CO$_2$(5\%) being quite
attractive in balancing safety concerns, neutron-background quenching,
and drift velocity.  Positive ions feeding back from endplane gas
amplification can be mitigated by the installation of a gating grid.

A silicon intermediate tracking detector just inside the TPC inner radius
provides nearly a factor of two improvement in momentum resolution for
high-$p_t$ tracks and offers a pattern recognition bridge between the
TPC and the vertex detector.  Two silicon options are under
consideration:  a silicon drift detector and a double-sided silicon
microstrip layer.  In each case the layer would have a half-length of
29.5 cm and an average radius of 48 cm.  The estimated space-point
resolutions in $r$--$\phi$ and $r$--$z$ are 7 $\mu$m and 10 $\mu$m,
respectively, for the silicon drift detector option, with both at
7$\mu$m for the double-sided microstrip option.

An additional or alternative intermediate tracker constructed from
scintillating fibers offers high-precision timing to allow the matching of
tracks to individual beam bunches.  The current NLC accelerator design
provides beams composed of trains of bunches with bunch
spacings of  1.4 ns.  Large rates of two-photon
interactions are expected both from interactions of virtual photons and from
 real photons created by  beamstrahlung.  The
overlap  of the two-photon events with $\ee$ annihilation events results
in additional `mini-jets', which can be a problem if tracks created in
different bunch crossings are not separated.
A scintillating-fiber intermediate tracker, coupled
by clear fiber to visible light photon counters and 
 read out by the SVXIIe chip~\cite{svxii} can achieve time
resolutions on the order of 1~ns to associate tracks with individual
bunches, as well as to complement time measurements in the TPC. 
Appropriate Si:As devices
manufactured by Boeing~\cite{VLPC} have a fast response time of less
than 100~ps.  One
possible system consists of two axial layers and two $3^{\circ}$-stereo
layers with a half-length of 29.5~cm at an average radius of 48~cm,
supported by a carbon fiber cylinder.  Scintillating fibers of diameter
800~$\mu$m would provide individual measurements to 230~$\mu$m and a
combined point measurement with a precision of $\sim 100$~$\mu$m,
resulting in a system with 15,000 channels.

As currently envisioned, the five layers of the L detector forward disk
system are double-sided silicon microstrips, at distances of 30 cm to
270 cm from the interaction point, with fixed outer radii at 48 cm.  Each
side provides counterposing $\pm$ 20 mrad $r$--$\phi$ stereo
information, with a point resolution of 7 $\mu$m.  For high-momentum
tracks at $\theta = 300$ mrad ($|\cos\theta| = 0.955$), this small-angle
stereo geometry provides a resolution in $\theta$ of about $\pm 300$
$\mu$rad.  If large-angle ($90^{\circ}$) stereo were used instead, the
$\theta$ resolution would improve to about $\pm 100$ $\mu$rad.  Although
the layout of silicon strip detectors is more naturally suited to
small-angle stereo, the demands placed on the $\theta$ resolution by the
determination of the differential luminosity spectrum may force the
consideration of large-angle stereo.

The performance of the L detector tracking system, including the CCD
vertex detector, is summarized in Fig.~\ref{fig:trackpresvsp}, which
shows fractional momentum resolution vs. momentum for tracks transverse
to the beam direction ($\cos\theta$=0).  Figure~\ref{fig:trackpresvscos}
shows the fractional momentum resolution \vs\ $\cos\theta$ for tracks of
momentum 100~GeV.  In the limit of high-momentum tracks, the L
tracking resolution in ${1/p_t}$ is $\sci{3}{-5}$ GeV$^{-1}$.
Figure~\ref{fig:recoil} shows the expected distribution in recoil mass
from dimuons in the Higgsstrahlung process $e^+e^-\rightarrow hZ
\rightarrow X\mu^+\mu^-$ at $\sqrt{s}$ = 350 GeV for the nominal L
detector baseline and for several globally rescaled resolutions in
${1/p_t}$.

\begin{figure}
\begin{minipage}[htbp]{2.8in}
\begin{center}
\vspace{-0.5in}
\mbox{\epsfig{file=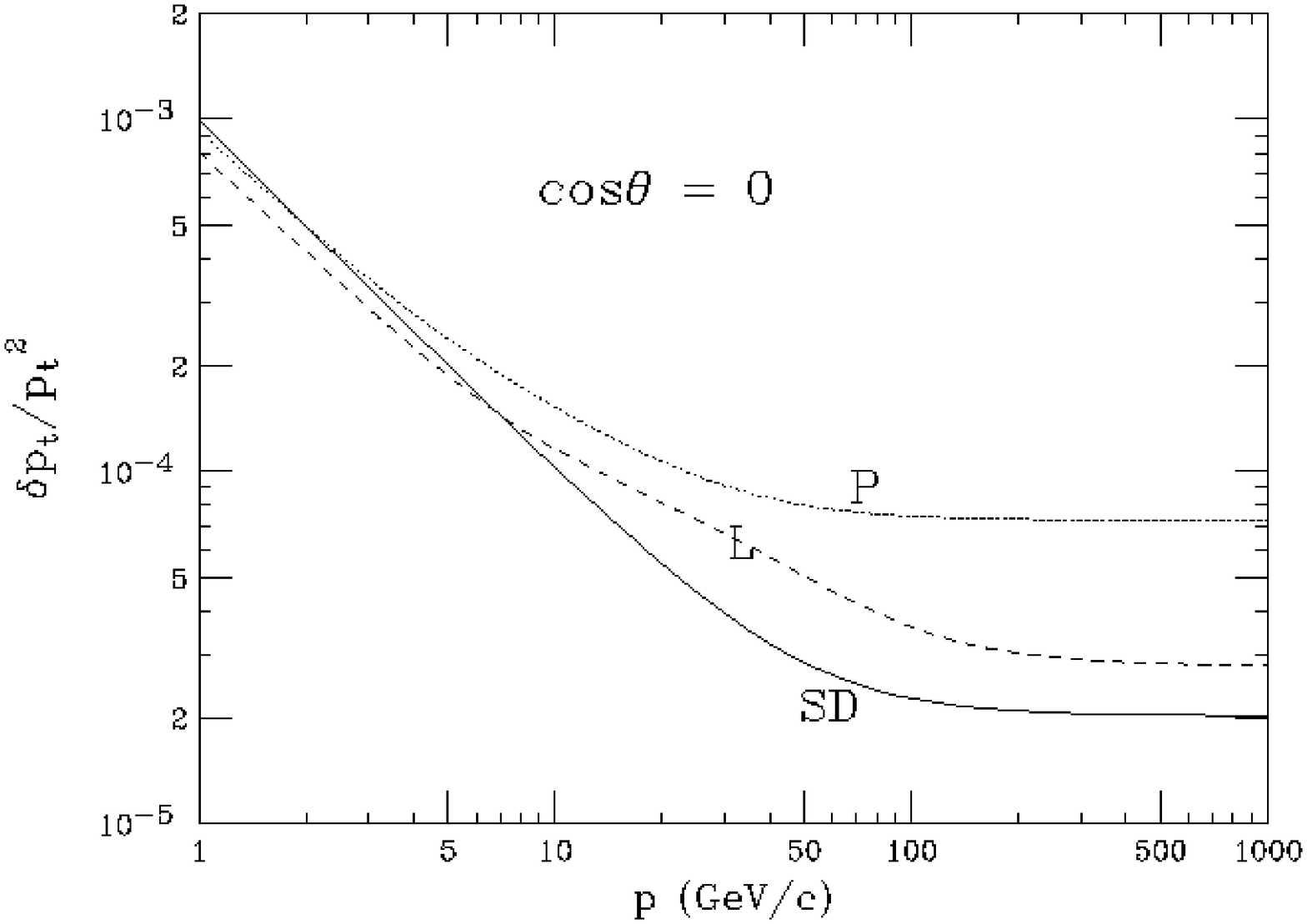,width=2.5in}}
\end{center}
\caption[frac]{
\label{fig:trackpresvsp}
Expected fractional momentum resolution \vs\ momentum for the
L, SD, and P central trackers for tracks transverse to the
beam direction.
}
\end{minipage}
\begin{minipage}[t]{0.2in}~~\end{minipage}
\begin{minipage}[t]{2.8in}
\begin{center}
\mbox{\epsfig{file=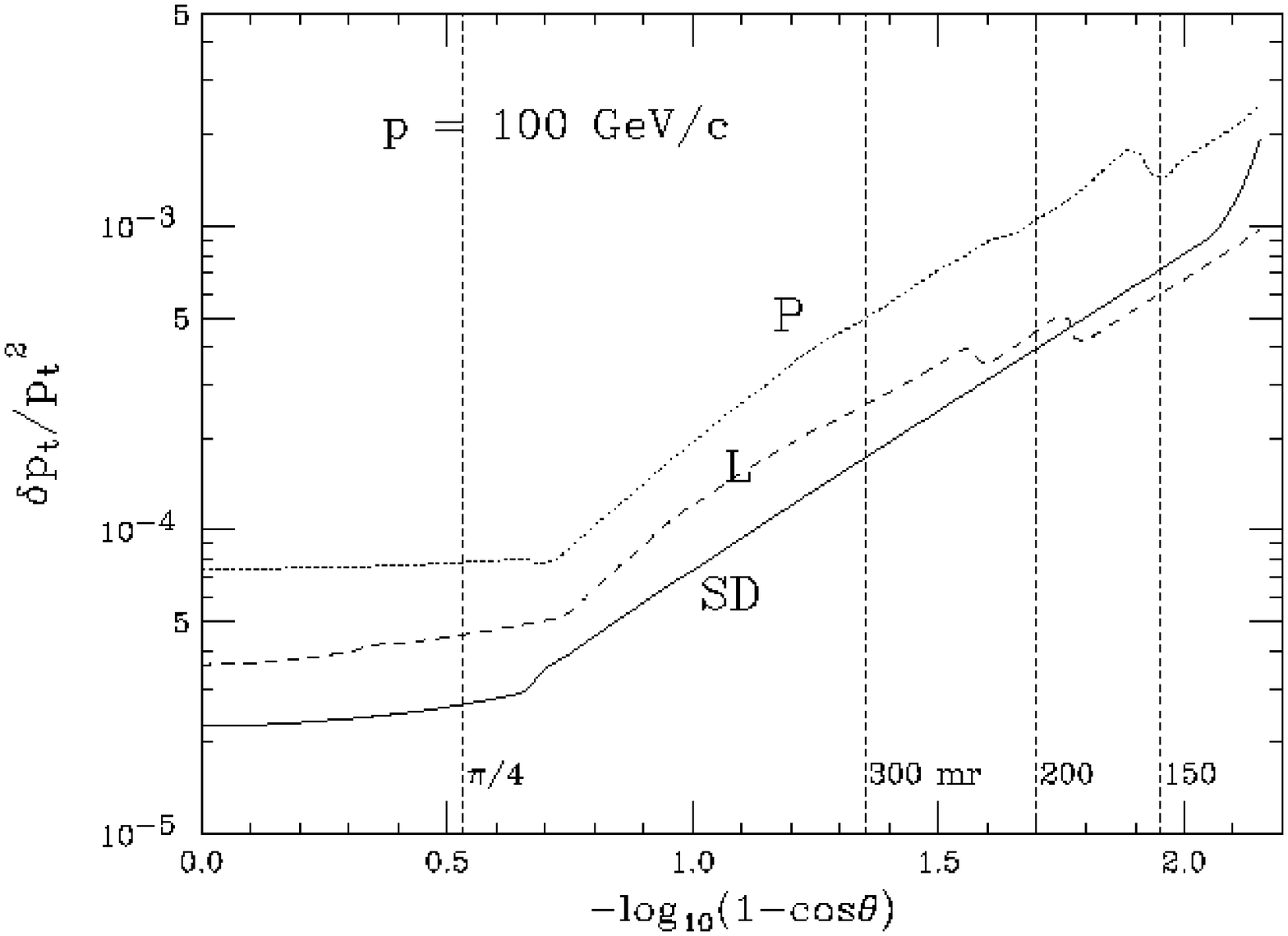,width=2.5in}}
\end{center}
\caption{Expected fractional momentum resolution \vs\ $\cos\theta$ for the
L, SD, and P central trackers for 100 GeV tracks.
\label{fig:trackpresvscos}
}
\end{minipage}
\end{figure}

\begin{figure}
\vspace{-.5 in}
\begin{minipage}[t]{2.8in}
%\begin{center}
\mbox{\epsfig{file=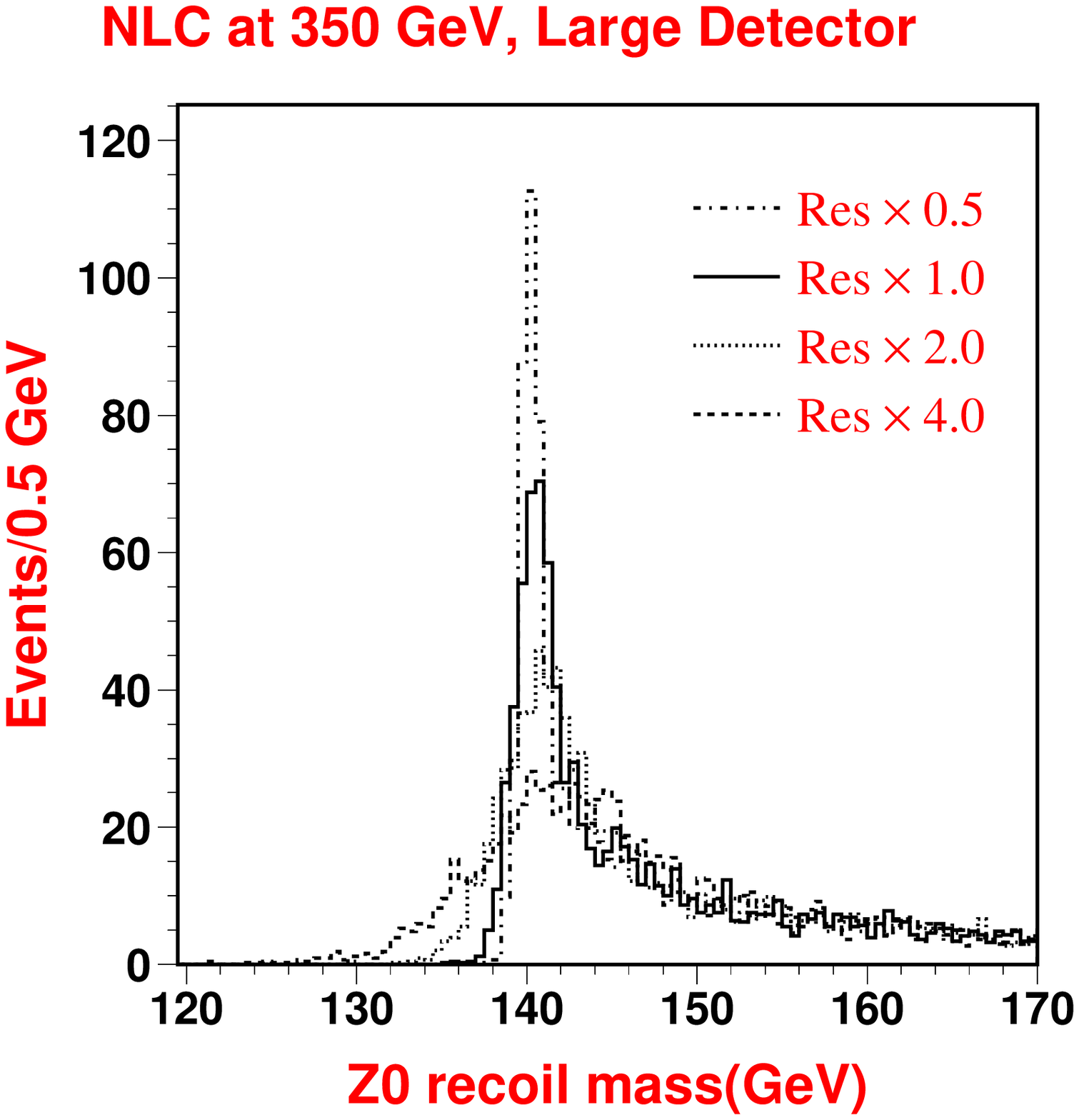,width=2.6in}}
%\end{center}
\caption{Expected recoil mass distribution in recoil mass from dimuons in the
Higgsstrahlung process $e^+e^-\rightarrow hZ \rightarrow X\mu^+\mu^-$ at
$E_{cm}$ = 350 GeV for the nominal L detector baseline and for several
globally rescaled resolutions in $1 / p_t$.
\label{fig:recoil}
}
\end{minipage}
\begin{minipage}[t]{0.2in}~~\end{minipage}
\begin{minipage}[htbp]{2.7in}
\begin{center}
\vspace{-1.1in}
\mbox{\epsfig{file=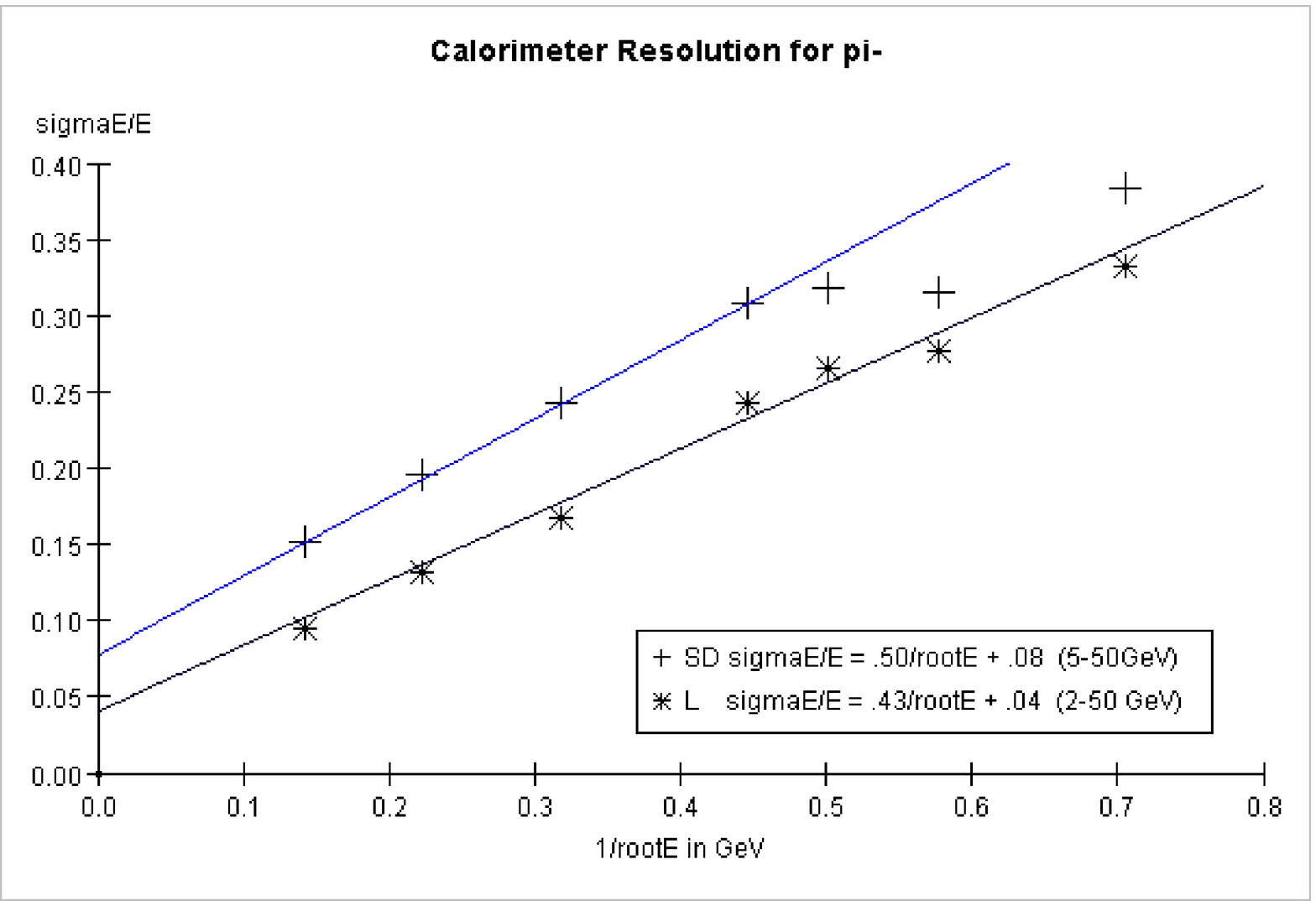,width=2.5in}}
\end{center}
\caption[frac]{
\label{fig:hadres}
Expected $\pi^-$ energy resolution in the L
($\sigma_E/E = 0.43 /\sqrt{E} +  0.04$)
 and SD
($\sigma_E/E = 0.50 /\sqrt{E} +  0.08$)
 Detectors.
}
\end{minipage}
\end{figure}

The electromagnetic calorimeter of the L Detector is a lead-scintillator
laminate with 4 mm lead followed by 1 mm scintillator for 40 layers.
This results in 28.6 radiation lengths with a 2.1 cm Moliere radius.
One layer of 1 cm$^2$ silicon pads is foreseen near shower maximum.
The transverse segmentation of the scintillator is $5.2$~cm~$\times$
$5.2$ cm.  The barrel of the electromagnetic calorimeter has an inner
radius of 200 cm.  The electromagnetic energy resolution is expected to
be $17\% /\sqrt{E}$.

The hadronic calorimeter is 120 layers of 8 mm lead layers with 2 mm
scintillator sampling.  The entire calorimeter comprises 6.6 interaction
lengths.  The transverse segmentation of the scintillator in the
hadronic calorimeter is 19 cm $\times$ 19 cm.  Figure \ref{fig:hadres}
presents the expected $\pi^-$ energy resolution.

The hope is that the large $BR^2$ of the L design  will allow jet
reconstruction using energy flow at a more modest cost than
Si/W, overcoming the
limited transverse segmentation possible with scintillator and the
larger Moliere radius of lead.
But,  since the transverse segmentation of the EMCal is
much larger than the Moliere radius, it is not clear whether energy flow can
be effectively carried out for L. This is in contrast to the SD case,
where the fine segmentation allows one to have some confidence that an
efficient EF reconstruction can be carried out.  This is
clearly an area where additional work with full shower simulations is
required.

Since shower reconstruction for an EF algorithm for the American
detectors is still in its infancy, one can in the meantime use
parameterizations of calorimeter performance using a fast simulation.
One would expect that the performance from full reconstructions will
eventually approach that of the fast simulation.  Therefore, for the
following performance plots we apply the energy flow technique, but
assume a perfect charged-neutral separation in the calorimeters.  The 
appropriate charged track resolutions and EMCal resolutions are
then applied.  This
assumption is not unreasonable for SD,  but for L it is probably too
idealized.  In any case, our method should indicate the asymptotic limit
of performance.

To examine jet energy resolution, we used $e^+e^-\rightarrow q\bar{q}$
events without ISR or beamstrahlung, and demanded that exactly two jets be
reconstructed.  Hence, $E_{\rm jet}=\sqrt{s}/2=E_{\rm beam}$.  An
example distribution of the reconstructed jet energy, for $\sqrt{s}=100$
GeV is given in Fig.~\ref{fig:calE_L}.  Only events with
$|\cos\theta_{\rm thrust}|<0.8$ are included.  The tail of the
distribution is due to QCD and jet-finding effects, whereas the
resolution we are interested in here is given by the Gaussian
distribution near $E_{\rm beam}$, and we take the $\sigma$ of this fit
as the resolution.  Figure ~\ref{fig:calEres_L} gives the resolution
(the asymptotic limit of performance without accounting for 
non-Gaussian tails, as described above) as a function
of $E_{\rm jet}$.  A fit to these data gives 
\beq
{{\sigma_{E_{\rm jet}}}\over{E_{\rm jet}}} =
            {{0.18}\over{\sqrt{E_{\rm jet}}}}\ .
\eeq{d-singlejet}
One should not expect to actually achieve this idealized resolution with
the L calorimeter.

\begin{figure}
\begin{minipage}[htbp]{2.8in}
\begin{center}
\vspace{-.5in}
\mbox{\epsfig{file=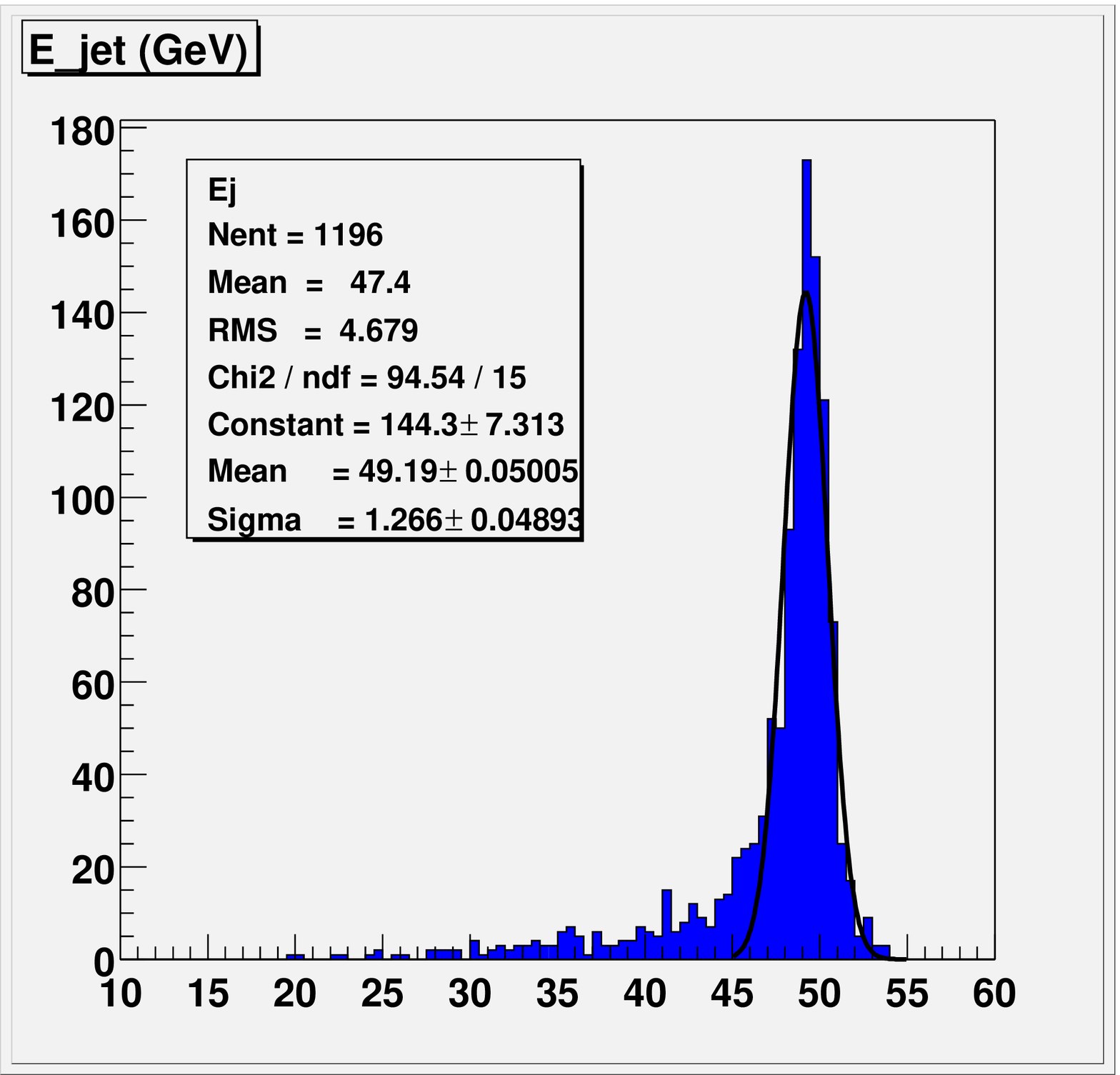,width=2.5in}}
\end{center}
\caption[frac]{
\label{fig:calE_L}
Reconstructed jet energy with the L detector for 50 GeV beam energy. }
\end{minipage}
\begin{minipage}[t]{0.2in}~~\end{minipage}
\begin{minipage}[t]{2.8in}
\begin{center}
\mbox{\epsfig{file=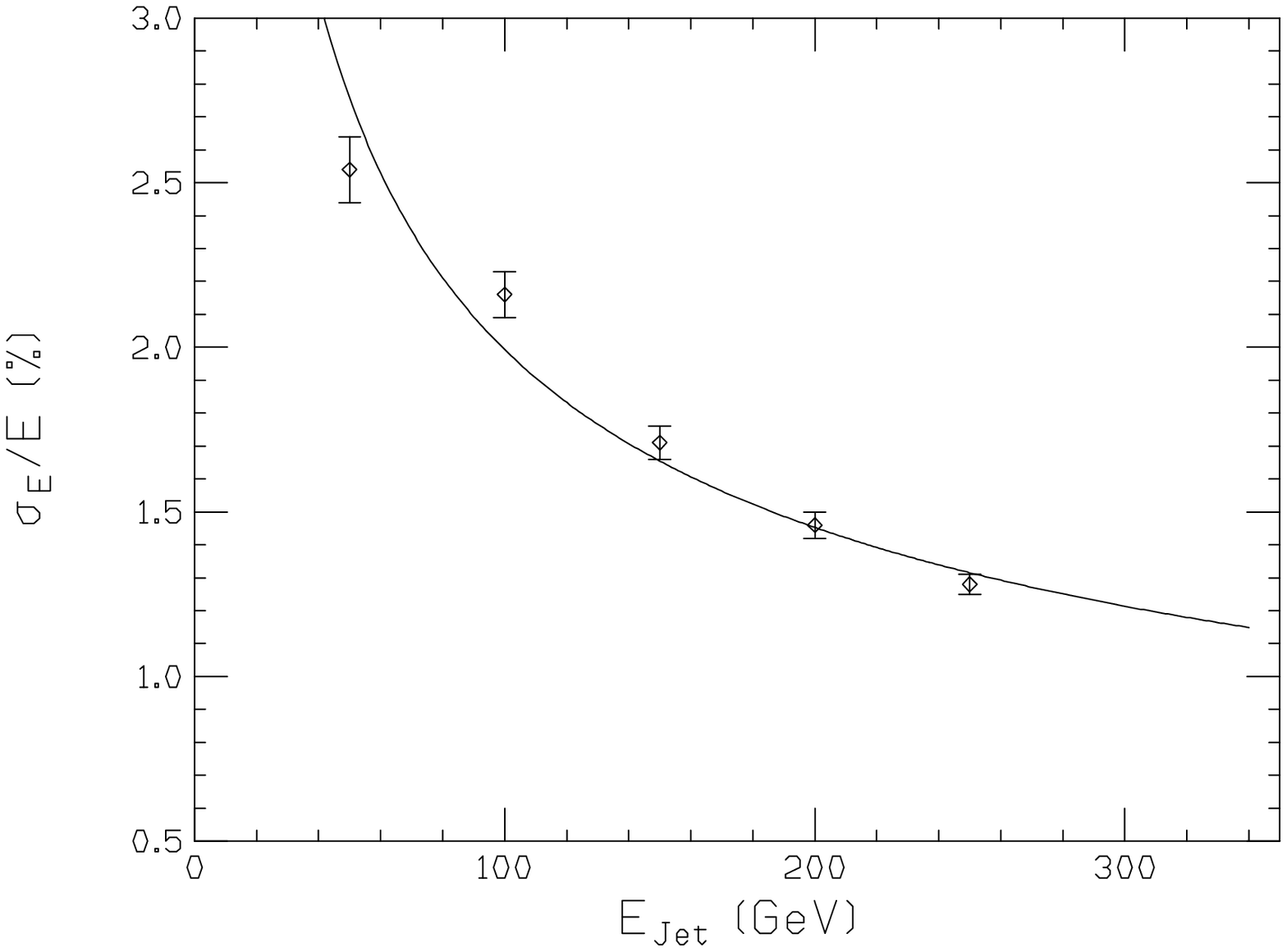,width=2.5in}}
\end{center}
\caption{Jet energy resolution (in \%) vs. jet energy for the L detector.
The curve is the fit described in the text.
\label{fig:calEres_L}
}
\end{minipage}
%\end{figure}
\vskip.41in
%\begin{figure}
\begin{minipage}[htbp]{2.8in}
\begin{center}
\vspace{-0.5in}
\mbox{\epsfig{file=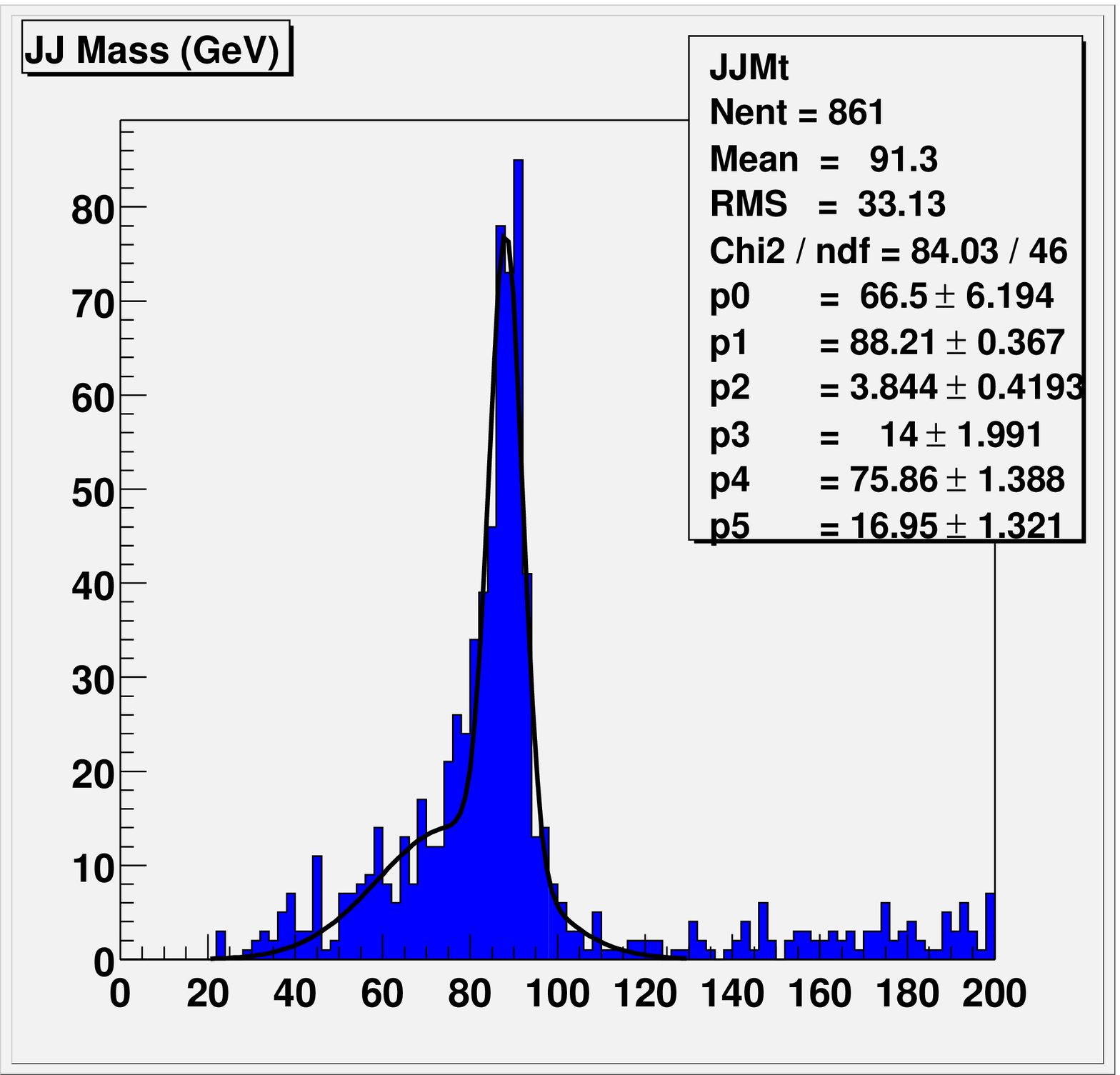,width=2.5in}}
\end{center}
\caption[frac]{
\label{fig:calM_L}
Reconstructed jet-jet mass for $Z$ candidates in
$e^+e^-\rightarrow ZZ\rightarrow$ hadrons at 350 GeV for the L detector.
}
\end{minipage}
\begin{minipage}[t]{0.2in}~~\end{minipage}
\begin{minipage}[t]{2.8in}
\begin{center}
\mbox{\epsfig{file=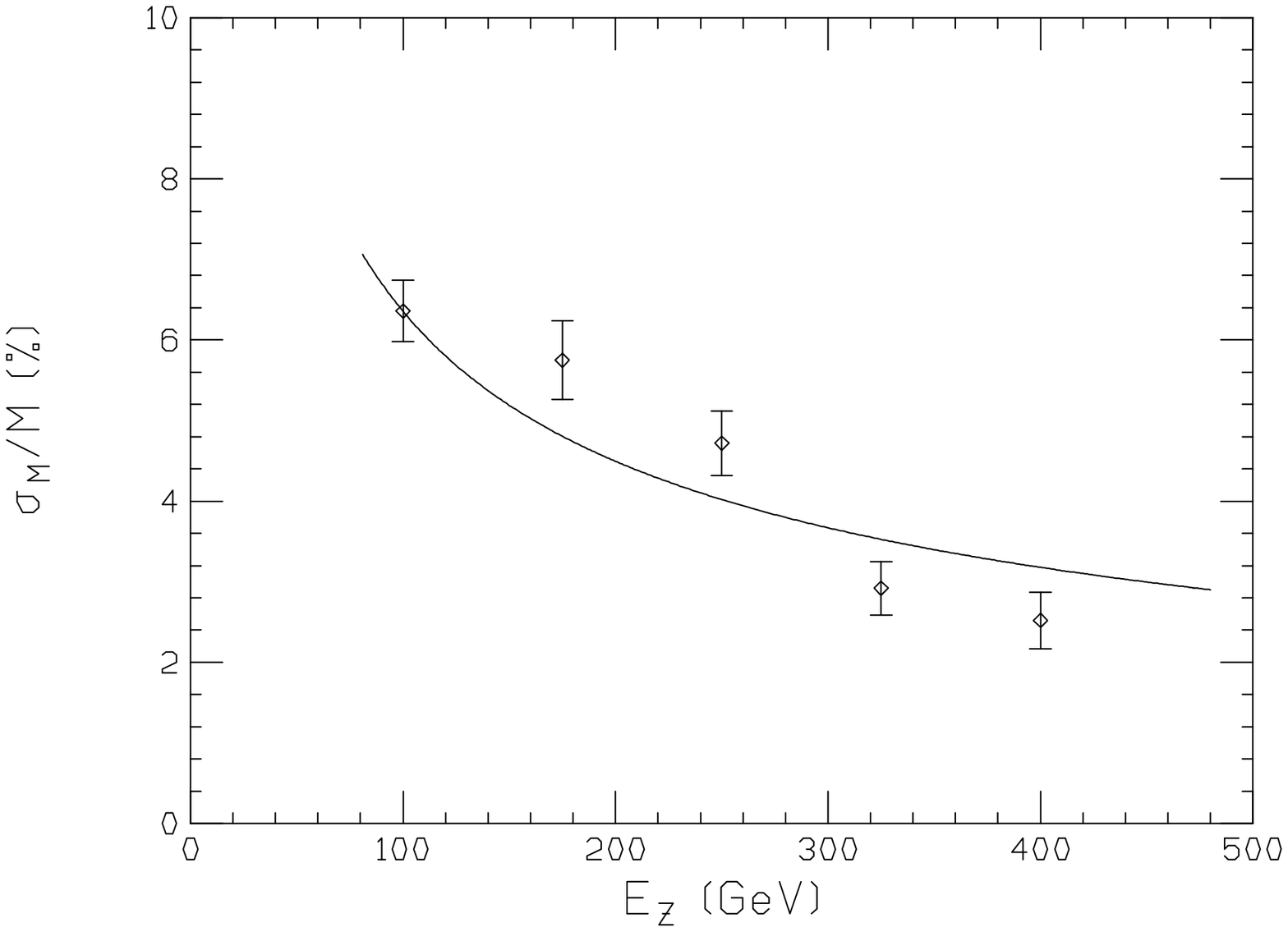,width=2.5in}}
\end{center}
\caption{Jet-jet mass resolution (in \%) for $Z\rightarrow$ 2 jets vs.
$Z$ energy for the L detector in $e^+e^-\rightarrow ZZ\rightarrow$ hadrons
events. The curve is the fit described in the text.
\label{fig:calMres_L}
}
\end{minipage}
\end{figure}

Another important and general measurement of performance is the jet-jet
mass resolution.  To examine this, we examine the process
$e^+e^-\rightarrow ZZ\rightarrow$ hadrons.  Exactly four final-state
jets were required.  To get a distribution with little background, we
require that one 2-jet combination have mass near $M_Z$, then plot the
mass of the other jet pair, $M_{jj}$.  An example $M_{jj}$ distribution
is given in Fig.~\ref{fig:calM_L} ~for $\sqrt{s}=350$ GeV.  Again, we
fit a Gaussian to the distribution near $E_Z=E_{\rm beam}$ to extract a
measure of the $M_{jj}$ resolution.  This resolution,
$\sigma_{Mjj}/M_{jj}$, is plotted vs.  $E_Z=\sqrt{s}/2\approx E_{jj}$ in
Fig.~\ref{fig:calMres_L}.  A fit to the data of the form
$(A/\sqrt{E_Z})\oplus B$ gives 
\beq
 {{\sigma_{M_{jj}}}\over{M_{jj}}}=
{{0.64}\over{\sqrt{E_Z}}} \ ,
\eeq{d-dijet}
 with negligible constant term.  To the extent
that the dijets from a $Z$ are perfectly identified and that no color 
connection or jet merging effects occur, the sampling term constant 
here should approach that for the single jet energy resolution 
given in \leqn{d-singlejet}.  The degradation of dijet mass resolution 
from this ideal limit requires more study.

The 3~T solenoidal coil is located outside the hadronic calorimeter
to optimize calorimeter performance.  The inner radius of the solenoidal
coil is 370 cm.

The muon system consists of 24 layers of 5 cm iron plates, with 3 cm
gaps for RPC detectors.  Axial strips of 3 cm pitch measure the $\phi$
coordinate to 1 cm precision in all 24 gaps, and every sixth gap
provides azimuthal strips for a measurement of the $z$ coordinate to 1 cm
precision.  The barrel muon system begins at a radius of 420 cm.  Figure
~\ref{fig:muhits} illustrates the expected performance for the L
detector.

\subsection{SD detector for the high energy IR}
\label{sec:sddetector}

The strategy of the `Silicon Detector' (SD) is based on the assumption
that energy flow calorimetry will be important.  While this has not yet
been demonstrated in simulation by the American groups, the TESLA
Collaboration has accepted it.  This assumption then leads directly to a
reasonably large value of ${BR}^2$ to provide charged-neutral separation
in a jet, and to an electromagnetic calorimeter (EMCal) design with a
small Moliere radius and small pixel size.  Additionally, it is
desirable to read out each layer of the EMCal to provide maximal
information on shower development.  This leads to the same nominal
solution as TESLA:  a series of layers of about 0.5 $X_0$ tungsten
sheets alternating with silicon diodes.  Such a calorimeter is
expensive; its cost is moderated by keeping the scale of the inner
detectors down.  This has two implications:  the space point resolution
of the tracker should be excellent to meet momentum resolution
requirements in a detector of modest radius, and the design should admit
high-performance endcaps so that the barrel length (or cos $\theta
_{\rm Barrel}$) will be small.  Obviously it is desirable to minimize
multiple scattering in the tracker, but compromises will be needed and must
be tested with detailed simulation.  The last real strategic question is
whether the Hadronic Calorimeter (HCal) will be inside or outside the
coil.  Locating the HCal inside the coil permits reasonably hermetic
calorimetry, but requires a larger, more expensive coil and more iron to
return the flux.  It is assumed that the detector will have an
 ultra-high-performance vertex detector based on CCD's or an
equivalent thin, small pixel technology, as we have discussed for the 
L detector.  A  muon tracker will
be interleaved in the iron flux return utilizing reliable RPC's or
equivalent.

These considerations lead to a trial design with a tracking radius of
1.25 m and a field of 5 T. This is a $BR^2$ of 8, compared to 10 for
TESLA and 12 for the L detector.  The tracker is 5 layers of silicon
strips with a $\cos\theta _{\rm Barrel}$ of 0.8.  Sets of five disks
with silicon strips are arranged as endcaps to complete the acceptance.  
The HCal is
inside the coil.  The quadrant view is shown in Fig.
~\ref{fig:sdetector}, and the major dimensions are tabulated in Table
~\ref{tab:dimen}.

\begin{figure}[tb]
\begin{center}
\epsfig{file=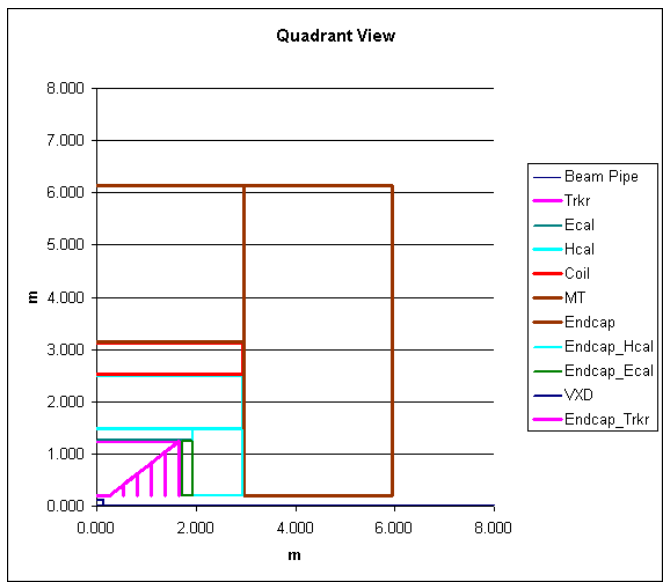,height=3.0in}
\caption{Quadrant view of the SD detector.}
\label{fig:sdetector}
\end{center}
\end{figure}

The SD detector relies entirely upon silicon tracking in a 5~T
solenoidal field in the central and forward regions.  Its central and
forward trackers consist of a 5-layer silicon barrel---a silicon drift detector
(SDD) or
microstrips---and five layers of double-sided silicon microstrip forward
disks.  Figure~\ref{fig:stracker} shows a sketch of the SD detector
tracking system.  The inner/outer radii of the barrel layers are 20/125
cm.  The inner and outer disks are at 40 cm and 167 cm from the
interaction point.  The boundary between the barrel and disk system lies
at $|\cos\theta|$ = 0.8.

\begin{figure}[htb]
\begin{center}
\epsfig{file=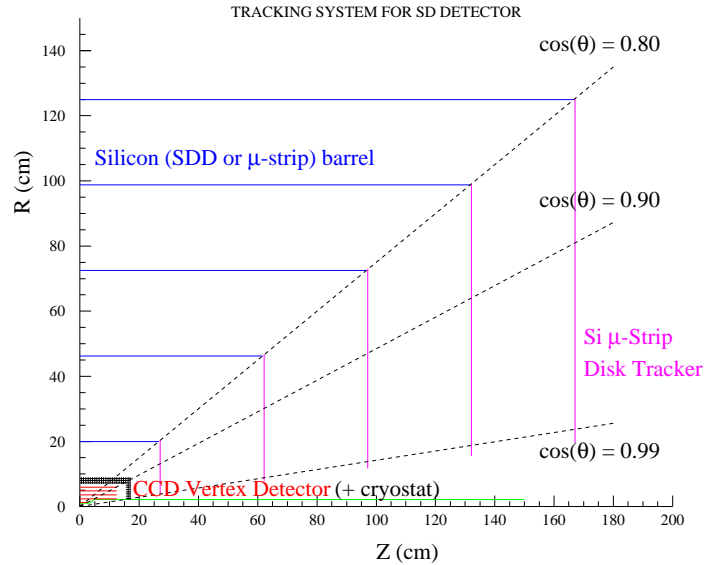,height=3.0in}
\caption{Sketch of SD detector tracking system.}
\label{fig:stracker}
\end{center}
\end{figure}

The SDD option provides a solid-state analog to a time projection
chamber.  A potential gradient is applied via implanted cathodes in the
silicon in order to force the generated electron cloud to drift through
the bulk of the silicon to a collection anode.  The highest voltage
supplied to a single cathode can be up to 2500 V. By measuring the cloud
distribution across the collection anodes and the drift time to the
anodes in parallel, one records three-dimensional position information
with a one-dimensional electronics readout.  Three-dimensional position
resolutions below 10 $\mu$m in each dimension can be achieved with an
anode spacing between 200 and 300 $\mu$m.  Thus, the electronics cost is
considerably reduced compared to other semi-conductor detector options.
Recently, a three-barrel SDD Tracker, using 216 large-area Silicon Drift 
wafers, was successfully completed and has been
installed in the STAR experiment at RHIC.

Compared to the STAR detector the following modifications would be made
to build a linear collider tracker:  1) increase the wafer size to 10 cm
$\times$ 10 cm; 2) reduce the wafer thickness from 300 to 150 $\mu$m; and 3)
redesign the front-end electronics for lower power to eliminate water
cooling.  The detector contains 56~m$^2$ of active silicon, requiring
about 6000 wafers and 4.4 million channels distributed over 229 ladders
constructed from carbon-fiber material.

The silicon strip detector (SSD) option makes use of what is at this
time a very mature tracking technology.  Nevertheless, several avenues
for further R\&D are discussed below.  It should be possible to exploit
the small (order $10^{-5}$) duty cycle of the linear collider to reduce
the power dissipated by the readout electronics by switching to a
quiescent state in between trains.  This would substantially reduce the
heat load, leading to a great reduction in the complexity and material
burden of the mechanical structure.

In order to improve the robustness of the detector against linear
collider backgrounds, it should be possible to develop a
microstrip readout with a short shaping time, 
with timing resolution of order 5--10
nsec.  This would allow out-of-time background hits to be eliminated
from the bunch train with a rejection factor of better than 10:1.

On the other hand, the high granularity of microstrip detectors would
make an SSD central tracker fairly robust against backgrounds even in
the absence of intra-train timing.  If instead it is felt that low- and
intermediate-momentum track parameter resolution is more important than
timing resolution, the use of a readout with a very  long shaping-time
should make
it possible to implement detector ladders of substantially greater
length than that of the 10--20 cm ladders of conventional strip detector
systems.  The AMS collaboration has developed a slow readout \cite{AMS}
with 6 electrons equivalent noise per cm of detector length.  This may
allow single ladders to stretch the entire half-length of the outermost
silicon layer, and for the inner layers to be thinned.  This, combined
with a space frame that derives much of its support from the ladders
themselves, would lead to a substantial reduction in the material
burden, and give an overall low-momentum track parameter resolution on
par with that of the L detector.

The forward disks for the SD tracker would have the same intrinsic
performance as those described above for the L detector.

The performance of the SD detector tracking system, including the CCD
vertex detector, is summarized in Fig.~\ref{fig:trackpresvsp} and
Fig.~\ref{fig:trackpresvscos}.  In the limit of high-momentum tracks,
the SD tracking resolution in ${1/p_t}$ is $\sci{2}{-5}$ GeV$^{-1}$.

The EMCal consists of layers of tungsten with gaps sufficient for arrays
of silicon diode detectors mounted on G10 mother boards.  The thickness
of these gaps is a major issue, in that it drives the Moliere radius of
the calorimeter.  A thickness of 4 mm seems quite comfortable,
accommodating a 0.3-0.5 mm silicon wafer, a 2 mm G10 carrier, and 1.5 mm
of clearance.  Conversely, 1.5 mm seems barely plausible, and probably
implies a stacked assembly rather than insertion into a slot.  For now,
we assume a 2.5 mm gap.

It is expected that the readout electronics from preamplification
through digitization and zero suppression can be integrated into the
same wafer as the detectors.  A fallback would be to bump- or 
diffusion-bond a separate chip to the wafer.  Thus it is expected that the pixel
size on the wafer will not affect the cost directly.  A pixel size
between 5 and 10 mm on a side is expected.  Shaping times would be
optimized for the (small) capacitance of the depleted diode, but will
probably be too long to provide any significant bunch localization
within the train.

The HCal is chosen to lie inside the coil.  This choice permits much
better hermeticity for the HCal, and extends the solenoid to the endcap
flux return.  This makes a more uniform field for the track finding, and
simplifies the coil design.  The HCal absorber is a non-magnetic metal,
probably copper or stainless steel.  Lead is possible, but is
mechanically more difficult.  The detectors could be `digital', with
high-reliability RPC's assumed.  The HCal is assumed to be 4 $\lambda$
thick, with 34 layers of radiator 2 cm thick alternating with 1 cm gaps.

We have examined performance for the SD detector model in the same way
as the L detector, calculating the asymptotic limit of performance. 
(See the corresponding discussion in Section 4.1 for the
limitations of this analysis.)  The
electromagnetic energy resolution is expected to be $18\% /\sqrt{E}$.
Figure \ref{fig:hadres} presents the expected $\pi^-$ energy resolution.
The resolution for jet energy reconstruction is given in Fig.
\ref{fig:calEres_SD}.  A fit to these data gives for the asymptotic limit
\beq
 {{\sigma_{E_{\rm jet}}}\over{E_{\rm jet}}} 
                  = {{0.15}\over{\sqrt{E_{\rm jet}}}} \ .
\eeqn

As previously, we fit a Gaussian to the distribution near $E_Z=E_{\rm
beam}$ to extract a measure of the $M_{jj}$ resolution.  This
resolution, $\sigma_{Mjj}/M_{jj}$, is plotted vs.  $E_Z$ in Fig.
~\ref{fig:calMres_SD}.  A fit to the data of the form
$(A/\sqrt{E_Z})\oplus B$ gives
\beq
 {{\sigma_{Mjj}}\over{M_{jj}}}=
{{0.72}\over{\sqrt{E_Z}}} \ ,
\eeqn
 with negligible constant term.  These idealized
studies are not yet precise enough to conclude that this is significantly
worse than the L Detector performance.

\begin{figure}
\begin{minipage}[htbp]{3.0in}
\begin{center}
\vspace{-1.2in}
\mbox{\epsfig{file=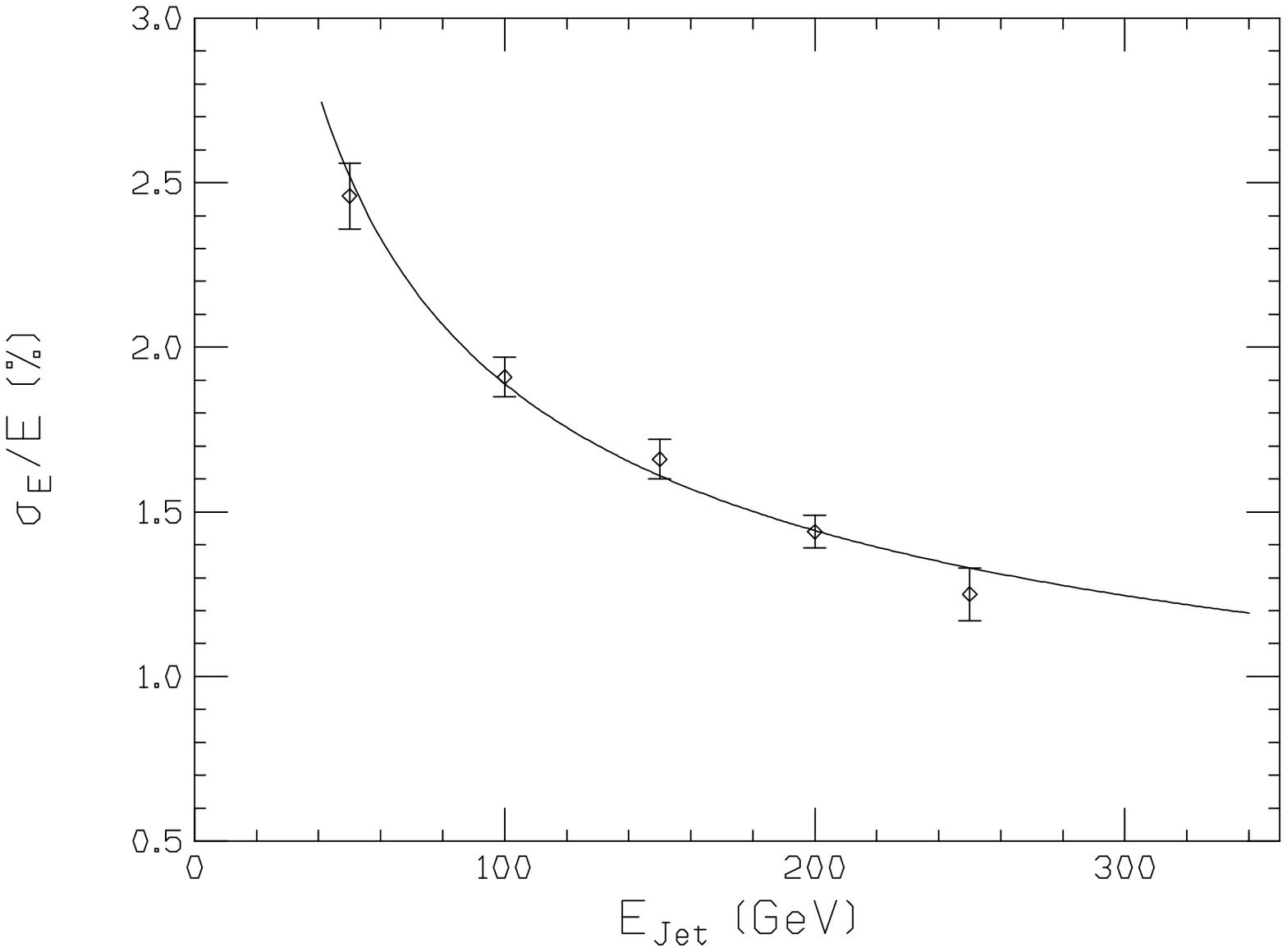,width=3.0in}}
\end{center}
\caption[Z jet-jet mass]{
\label{fig:calEres_SD}
Jet energy resolution (in \%) {\it  vs.}  jet energy for the SD detector.
The curve is the fit described in the text.
}
\end{minipage}
\begin{minipage}[t]{0.2in}~~\end{minipage}
\begin{minipage}[t]{3.0in}
\begin{center}
\mbox{\epsfig{file=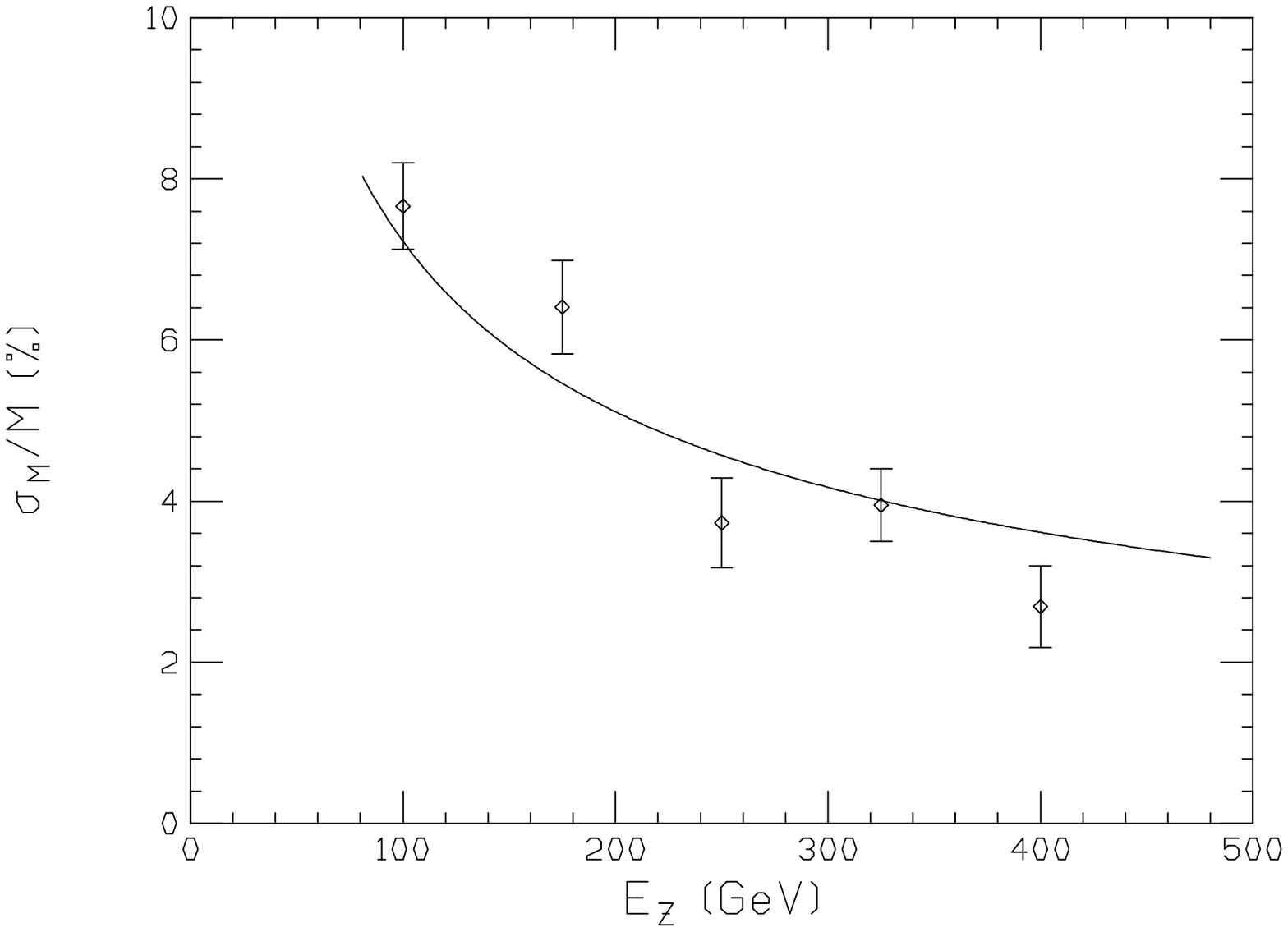,width=3.0in}}
\end{center}
\caption{Jet-jet mass resolution (in \%) for $Z\rightarrow$ 2 jets {\it vs.}
$Z$ energy for the SD detector in $e^+e^-\rightarrow ZZ\rightarrow$ hadrons
events. The curve is the fit described in the text.
\label{fig:calMres_SD}
}
\end{minipage}
\end{figure}

The coil concept is based on the CMS design, with two layers of
superconductor and stabilizer.  The stored energy is 1.4 GJ, compared to
about 2.4 GJ for the TESLA detector and 1.7 GJ for the L detector.  The
coil thickness is 60 cm, which is probably conservative.

The flux return and muon tracker is designed to return the flux from the
solenoid, although the saturation field for the iron is assumed to be
1.8 T, which may be optimistic.  The iron is laminated in 5 cm slabs
with 1.5 cm gaps for detectors.

\subsection{P detector for the lower-energy IR}
\label{sec:pdetector}

The P Detector is proposed as a lower-cost detector for the  second IR, capable
of the performance required for lower-energy operation, including the
$Z$-pole physics.

The P detector is illustrated in Fig.~\ref{fig:pdet}.  The dimensions
of the P Detector are presented in Table~\ref{tab:dimen}.

\begin{figure}[htbp]
\begin{center}
\epsfig{file=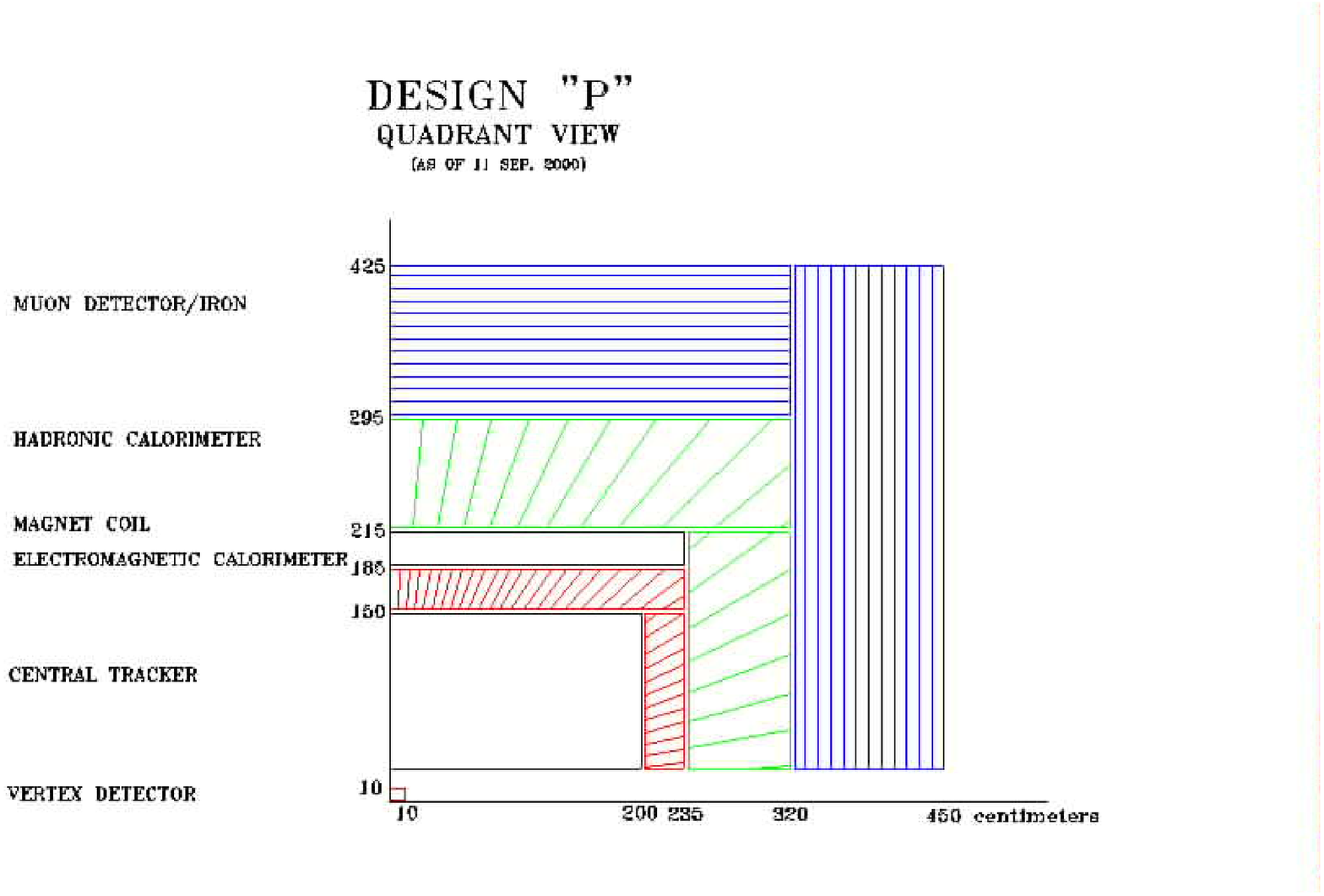,height=3.5in}
\caption{Quadrant view of the P detector.}
\label{fig:pdet}
\end{center}
%\end{figure}
\vspace{.1in}
%\begin{figure}[htb]
\begin{center}
\epsfig{file=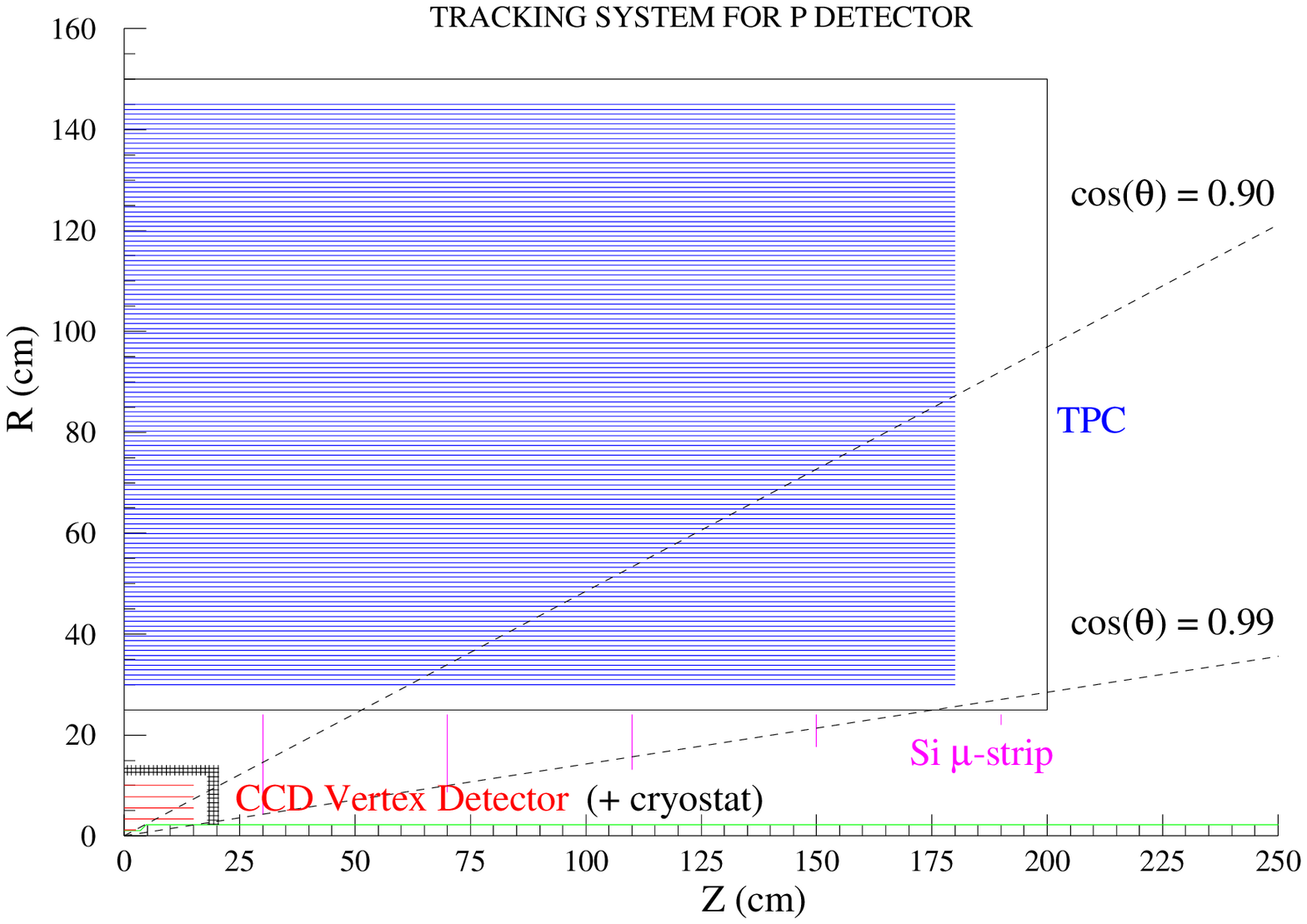,height=3.0in}
\caption{Sketch of P detector tracking system.}
\label{fig:ptracker}
\end{center}
\end{figure}

The P detector employs the same CCD vertex detector design described for
the L detector above, illustrated in Fig.~\ref{fig:vxd}.

The P detector's tracker design is modelled very closely upon that of
the L detector.  Since it is 
meant to operate at lower center-of-mass energies, its
required resolution in ${1/p_t}$ is correspondingly less severe,
allowing for a smaller tracking system and therefore a smaller, cheaper
overall detector design.  Figure~\ref{fig:ptracker} shows a sketch of
the P detector tracking system.

Briefly, the P central tracker consists of a 120-layer TPC, of
inner/outer radii = 25/150 cm and half-length 200 cm.  Again, one or
more intermediate tracking layers of silicon or scintillating fiber just
inside the inner TPC radius may be desirable.  The forward tracker
consists of five silicon microstrip disks similar to those in the L and
SD detectors.  The performance of the P detector tracking system in a 3 T
solenoidal field, including the CCD vertex detector, is summarized in
Fig.~\ref{fig:trackpresvsp} and Fig.~\ref{fig:trackpresvscos}.  In
the limit of high-momentum tracks, the P tracking resolution in ${1/p_t}$
 is $\sci{6}{-5}$ GeV$^{-1}$.

The 3 Telsa solenoidal coil is located outside the electromagnetic
calorimeter and inside the hadronic calorimeter.  This compromise (over
the desire to move the coil outside the hadronic calorimeter) contains
the cost of the P detector.  The inner radius of the solenoid is 185 cm.

The electromagnetic calorimeter of the P Detector consists of 32 layers
of lead-scintillator laminate, with 4 mm lead layers followed by 3 mm
scintillator, for 22.8 radiation lengths.  These layers are ganged in
pairs, giving 16 readout layers.  One layer of 1 cm$^2$ silicon pads
is forseen near the EMCal shower maximum.  The transverse segmentation of the
scintillator is 2 degrees $\times$ 2 degrees.  It has an inner radius of
150 cm.

The hadronic calorimeter is 65 layers of 8 mm lead layers with 3 mm
scintillator sampling.  These layers are ganged to produce 8 independent
samples.  The inner radius of the hadronic calorimeter barrel is 215 cm.
The entire calorimeter (electromagnetic and hadronic) comprises 3.9
interaction lengths.  The transverse segmentation of the scintillator in
the hadronic calorimeter is 4 degrees $\times$ 4 degrees.

Given its segmentation, the P detector would not be well-suited for  using
energy flow in jet reconstruction.  Unlike L and SD, the segmentation
is organized as towers of constant $\theta$ and $\phi$.  For running at
the $Z$, excellent jet reconstruction is probably not an important
issue.  However, at higher energy, for light Higgs or $W$-pair physics,
for example, this conclusion is less clear.  Jet reconstruction for P
would most likely be carried out using the calorimeter alone (or the  tracker
alone).  Note, however, that the Pb-scintillator ratio, as currently
proposed, would not be expected to give good compensation of
electromagnetic and hadronic energy depositions.  Performance results
for jet reconstruction, similar to those given for L and SD, have not
yet been carried out.  The results would provide an interesting point of 
comparison to the energy flow performance of SD.

The muon system consists of 10 layers of 10 cm iron plates, with 3 cm
gaps for RPC detectors.  Axial strips of 3 cm pitch measure the $\phi$
coordinate to 1 cm precision in all 10 gaps, and two gaps (5 and 10)
provide azimuthal strips for a measurement of the $z$ coordinate to 1 cm
precision.

\subsection{Cost estimates}

The costs of the subsystems of each of the three detectors have been
estimated based on past experience and escalation to FY01.  The three
cost estimates are shown in
Table~\ref{tab:costs}.  Approximately 40\% contingency is assumed for
each of the detectors, resulting in a total cost estimate of \$359
million for the L detector, \$326 million for the SD detector, and \$210
million for the P detector.

\begin{table}[htb]
\begin{center}
\begin{tabular}{l|ccc}\hline\hline
Detector & L &  SD & P \\ \hline
1.1 Vertex	&	    4.0	 &   4.0 & 	    4.0 \\
1.2 Tracking	&	  34.6	 & 19.7	  &  23.4 \\
1.3 Calorimeter	 & 48.9	 & 60.2	 &   40.7\\
\ \  1.3.1 EM	&	(28.9)	 & (50.9)&	(23.8)\\
\ \   1.3.2 Had	&	(19.6)		& (8.9)& (16.5)\\
\ \   1.3.3 Lum	&	  (0.4)	 & (0.4) &	  (0.4)\\
1.4 Muon	&	  16.0	 & 16.0	 &     8.8\\
1.5 DAQ		&  27.4 & 	52.2 & 	  28.4\\
1.6 Magnet \& support& 110.8 & 75.6 & 30.5\\
1.7 Installation	  &  7.3	 &   7.4	 &     6.8\\
1.8 Management	   & 7.4 & 	  7.7	 &     7.4\\ \hline
    SUBTOTAL		 &  256.4 & 	242.8	 &   150.0\\
1.9 Contingency	& 102.6	 &  83.4	 &   60.0\\ \hline
Total		&	359.0	& 326.2 &	210.0 \\ \hline
\hline

\end{tabular}
\caption{$e^+e^-$ linear collider detector budgets
         (WBS to subsystem level) in M\$ FY01.}
\label{tab:costs}
\end{center}
\end{table}

 \emptyheads

\addtocontents{toc}
{\bigskip\noindent{\huge Questions for Further Study}}

\begin{center}\begin{large}{\Huge \sffamily Questions for Further Study}
\end{large}\end{center}

\blankpage \thispagestyle{empty}
\fancyheads

\setcounter{chapter}{15}

\chapter{Suggested Study Questions on LC Physics and Experimentation}
\fancyhead[RO] {Suggested Study Questions on LC Physics and Experimentation}

\section{Physics issues}

\subsection{Higgs physics}

For further information on this section, consult with: 
Jack Gunion, Howard Haber, Andreas Kronfeld, Rick van Kooten.

\begin{enumerate}
\item  Perform a fully simulated study of the precision to which
        Higgs branching ratios can be determined for $m_h$ = 115 GeV;
        for $m_h$ = 140 GeV; for $m_h$ =  200 GeV.   How do these precisions
        depend on CM energy?
\item   Is $\gamma\gamma$  needed to measure the total Higgs width, for
        low mass Higgs?
\item Outline the necessary experimental program to determine
        the spin/parity of a putative Higgs state.
\item   Optimize a program for determination of the Higgs self-couplings.
          What requirements does this study impose on the dijet invariant
           mass  resolution?
\item  What is the utility of positron polarization for Higgs
        measurements?
\item  From knowledge of measured Higgs branching ratios (fermion
        pairs, $ZZ$, $WW$, $gg$, $\gamma\gamma$), the total width, 
and the couplings $g_{ZZh}$, $g_{WWh}$, what reach is available to 
       detect the presence of the SUSY states $H$, $A$?
        What is the relative importance of errors in each measurement?
\item To what extent can one measure $\tan\beta$ for the SUSY Higgs
                from Higgs sector
        measurements alone?  Is it possible to do so in a truly 
         model-independent way for the most general sets of MSSM parameters?
\item   How will one disentangle $H^0$ and $A^0$ in the decoupling limit where
        the masses are nearly degenerate?

\item Contrast the  use of $\ee$ and $e^-e^-$ beams for the
            $\gamma\gamma\to h$ measurement.  The use of $\ee$ admits
         numerous physics backgrounds that are absent for $e^-e^-$.
           Is it critical to avoid these backgrounds?  Can the advantage
          of $e^-e^-$ over $\ee$ be compensated by higher integrated
             luminosity?
\item  The dominant backgrounds to $\gamma\gamma \to h\to b\bar b$
 are $\gamma\gamma\to b\bar b(g)$ and  $\gamma\gamma\to c\bar c(g)$. 
  The production
   cross section for $c\bar c(g)$ is about 25 times larger than for $b\bar b(g)$.
    The background can be suppressed, first, by improved $b$ tagging, and
     second, by improved Higgs (two-jet) mass resolution.  With this in 
     mind, what is the optimal strategy for isolating the Higgs peak
     from the background?

\item   Contrast the  use of $\ee$ and $e^-e^-$ beams, in the same
          way, for a broadband search for a heavy Higgs $s$-channel resonance
            in $\gamma\gamma$.
\end{enumerate}

\subsection{Supersymmetry}

For further information on this section, consult with: Jonathan Feng,
Uriel Nauenberg, Frank Paige, James Wells.

\begin{enumerate}
    
 \item  Develop a plan for  measuring the chargino mass matrix, 
           including mixing,  for the most general sets of MSSM parameters.

\item   Do the same for the neutralino, stau and stop mixing matrices.

 \item  Is there a program by which one could, at least in principle,
            measure all 105 independent MSSM parameters?

 \item   What can LC measurements tell us, and with what precision,
        about the nature of the SUSY model and the SUSY breaking
        mechanism and scale?   What can be learned about the scale and
        physics of grand unification?

 \item   Evaluate the benefit of positron polarization for SUSY
        measurements.

 \item  For what questions of SUSY spectroscopy are $\gamma\gamma$,
                 $e\gamma$, and $e^-e^-$ beams of special importance?

  \item  How well can CP-violating effects be studied in supersymmetry?
        How do these compare and connect to those made in the $B$
        factories or $K$ decays?

  \item  What limits can be set on lepton flavor violation in slepton
        reactions?  Is it possible to measure quark flavor
          violation effects that are
           associated with SUSY parameters and independent of 
              CKM mixing?

   \item  What measurements from the LC would be required to verify the
           neutralino origin of cosmological dark matter?

  \item  What information encoded in the SUSY parameters can provide 
             information about the nature of string/M theory?

\end{enumerate}

\subsection{New physics at the TeV scale}

For further information on this section, consult with: Tim Barklow, 
Bogdan Dobrescu, JoAnne Hewett, Slawek Tkaczyk.

\begin{enumerate}

 \item  What precision can eventually  be 
          reached on anomalous $WWV$, $ZZV$ and $t\bar tV$
        couplings?   What machine parameters are needed?

   \item  For the broad range of strong coupling models that obey
        existing precision EW constraints, what are the observable
        consequences at a 500 GeV LC?    At 1000 GeV?   At 1500 GeV?  
       Are there models of strong coupling for which there are no
        observable consequences at 500 GeV?

   \item  Is it possible for models of a strong-coupling Higgs sector
             to mimic predictions of supersymmetry or extended Higgs models
            in a way that these models cannot be distinguished at the LHC?
            What $\ee$ measurements would be most important in these cases?

  \item   What is the utility of $\gamma\gamma$ or $e^-e^-$ operation for
        probing the strong coupling models?

  \item   Develop general classification of models with large extra
             dimensions.

   \item   How can measurements at the TeV scale constrain
        string/M-theory models with string or quantum gravity
           scales much less than $10^{19}$ GeV?

   \item   Describe the reach of a LC for seeing large extra dimensions 
        as a function of energy and luminosity in various
         scenarios.  To what extent does the higher
        precision of a 500 GeV LC complement the higher energy reach
        of the LHC? 

   \item What is the role of $\gamma\gamma$, $e \gamma$, and $e^-e^-$ 
            experiments in probing models with extra dimensions?

   \item    What would be the role of the LC in understanding the 
              nature of cosmological dark matter in models not related
             to supersymmetry?

   \item   In what way can LC measurements constrain gravitational
        effects such as Hawking black hole radiation?
\end{enumerate}

\subsection{Top quark physics}

For further information on this section, consult with: Ulrich Baur,
David Gerdes.

\begin{enumerate}

\item How well can the top quark width be determined from threshold
         measurements?  A full analysis should include the threshold shape,
      the top quark momentum distribution, and the forward-backward
       asymmetry from S--P mixing.  Are there additional effects that 
        can contribute to this determination?

  \item  Can one determine the top quark Yukawa coupling at the $t\bar t$
        threshold?  With what precision?

  \item Can CP violation associated with the top quark be probed at the
            $t \bar t$ threshold?

  \item  Can a high-precision top quark mass be obtained from continuum
             $t\bar t$ production?  Is there an infrared-safe definition of
            $m_t$ that can be applied to this analysis?

 \item  How well can the top quark Yukawa coupling be determined in 
           $\ee\to t\bar t h$?  What backgrounds arise from other top 
            quark production processes (\eg,  $\ee\to t\bar t g$)?
           Are spin correlations derived from kinematic fitting useful
             in this analysis?

  \item   How well can one measure the vector and axial $t\bar tZ$ couplings?

  \item  How well can one measure the $t\bar t\gamma$ form factors and the
               top anomalous magnetic moment?
  
  \item How well can one measure the $(V+A)$ decay of the top quark?
           
  \item What ambiguities arise when one fits for more than one anomalous
             coupling at a time?  Can polarization or spin correlation 
             measurements resolve these ambiguities?

\end{enumerate}

\subsection{QCD and two-photon physics}

For further information on this section, consult with: Bruce Schumm,
 Lynne Orr.

\begin{enumerate}

\item  What is the precision that can be obtained for 
$\alpha_s$ from $\ee$ annihilation?  In particular,
can
  it be definitively demonstrated that detector systematics are less than
  $\pm 1\%$?

   \item  What is the precision that can be obtained for $\alpha_s$ from 
              measurements on the top quark?
 \item  Outline the program for obtaining the photon structure
        functions.  What energies of operation are desired, and are
         special beam conditions required?

   \item  How can the LC make definitive studies of all-orders BFKL
        resummation?

\end{enumerate}

\subsection{Precision electroweak measurements}

For further information on this section, consult with: Lawrence Gibbons,
       Bill Marciano.

\begin{enumerate}

  \item Evaluate the need for Giga-Z in various scenarios in which there
            do or do not exist light Higgs particles.

   \item Evaluate the need for Giga-Z in scenarios in which new light 
              particles from supersymmetry or other new physics are discovered.

 \item  Are there strategies for further improving 
             the precision for measuring $\sstw$ using
        $Z$-pole observables?  How can the various systematics limits
           described in the text be avoided?

 \item  Evaluate the precision of $W$ and top quark mass measurements.  What
         special measurements of the accelerator parameters will be needed
            to achieve this precision?

 \item   What are the systematic limits on $B$ physics measurements, including
         CKM parameters and  rare $B$ decay rates, at a polarized
           $Z$ factory?

\end{enumerate}

\section{Accelerator issues}

\subsection{Running scenarios}

For further information on this section, consult with: Joel Butler, 
Paul Grannis, Michael Peskin.

\begin{enumerate}
 \item    What elements should be present in a charge to a future
    international technical panel established to compare linear
    collider technical proposals?  What emphasis should be given to
    risk analysis, needed R\&D, upgradability in energy or luminosity,
    cost comparison?

\item     For a physics-rich scenario (\eg, low mass Higgs and SUSY with
    observable $\neu1$, $\neu2$, $\ch1$, $\s t$, $\s \tau$) outline the
    desired run plan,  giving  the required
         integrated luminosity for all necessary beam
    energies, beam polarizations, beam particles.   What compromises
    can be envisioned to limit the number of distinct machine parameters
    without undue effect on the physics results?

\item  Do the same for a thinner physics scenario (\eg, with  Higgs mass of 180
    GeV and no supersymmetry or other new particle observation).
\end{enumerate}

\subsection{Machine configuration}

For further information on this section, consult with: Charles Prescott, 
     Tor Raubenheimer, Andre Turcot.

\begin{enumerate}

 \item  Evaluate an IR scheme with IR1 capable of operation
        at $E_{CM} \leq 250$ GeV and IR2 capable of operation at
        $E_{CM} < 500$ (1000) GeV.  Contrast this configuration
        with one in which two detectors share an IR in push-pull
        mode.

  \item  How important is it that the LEIR be able to operate
            at energies of 500 GeV or higher?

  \item  Evaluate the benefits from simultaneous operations
        at two IRs (with interleaved pulse trains).  What are the 
        constraints on the collider design?

  \item What are the requirements imposed on the first-phase
        accelerator design to permit upgrade to multi-TeV energies?

  \item  What constraints and opportunities are brought by including a
        free electron laser facility with the NLC?  Are there other
        non-HEP uses of the linear accelerator that could be contemplated?

\end{enumerate}
\subsection{Positron polarization}

For further information on this section, consult with: John Jaros, 
        Steve Mrenna, Mike Woods.
\begin{enumerate}

  \item  Evaluate the need for positron polarization in accomplishing
        the physics program.  What polarization (and error), energy
        (and error), luminosity are required for the relevant physics
          topics?
\end{enumerate}
\subsection{Photon collider}

For further information on this section, consult with: Jeff Gronberg, Adam
    Para, Tom Rizzo, Karl van Bibber.
\begin{enumerate}

  \item  Compile the list of physics topics for which $\gamma\gamma$
        operation is essential or desirable.

 \item Typically $\gamma\gamma$ luminosity and $e\gamma$
                 luminosity are comparable
   at a $\gamma\gamma$ collider. Identify $e\gamma$ processes that might be
            problematic
   backgrounds for $\gamma\gamma$ physics analyses.

   \item  How can a detector be made compatible with both  $\gamma\gamma$ and
        $\ee$ operation?

  \item Is it sufficient to provide $\gamma\gamma$ collisions only for
       $ E_{CM}(\gamma \gamma) < 400$ GeV (\ie, at the low energy IR)?

 \item  Evaluate the prospects for high-power lasers and the configuration of 
            the $\gamma\gamma$ IR.  Is  R\&D needed on the most important
             IR components (\eg, mirrors, masking, beam stability)?

\end{enumerate}
\subsection{$e^-e^-$}

For further information on this section, consult with: Jonathan Feng,
        Clem Heusch.
\begin{enumerate}
        
  \item  Compile the list of physics topics for which $e^-e^-$
        operation is essential or desirable.

\end{enumerate}

\subsection{Fixed Target}

For further information on this section, consult with: Mike Woods.

\begin{enumerate}

\item
 What experiments could be done using the $e^-$ or $e^+$ beam
         of a linear collider for fixed target experiments?
    For example, can
          M\o ller scattering of a fixed target beam be used to obtain
         $\sstw$ with very high precision? Can the spent
          beams that have passed through the interaction region be used
          in these experiments?

\item
 What are the relative
          advantages of $e^-$ {\em vs.} $e^+$ beams?

\item
   What experiments could be done using the polarized $\gamma$
             beams from laser backscattering for fixed target experiments?
          Can fixed target experiments be done with the spent beams
                 while the collider  is operating in  $\gamma\gamma$ mode?
\end{enumerate}

\section{Detector issues}

\subsection{Detectors}

For further information on this section, consult with: Jim Brau, 
Marty Breidenbach, Gene Fisk, Ray Frey, Tom Markiewicz, Keith Riles.
      
\begin{enumerate}

  \item  What are the physics reasons for wanting exceptional jet
        energy (mass) resolution?  How do signal/backgrounds and
        sensitivities vary as a function of resolution?  Is mass
        discrimination of $W$ and $Z$ in the dijet decay mode feasible,
        and necessary?

  \item  How does energy flow calorimetry resolution depend on such
        variables as Moliere radius, $\Delta \theta/ \Delta\phi$
        segmentation, depth segmentation, inner radius, $B$ field,
        number of radiation lengths in tracker, etc.?

  \item  What benefits arise from very high-precision tracking (\eg,
        silicon strip tracker)? What are the limitations imposed by
        having relatively few samples, and by the associated radiation
        budget?  What minimum radius tracker would be feasible?

  \item  Evaluate the dependence of physics performance on solenoidal
        field strength and radius.

\end{enumerate}

 \emptyheads
\blankpage \thispagestyle{empty}

\end{document}